\documentclass[usenatbib]{mnras}
\usepackage{graphicx}
\usepackage{subfig}
\usepackage{url}
\usepackage{amsmath}
\usepackage{array} 
\usepackage{threeparttable}
\usepackage{hyperref}
\usepackage{caption3}
\usepackage{pdflscape}
\usepackage[all]{hypcap}        
\usepackage[capitalise]{cleveref}
\crefformat{equation}{Eq.~#2#1#3}
\usepackage{soul}
\usepackage{xcolor} 
\definecolor{darkblue}{rgb}{0,0,0.5}
\hypersetup{
  citecolor    = darkblue   
}
\graphicspath{{figures/}} 

\newcommand{\Ha}{H$\alpha$}
\newcommand{\kms}{km~s$^{-1}$}                         
\newcommand{\simlt}{{\small\raisebox{-0.6ex}{$\,\stackrel{\raisebox{-.2ex}{$\textstyle <$}}{\sim}\,$}}}
\newcommand{\simgt}{{\small\raisebox{-0.6ex}{$\,\stackrel{\raisebox{-.2ex}{$\textstyle >$}}{\sim}\,$}}}

\newcommand{\msun}{\mbox{\,M$_\odot$}}        
\newcommand{\hst}{{\em HST}}
\newcommand{\fl}{F$_\text{light}$}
\newcommand{\sex}{{\sc SExtractor}}

\defcitealias{fruchter06}{F06}
\defcitealias{svensson10}{S10}
\defcitealias{bloom02}{BKD02}
\defcitealias{blanchard16}{BBF16}

\begin{document}

\title[LGRB hosts with HST]{The host galaxies and explosion sites of long-duration gamma ray bursts: {\em Hubble Space Telescope} near-infrared imaging}

\author[Lyman et al.]
{\parbox{\textwidth}{J. D. Lyman$^1$\thanks{E-mail: J.D.Lyman@warwick.ac.uk},
		     A. J. Levan$^1$,
		     N. R. Tanvir$^2$,
		     J. P. U. Fynbo$^3$,
		     J. T. W. M$^\text{c}$Guire$^2$,
		     D. A. Perley$^3$,
		     C. R. Angus$^1$,
		     J. S. Bloom$^4$,
		     C. J. Conselice$^5$,
		     A. S. Fruchter$^6$,
		     J. Hjorth$^3$,
		     P. Jakobsson$^7$,
		     R. L. C. Starling$^2$     
		     }
\vspace{0.4cm}\\
$^1$Department of Physics, University of Warwick, Coventry CV4 7AL, UK\\
$^2$Department of Physics and Astronomy, University of Leicester, Leicester, LE1 7RH, UK\\
$^3$Dark Cosmology Centre, Niels Bohr Institute, University of Copenhagen, Juliane Maries Vej 30, 2100 K\o benhavn \O, Denmark\\
$^4$Department of Astronomy, University of California, Berkeley, CA 94720, USA\\
$^5$School of Physics \& Astronomy, The University of Nottingham, University Park, Nottingham, NG7 2RD, UK\\
$^6$Space Telescope Science Institute, 3700 San Martin Drive, Baltimore, MD 21218, USA\\
$^7$Centre for Astrophysics and Cosmology, Science Institute, University of Iceland, Dunhagi 5, 107 Reykjavík, Iceland\\
}

\date{Accepted . Received ; in original form }

\pagerange{\pageref{firstpage}--\pageref{lastpage}} \pubyear{2016}

\maketitle

\label{firstpage}

\begin{abstract}
We present the results of a {\em Hubble Space Telescope} WFC3/F160W SNAPSHOT survey of the host galaxies of 39 long-duration gamma-ray bursts (LGRBs) at $z < 3$. We have non-detections of hosts at the locations of 4 bursts. Sufficient accuracy to astrometrically align optical afterglow images and determine the location of the LGRB within its host was possible for 31/35 detected hosts. In agreement with other work, we find the luminosity distribution of LGRB hosts is significantly fainter than that of a star formation rate-weighted field galaxy sample over the same redshift range, indicating LGRBs are not unbiasedly tracing the star formation rate. Morphologically, the sample of LGRB hosts are dominated by spiral-like or irregular galaxies. 
We find evidence for evolution of the population of LGRB hosts towards lower-luminosity, higher concentrated hosts at lower redshifts.
Their half-light radii are consistent with other LGRB host samples where measurements were made on rest-frame UV observations. In agreement with recent work, we find their 80~per~cent enclosed flux radii distribution to be more extended than previously thought, making them intermediate between core-collapse supernova (CCSN) and super-luminous supernova (SLSN) hosts. The galactocentric projected-offset distribution confirms LGRBs as centrally concentrated, much more so than CCSNe and similar to SLSNe. LGRBs are strongly biased towards the brighter regions in their host light distributions, regardless of their offset. We find a correlation between the luminosity of the LGRB explosion site and the intrinsic column density, N$_\text{H}$, towards the burst.
\end{abstract}


\section{Introduction}
\label{sect:intro}

Long-duration gamma-ray bursts (LGRBs) mark the spectacularly violent deaths of massive stars. Distinguished observationally from ordinary core-collapse supernovae (\mbox{CCSNe}), the star's death is accompanied by an intense burst of high energy photons lasting generally from a few seconds to minutes, followed by a power-law decay of radiation over a wide range of frequencies (X-ray/UV -- Radio), these are named the `prompt emission' and `afterglow' respectively. The association of LGRBs and the deaths of massive stars has been well established through observations of spatially and temporally coincident LGRBs and CCSNe. In a handful of cases there has been a spectroscopic confirmation of the accompanying SN \citep[e.g.][]{galama98,hjorth03,stanek03,malesani04,pian06,starling11,berger11,schulze14} with a growing number of cases where a photometric rebrightening of the decaying afterglow is consistent with a SN \citep[see][for reviews of the GRB-SN connection]{hjorth12, cano16}. In each case where a SN accompanying a LGRB has been classified, it has been a broad-lined SN~Ic (SN~Ic-BL). The mechanism producing LGRBs is generally accepted as the collapsar model \citep{woosley93,macfadyen99}, wherein a collapsing massive star produces a central engine capable of powering bipolar relativistic jets that penetrate the outer layers of the star, emerging to produce the eponymous gamma-rays. Beyond being massive stars stripped of their envelopes, the conditions of the progenitor required to produce a LGRB remain poorly defined \citep[][for a review]{levan16} -- for example it remains unclear whether LGRBs can be produced in star formation episodes at all metallicities. This uncertainty in the nature of the progenitors limits their potential use as cosmic beacons of star formation across the Universe. The transient nature of LGRBs makes follow up observations difficult -- even within the era of accurate and early localisation of the explosion thanks to dedicated high-energy observatories. As with other transients, further clues to their nature have been gleaned through analyses of their hosts and host-environments. 

The hosts of LGRBs are predominantly young and exclusively star-forming galaxies \citep[e.g.][]{christensen04,levesque14}, consistent with the picture of massive star progenitors. They are distinguished from field galaxy samples as being compact \citep{fruchter06,svensson10,kelly14}, although they still follow the size-luminosity relation of field galaxies \citep{wainwright07} as they also tend to be less luminous \citep[e.g.][]{lefloch03, perley16}. The low luminosities are indicative of low stellar mass \citep[e.g.][]{savaglio09,castro10,vergani15} and low metallicity \citep[e.g.][]{savaglio09,levesque10a} hosts compared to field galaxy samples. Analysis of $z \sim 6$ LGRB hosts based on the three recent detections of \citet{mcguire16} suggests they are compact and low metallicity, and comparable to Lyman-break galaxy samples at a similar redshift.
The underlying LGRB host-population, however, is not comparable to a volume-limited field galaxy population as the rate of production is expected to be linked to the ongoing SFR of a galaxy and so comparison must be made to hosts of other SFR-weighted tracers \citep[e.g. CCSNe][]{fruchter06,svensson10} or a SFR-weighted field sample. A study by \citet{fynbo08d}, based on observations of LGRB damped Ly$\alpha$ systems, found that their metallicity distribution is consistent with a scenario where LGRBs are drawn uniformly from a SFR-weighted luminosity function of star-forming galaxies at $z \sim 3$. Studies of $z < 1$ LGRBs however show their rates in massive, metal-rich hosts are significantly depressed \citep{stanek06,graham13,vergani15}, indicating that LGRB production is strongly suppressed at metallicities above $0.3-1 Z_\odot$ \citep[e.g.][]{kruhler15,graham15,perley16b}.

Beyond integrated host galaxy properties, high-resolution imaging coupled with good enough localisation of the bursts allows further investigations into the explosion sites of LGRBs, and the nature of these explosion sites within their hosts' morphologies and light profiles.

\citet{bloom02} investigated the radial offsets of LGRBs and found the offsets to trace well the hosts' UV light distributions. This association was discrepant with that expected for (then) potential GRB progenitor systems involving delayed mergers of compact BH-NS or NS-NS binaries\footnote{Such compact binary mergers are now the favoured progenitor model for short-duration gamma-ray bursts \citep[e.g. see reviews of][]{berger14, levan16}.}, but is in good agreement with expectations from the collapsar model. The offset distribution of LGRBs shows them to be more centrally concentrated than SNe type II, although consistent with the more centrally concentrated stripped-enveloped (type Ib/c and Ic-BL) subtypes of CCSNe \citet[][hereafter \citetalias{blanchard16}]{blanchard16}.

Statistical studies of the association between transients and their host light distribution can provide further constraints on the nature of GRB \citep[][hereafter \citetalias{fruchter06}]{fruchter06} and SN progenitors  \citep{james06} -- the formalism of these pixel-based diagnostics is presented in \cref{sect:pixstat}. For example, a statistical association between the locations of a transient type and star formation tracing light of their hosts strongly argues for short-lived, and therefore probably high-mass, progenitors. Conversely a lack of a statistical association would imply longer-lived and therefore probably lower-mass progenitors. LGRBs are more strongly associated with the UV light of their hosts compared to CCSNe (\citetalias{fruchter06}, \citealt{svensson10} [hereafter \citetalias{svensson10}]), pointing towards even more massive progenitors than CCSNe. When splitting by CCSN type, \citet{kelly08} find that LGRBs and SNe~Ic and SNe~Ic-BL have a similar very strong association to the $g$-band flux of their hosts, with other CCSN types displaying a lower degree of association. These similarities in the explosion sites of large samples of SNe~Ic-BL and LGRBs extends the evidence for an association between the two to include the more distant events, where direct observations of the accompanying SNe are infeasible. This strong bias for LGRBs to explode in the brightest regions of their hosts has been recently questioned by the analysis of \citep{blanchard16}. These authors argue the association is not as strong as previously found, and in particular that LGRBs offset from the central regions of their hosts have no preference for bright regions. They argue that the strong association between LGRBs and bright regions of their hosts is thus driven entirely by the small offset bursts.

In this paper we present the results of a \hst{} NIR imaging survey of the hosts of 39 LGRB hosts, performing astrometric alignment of the bursts' optical afterglows to additionally investigate the explosion sites of the GRBs within their hosts. Throughout the paper we adopt the cosmological parameters $H_0 = 73.24$~\kms{}~Mpc$^{-1}$ \citep{riess16} and $\Omega_\text{m} = 0.3$.

\section{Samples and Observations}
\label{sect:sample}

\begin{table*}
\begin{threeparttable}
 \caption{Sample of Long-GRBs. Where available, the redshifts in \citet{fynbo09} were used, superseding those in the original GCNs.}
 \begin{tabular}{llrlcl}
\hline
GRB     & Date Obs.  & Exp time (s)\tnote{a} & z    & Telescope\tnote{b}    & Refs \\
\hline
050315  & 2011-07-20 & 1209  & 1.949   & Blanco 4.0m/Mosaic II & \citet{kelson05} \\
050401  & 2010-10-01 & 1612  & 2.8983  & VLT/FORS2             & \citet{fynbo05a} \\
050824  & 2011-01-18 & 906   & 0.8278  & VLT/FORS2             & \citet{fynbo05b} \\
051016B & 2011-01-16 & 906   & 0.9364  & {\em Swift}/UVOT\tnote{c}   & \citet{soderberg05} \\ 
060124  & 2010-09-28 & 1612  & 2.297   & {\em Swift}/UVOT      & \citet{cenko06} \\
060218  & 2010-10-12 & 906   & 0.0334  & VLT/FORS2             & \citet{pian06} \\
060502A & 2010-10-11 & 1209  & 1.5026  & Gemini/GMOS-N         & \citet{cucchiara06} \\
060505  & 2011-08-03 & 906   & 0.089   & VLT/FORS1             & \citet{ofek06} \\
060602A & 2010-12-05 & 906   & 0.787   & NOT/ALFOSC            & \citet{jakobsson07a} \\
060614  & 2010-10-08 & 906   & 0.125   & VLT/FORS2             & \citet{price06} \\ 
060729  & 2010-09-15 & 906   & 0.5428  & \hst{}/ACS            & \citet{thoene06a} \\
060912A & 2011-09-23 & 906   & 0.937   & VLT/FORS1             & \citet{jakobsson06a} \\
061007  & 2011-07-08 & 1209  & 1.2622  & VLT/FORS1             & \citet{jakobsson06b} \\
061110A & 2010-09-30 & 906   & 0.7578  & VLT/FORS1             & \citet{thoene06b} \\
070318  & 2010-12-31 & 906   & 0.8397  & VLT/FORS1             & \citet{jaunsen07} \\
070521  & 2011-08-02 & 906   & 2.0865  &  --\tnote{c}          & \citet{kruhler15}\\ 
071010A & 2010-10-29 & 906   & 0.98    & VLT/FORS1             & \citet{prochaska07} \\
071010B & 2010-11-25 & 906   & 0.947   & Gemini/GMOS-N         & \citet{cenko07} \\
071031  & 2010-11-20 & 1612  & 2.6918  & VLT/FORS2             & \citet{ledoux07} \\ 
071112C & 2010-10-08 & 906   & 0.8227  & Gemini/GMOS-N         & \citet{jakobsson07b} \\
071122  & 2010-12-21 & 1209  & 1.14    & Gemini/GMOS-N         & \citet{cucchiara07} \\ 
080319C & 2010-09-19 & 1209  & 1.9492  & Gemini/GMOS-N         & \citet{wiersema08} \\
080430  & 2011-06-21 & 906   & 0.767   & NOT/MOSCA             & \citet{cucchiara08a} \\ 
080520  & 2011-02-08 & 1209  & 1.5457  & VLT/FORS2             & \citet{jakobsson08a} \\
080603B & 2011-08-06 & 1612  & 2.6892  & LT/RATCam             & \citet{fynbo08a} \\
080605  & 2012-02-22 & 1209  & 1.6403  & VLT/FORS2             & \citet{jakobsson08b} \\
080707  & 2010-10-31 & 1209  & 1.2322  & VLT/FORS1             & \citet{fynbo08b} \\
080710  & 2011-02-12 & 906   & 0.8454  & Gemini/GMOS-N         & \citet{perley08} \\ 
080805  & 2011-10-01 & 1209  & 1.5042  & VLT/FORS2             & \citet{jakobsson08c} \\
080916A & 2011-03-19 & 906   & 0.6887  & VLT/FORS1             & \citet{fynbo08c} \\
080928  & 2010-09-18 & 1209  & 1.6919  & VLT/FORS2             & \citet{vreeswijk08} \\ 
081007  & 2010-11-30 & 906   & 0.5295  & Gemini/GMOS-S         & \citet{berger08a} \\
081008  & 2011-09-04 & 1209  & 1.967   & Gemini/GMOS-S         & \citet{cucchiara08b} \\ 
081121  & 2011-01-04 & 1612  & 2.512   & {\em Swift}/UVOT\tnote{c} & \citet{berger08b} \\
090418A & 2010-10-02 & 1209  & 1.608   & Gemini/GMOS-N         & \citet{chornock09a} \\
090424  & 2011-05-03 & 906   & 0.544   & Gemini/GMOS-S         & \citet{chornock09b} \\
090618  & 2011-01-09 & 906   & 0.54    & WHT/ACAM              & \citet{cenko09}\\          
091127  & 2010-12-16 & 906   & 0.4903  & Gemini/GMOS-N         & \citet{cucchiara09} \\
091208B & 2010-10-10 & 1209  & 1.063   & Gemini/GMOS-N         & \citet{wiersema09} \\
\hline
\end{tabular}
\label{tab:sample}
\begin{tablenotes}
 \item [a] {Exposure time of \hst{} WFC3 F160W observations.}
 \item [b]{Telescope the afterglow imaging used for alignment was taken on.}
 \item [c]{No accurate alignment to the \hst{} image could be made (see \cref{sect:astrometry}).}
\end{tablenotes}
\end{threeparttable}
\end{table*}

The sample of bursts comprises those observed pseudo-randomly by the observing schedule from an initial list of targets in a SNAPSHOT \hst{} program (SNAP 12307, PI: Levan). The initial target list was formed from {\em Swift}-detected LGRBs\footnote{The nature of GRBs 060505 and 060614, and their place within the traditional short/long GRB dichotomy, has been questioned \citep[e.g.][]{fynbo06, dellavalle06, gehrels06, galyam06, zhang07}. We include these in our sample, however their exclusion does not significantly affect our results or discussion.} with a spectroscopic redshift of $z < 3$ and low Galactic extinction ($A_V < 0.5$~mag). The requirement of a spectroscopic redshift implies well-localised bursts with detected optical afterglows. The one exception in the observed sample is GRB 070521 -- for this burst the host assignment of \citet{perley09} was used by \citet{kruhler15} to obtain a spectroscopic redshift. We also adopt this host and its redshift as that of the burst.
The observations were taken at late times after the LGRB events and thus we expect no significant contamination from the GRB light in our sample. The rest-frame lag between the event and the \hst{} observations is more than a year for most events. The shortest is GRB 091208B, which had a lag of $\sim5$~months (rest-frame). There are cases of extremely luminous LGRBs with long-lasting afterglows (and associated SNe) that can be monitored for extended timescales, such as GRB 130427A \citep[e.g.][]{levan14,perley14}. An example in our sample is GRB 060729; despite its exceptionally long-lasting x-ray afterglow \citep{grupe10}, the optical/NIR afterglow faded below the detection limit of \hst{} F160W in less than 9 months \citep[rest-frame,][]{cano11}.
For the majority of LGRBs, the fading afterglow becomes negligible on much shorter timescales, and thus we do not consider the presence of any residual light to affect our results for the sample significantly.

All imaging was performed with the WFC3/IR instrument and filter F160W ($\lambda_\text{cen} \sim 15400$~\AA{}). These were drizzled using {\sc AstroDrizzle}\footnote{\url{http://drizzlepac.stsci.edu/}} to a pixel scale of 0.065 arcsec using \texttt{pixfrac} = 0.8, with all resulting weight images satisfying the pixel value rms/median $< 0.2$ criterion. For robust identification of the host and to perform environmental analyses of the locations of the GRBs within their hosts, follow-up observations of the optical afterglow were required. Optical afterglow imaging came from a wide variety of proposals by different groups on various telescopes acquired through the relevant data archives, where appropriate. The GRB names, date and exposure time of the \hst{} imaging, redshift of the burst, telescope used for the afterglow imaging and references are presented in \cref{tab:sample}. 

We show in \cref{fig:segmaps,fig:segmaps_noalign} stamps of the hosts. These stamps show visually the results of our analyses which are described in \cref{sect:methods}.

\begin{figure*}
\centering
\subfloat{\includegraphics[width=0.25\linewidth]{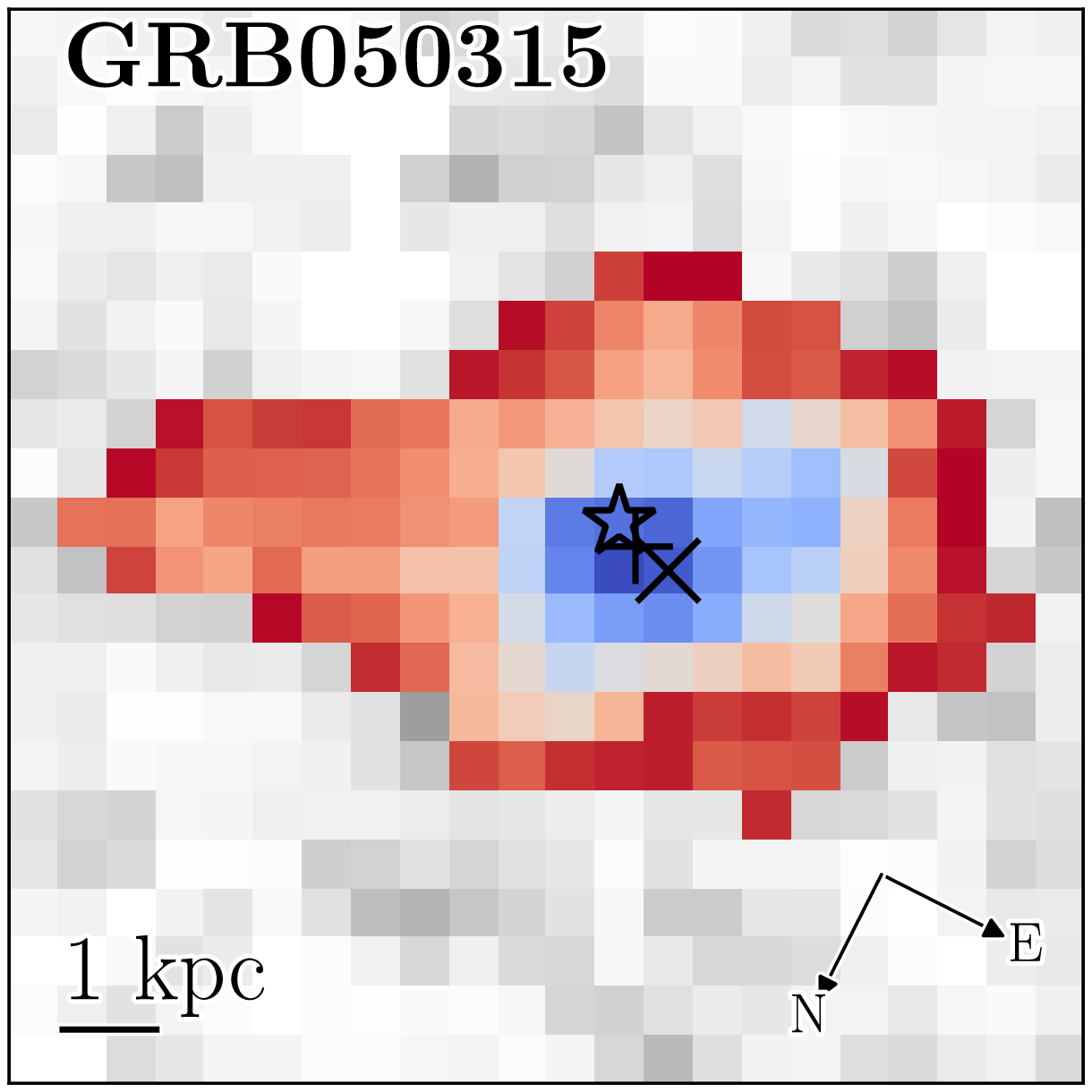}}\hspace*{-0.185cm}
\subfloat{\includegraphics[width=0.25\linewidth]{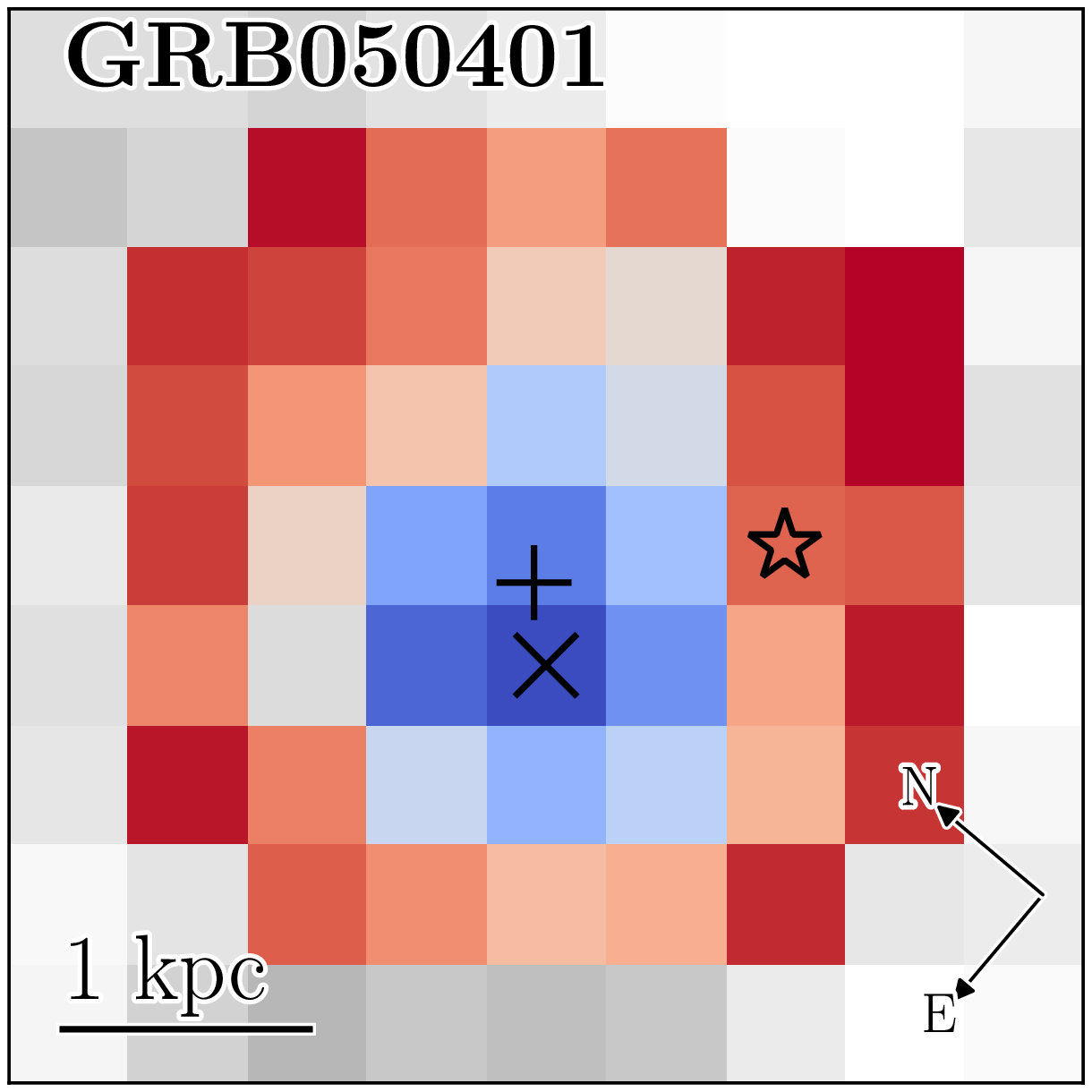}}\hspace*{-0.185cm}
\subfloat{\includegraphics[width=0.25\linewidth]{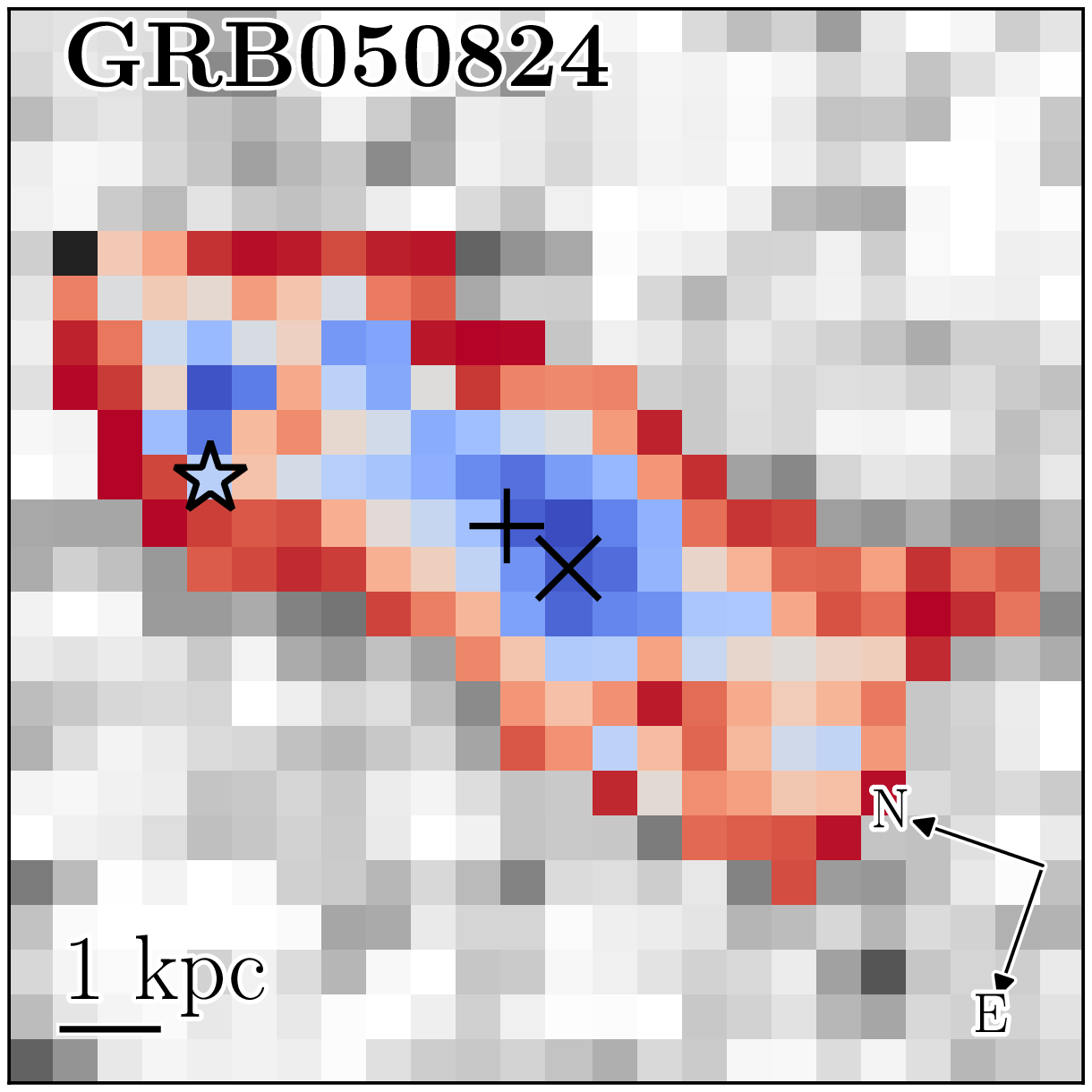}}\hspace*{-0.185cm}
\subfloat{\includegraphics[width=0.25\linewidth]{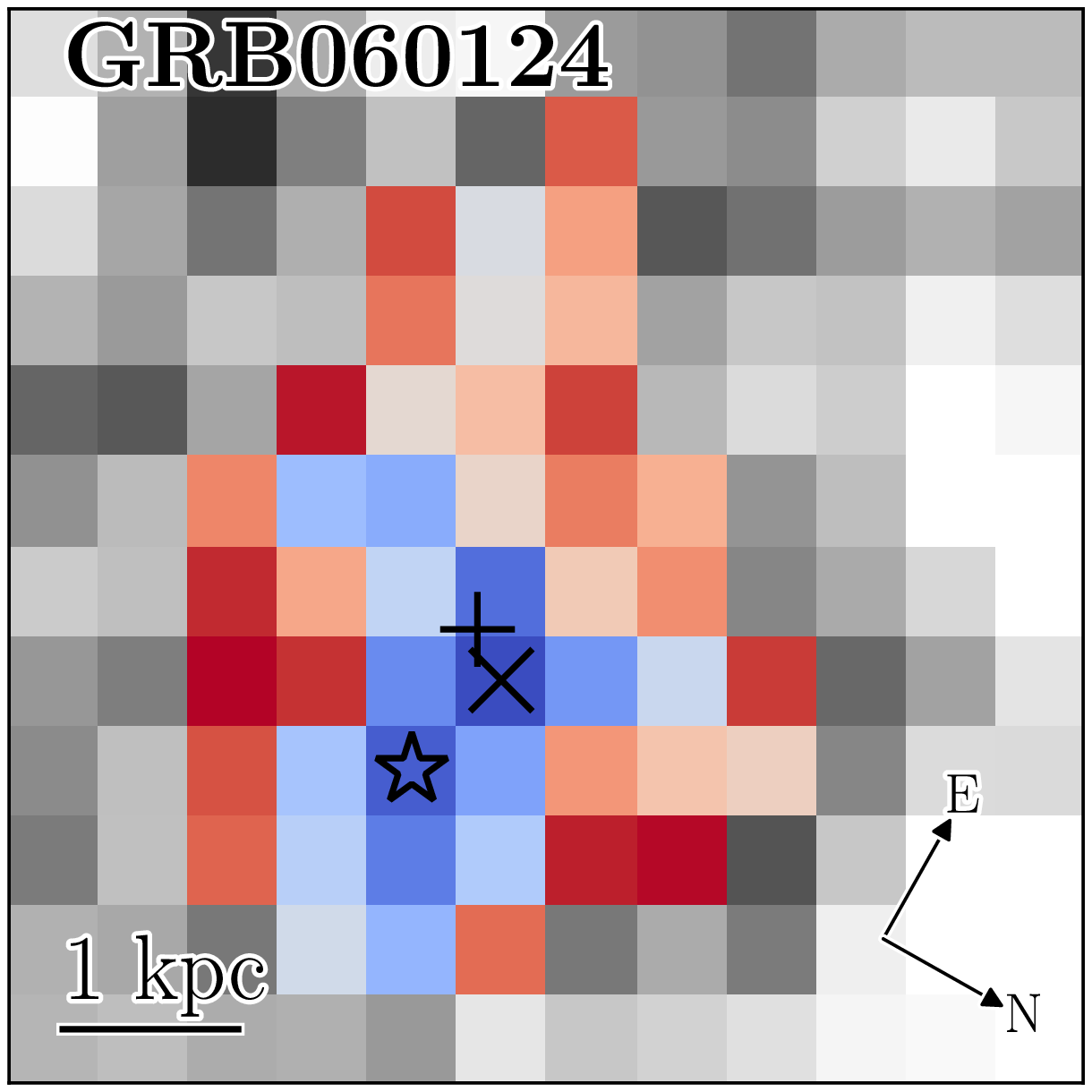}}\\
\vspace{-0.4cm}
\subfloat{\includegraphics[width=0.25\linewidth]{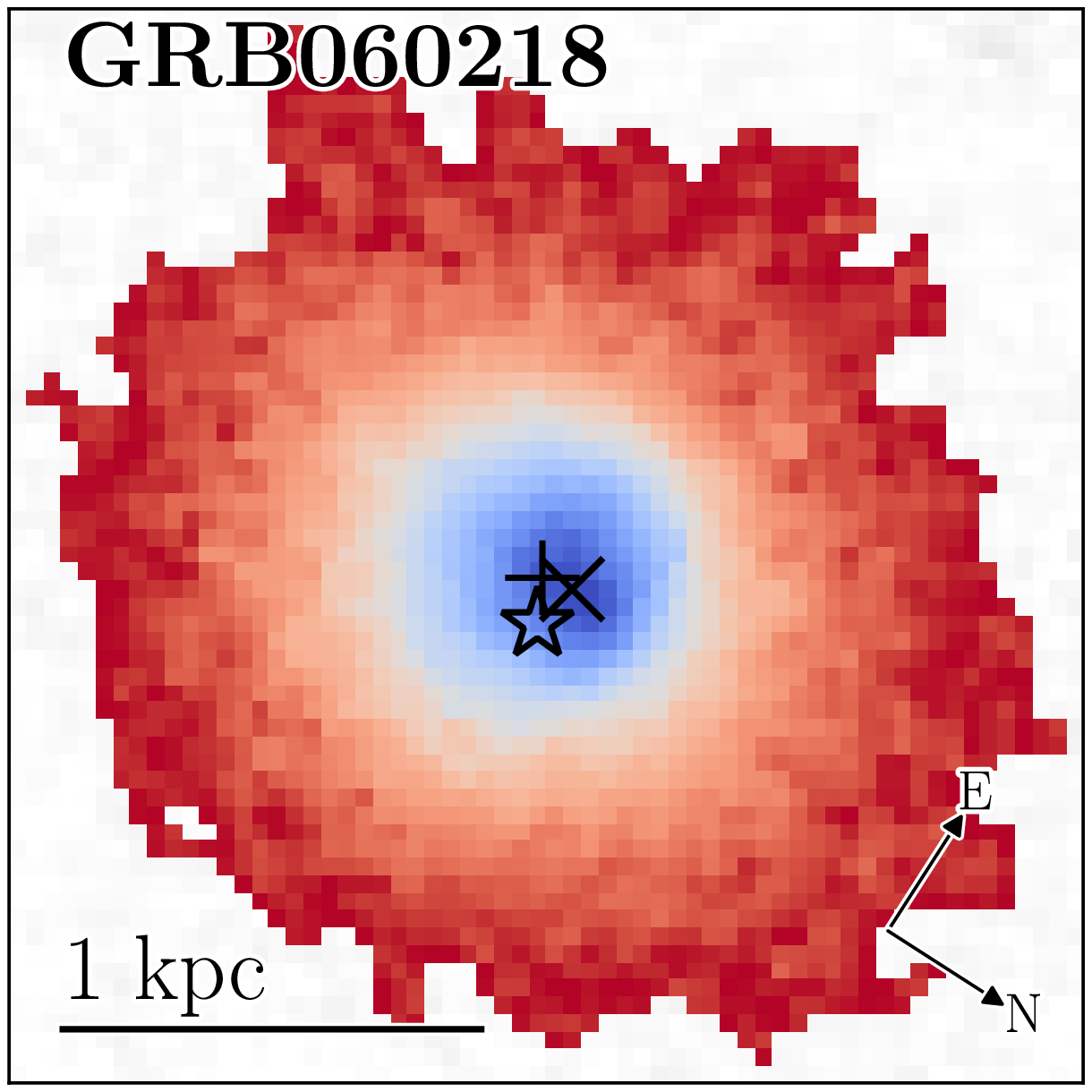}}\hspace*{-0.185cm}
\subfloat{\includegraphics[width=0.25\linewidth]{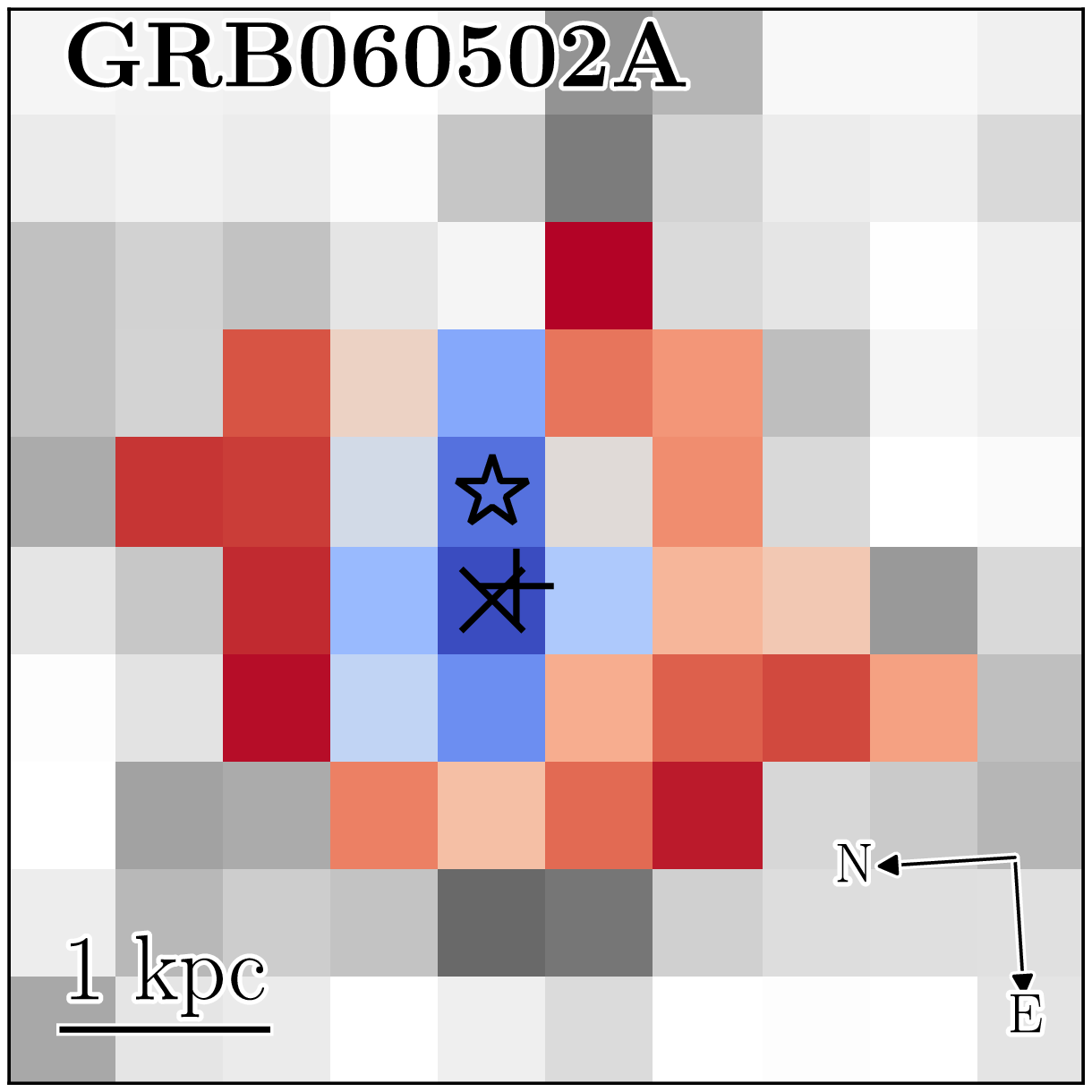}}\hspace*{-0.185cm}
\subfloat{\includegraphics[width=0.25\linewidth]{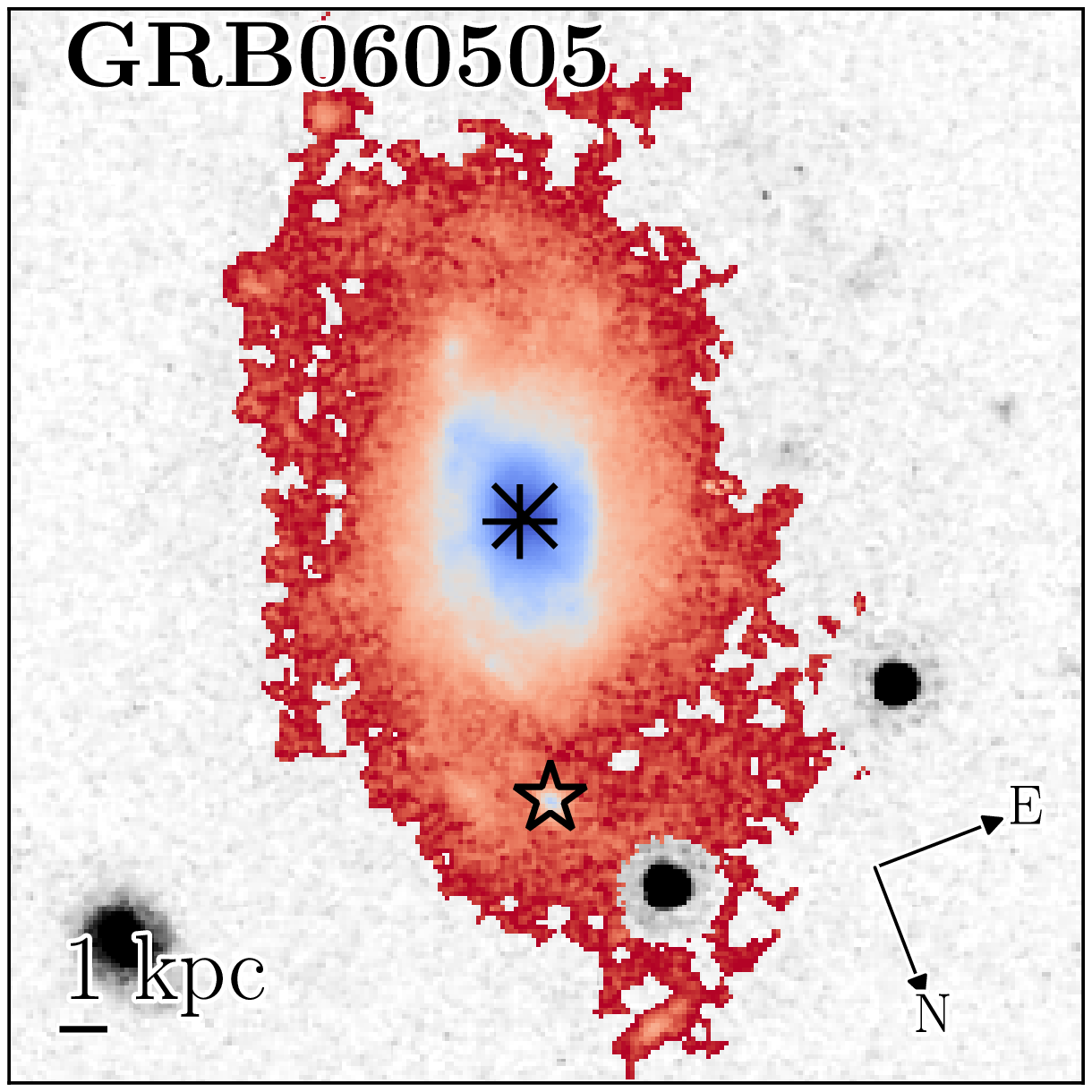}}\hspace*{-0.185cm}
\subfloat{\includegraphics[width=0.25\linewidth]{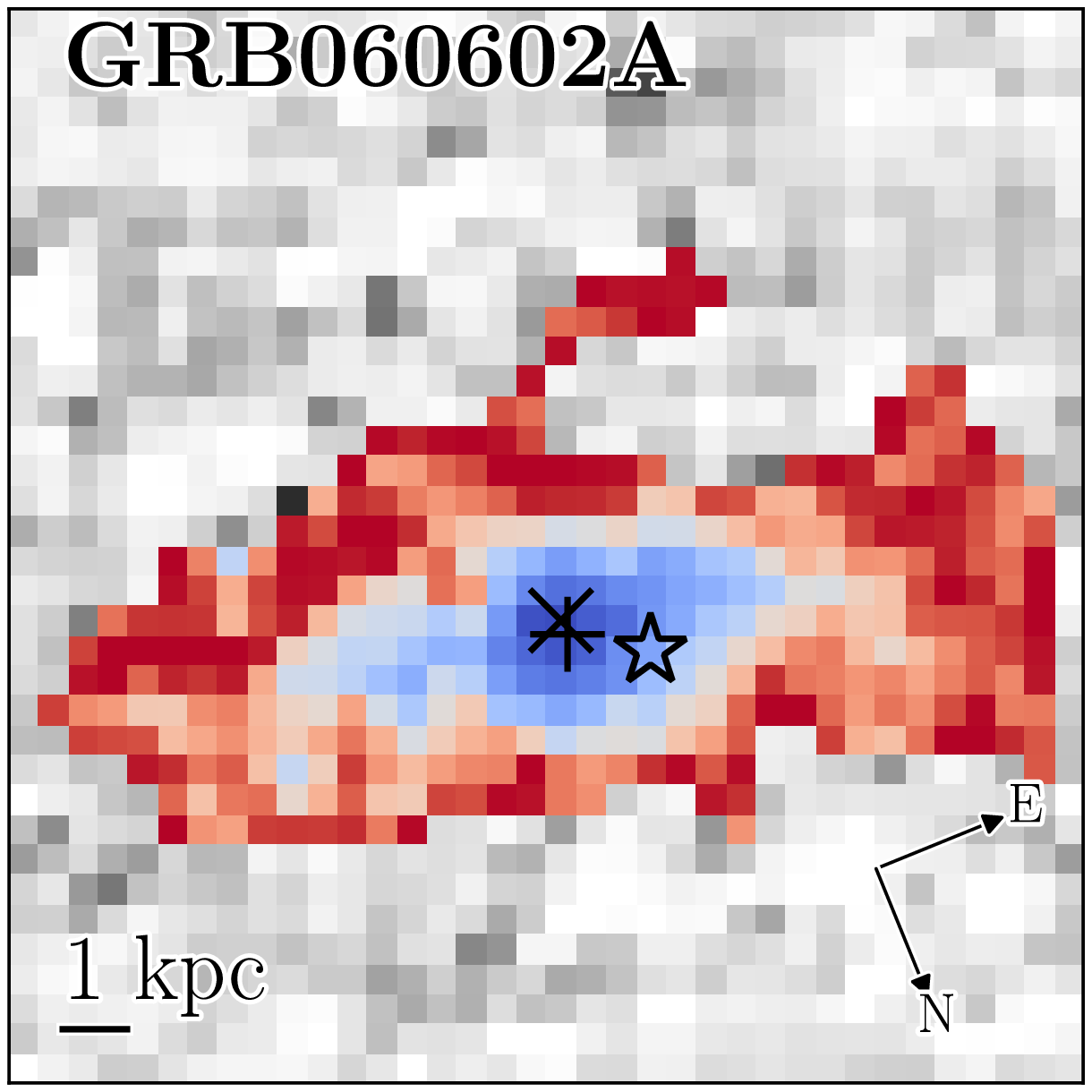}}\\
\vspace{-0.4cm}
\subfloat{\includegraphics[width=0.25\linewidth]{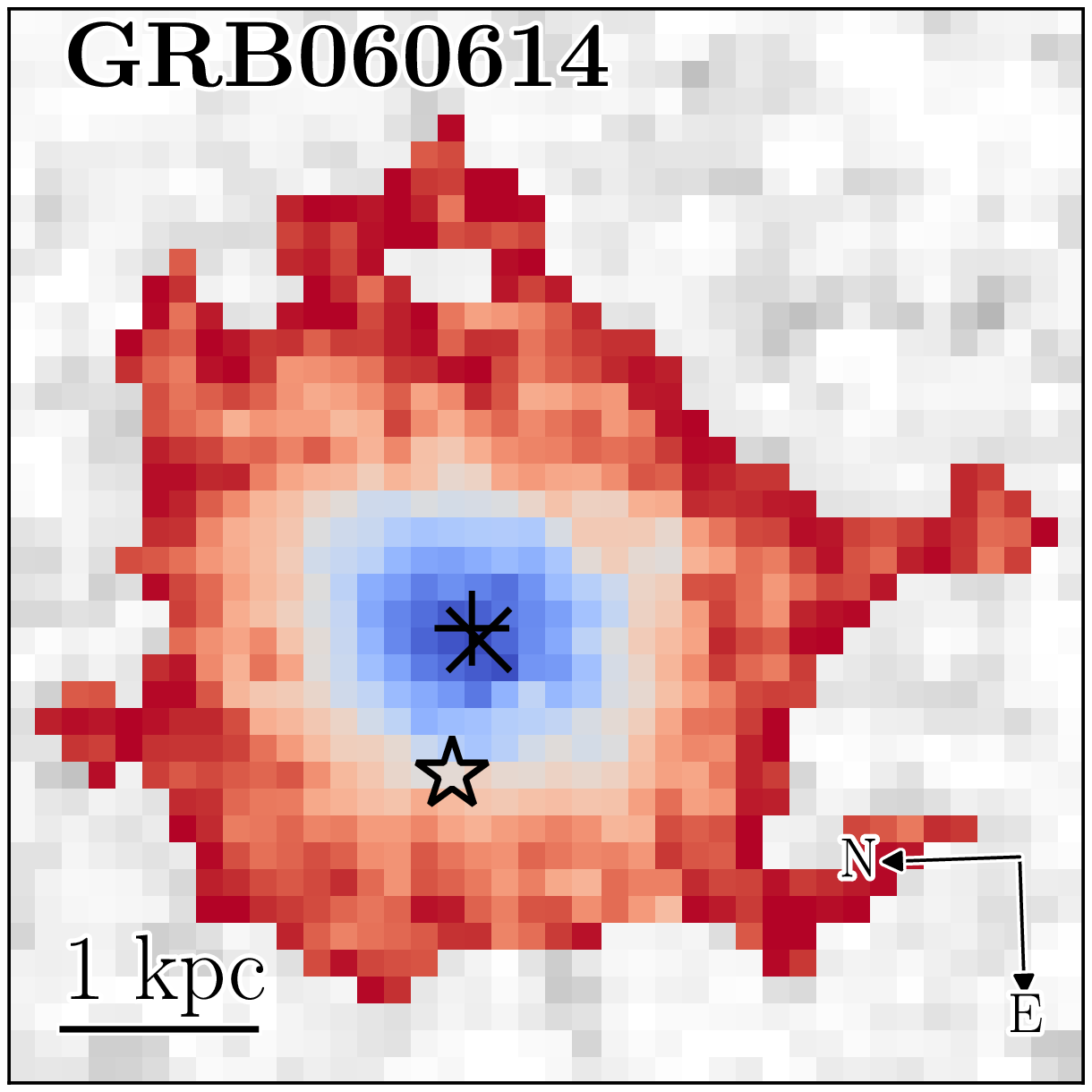}}\hspace*{-0.185cm}
\subfloat{\includegraphics[width=0.25\linewidth]{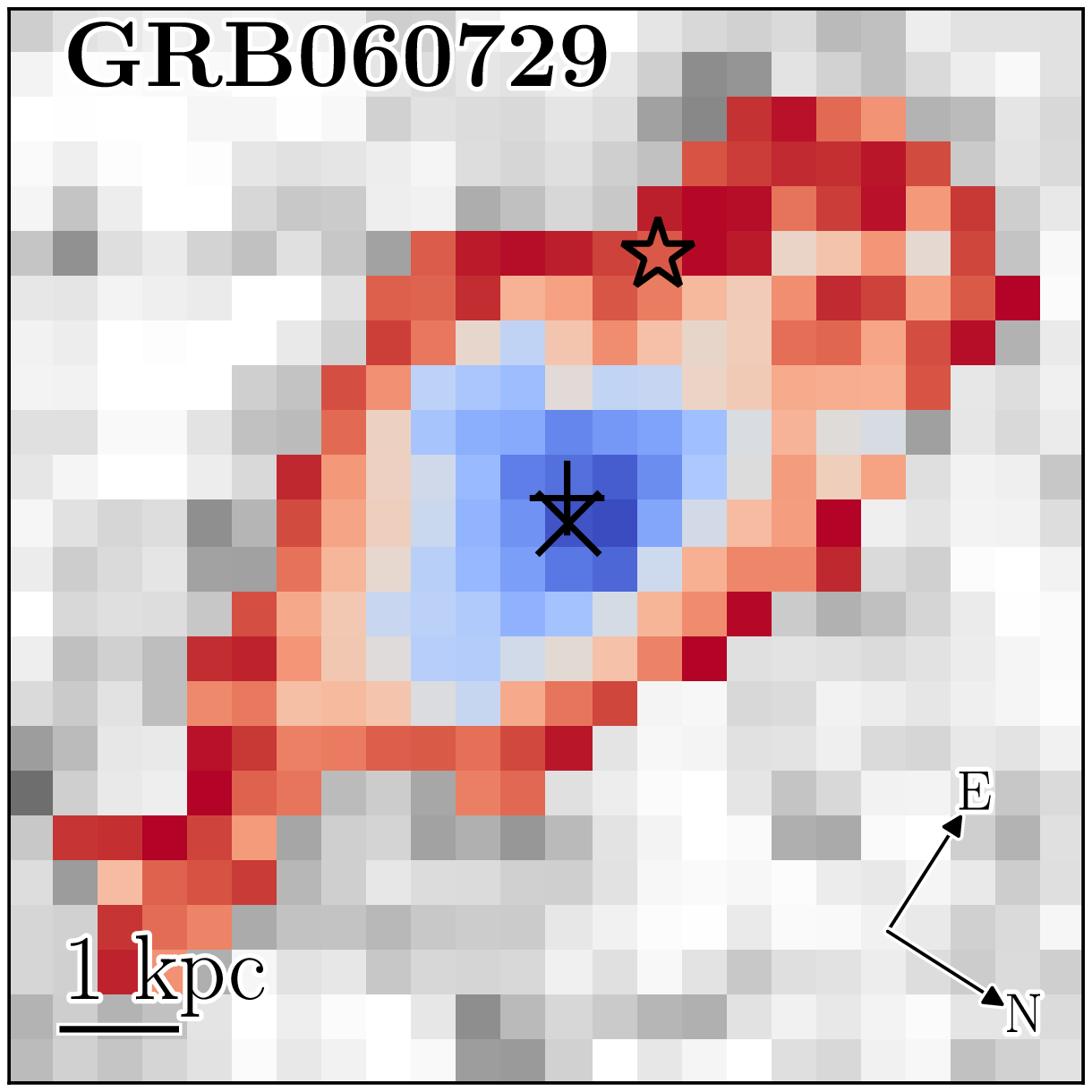}}\hspace*{-0.185cm}
\subfloat{\includegraphics[width=0.25\linewidth]{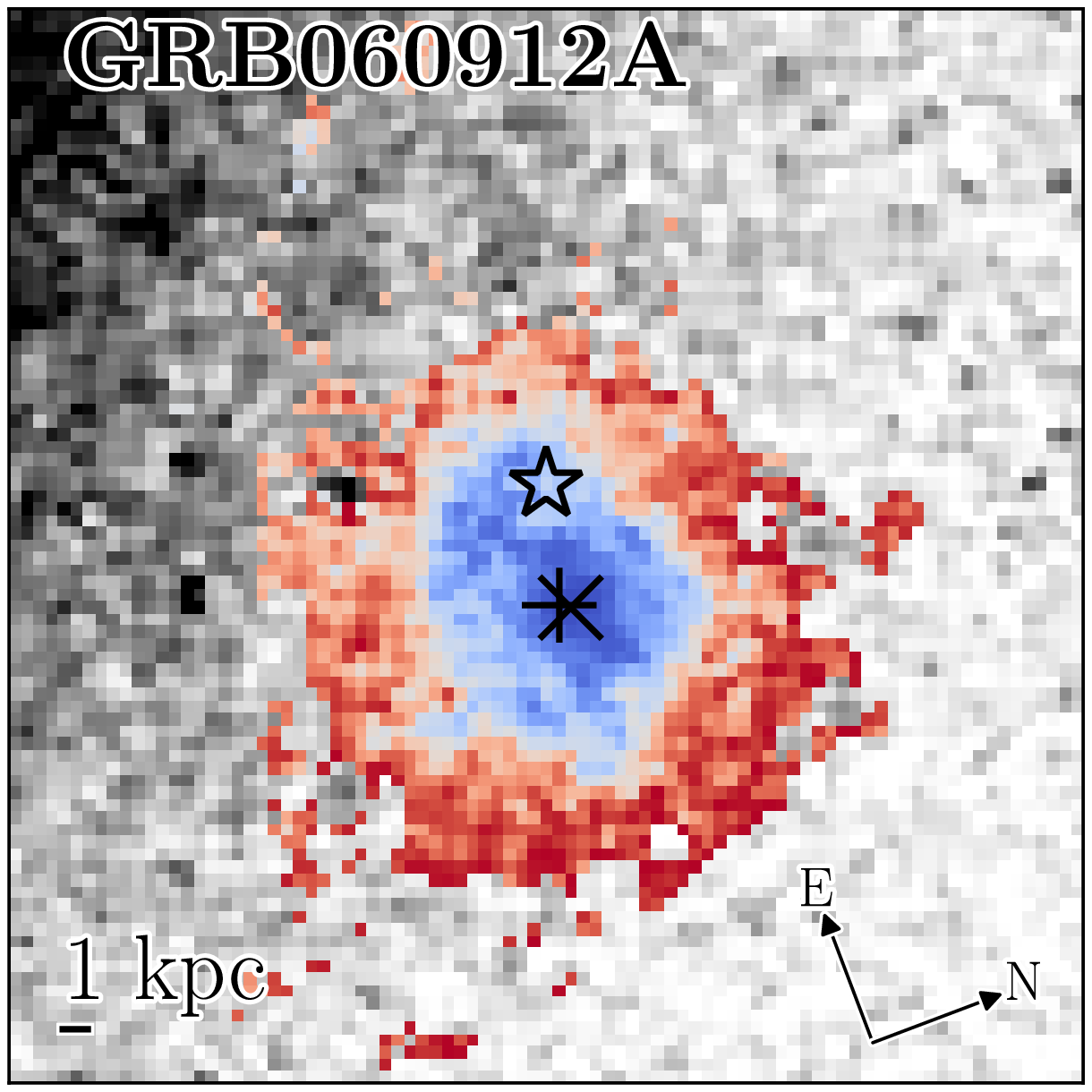}}\hspace*{-0.185cm}
\subfloat{\includegraphics[width=0.25\linewidth]{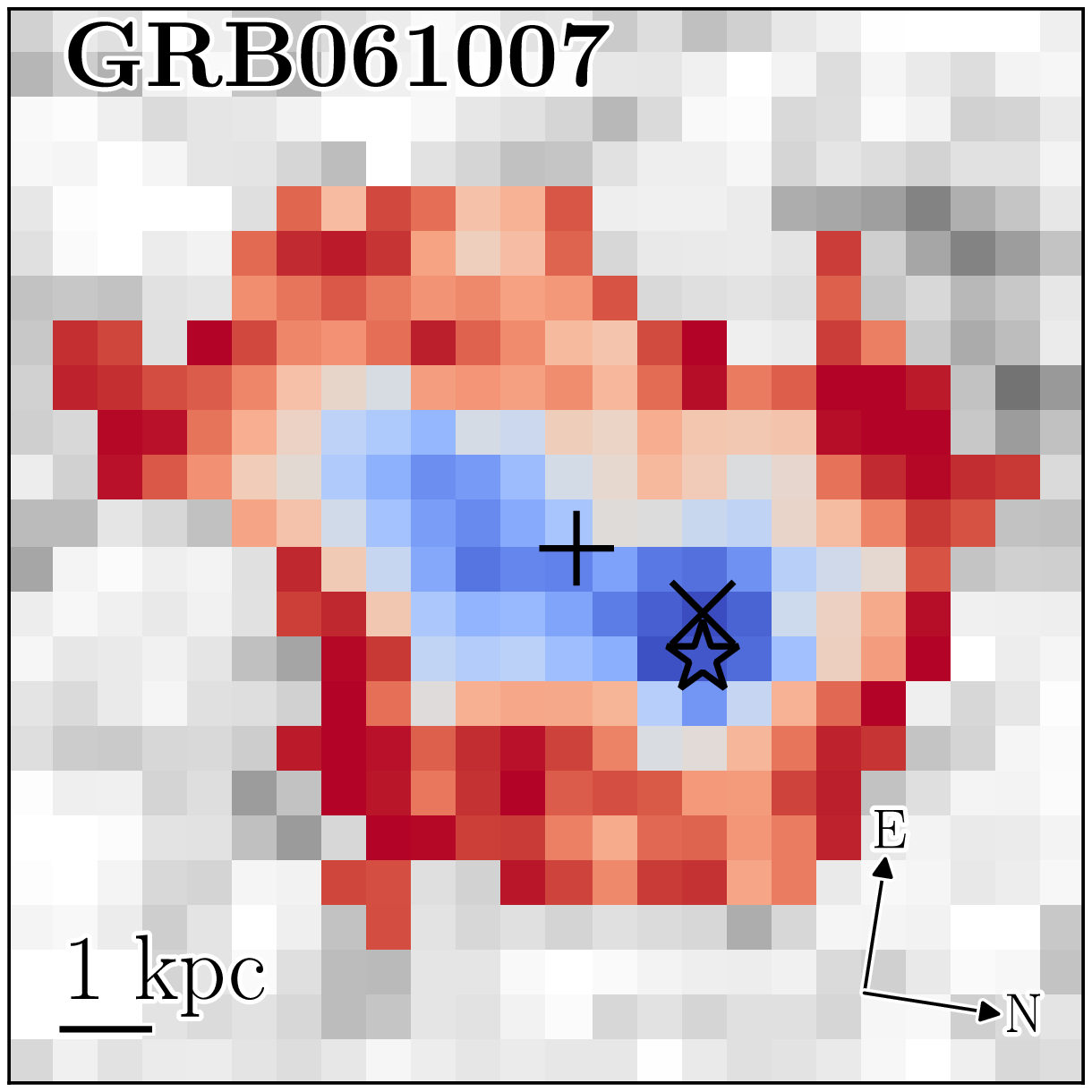}}\\
\vspace{-0.4cm}
\subfloat{\includegraphics[width=0.25\linewidth]{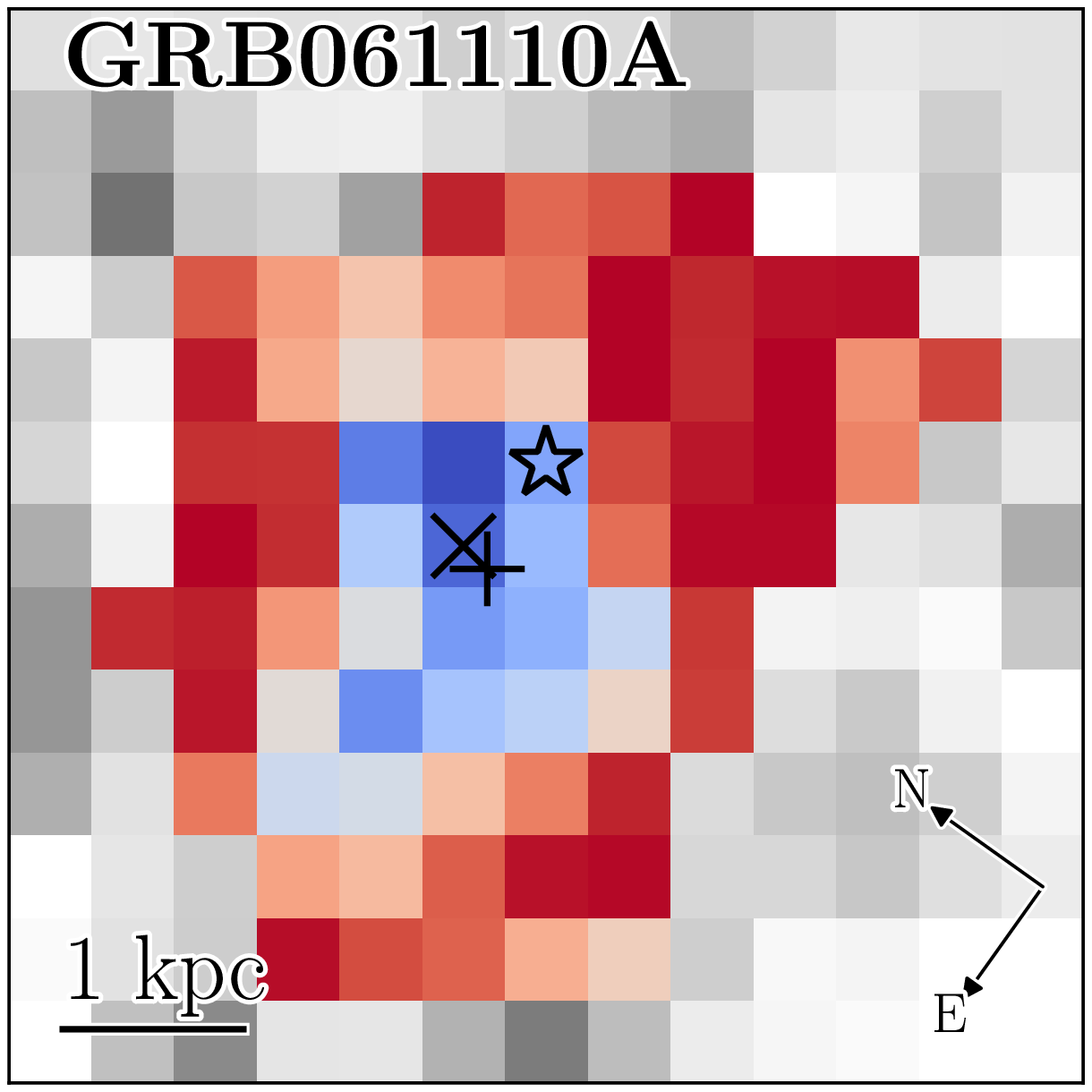}}\hspace*{-0.185cm}
\subfloat{\includegraphics[width=0.25\linewidth]{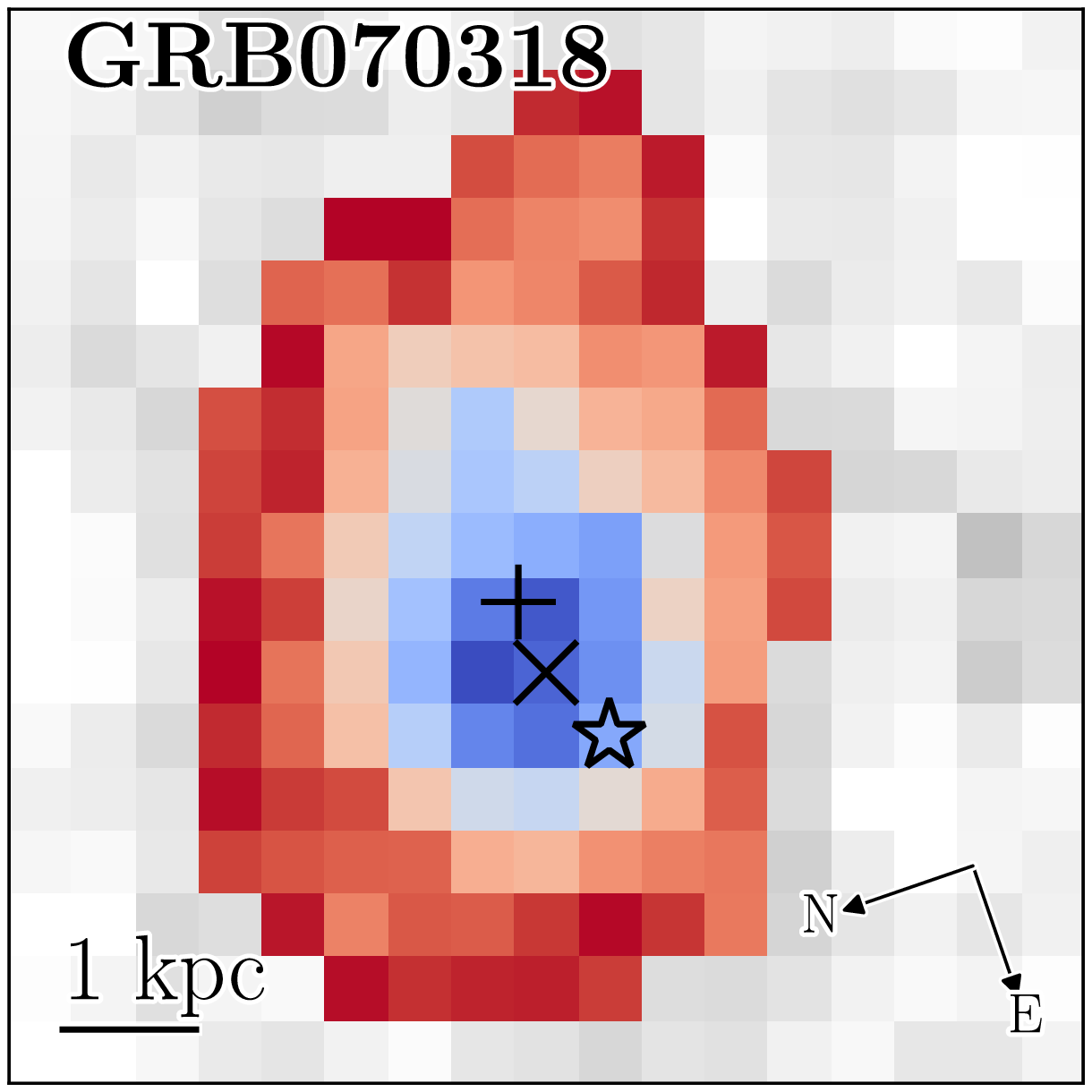}}\hspace*{-0.185cm}
\subfloat{\includegraphics[width=0.25\linewidth]{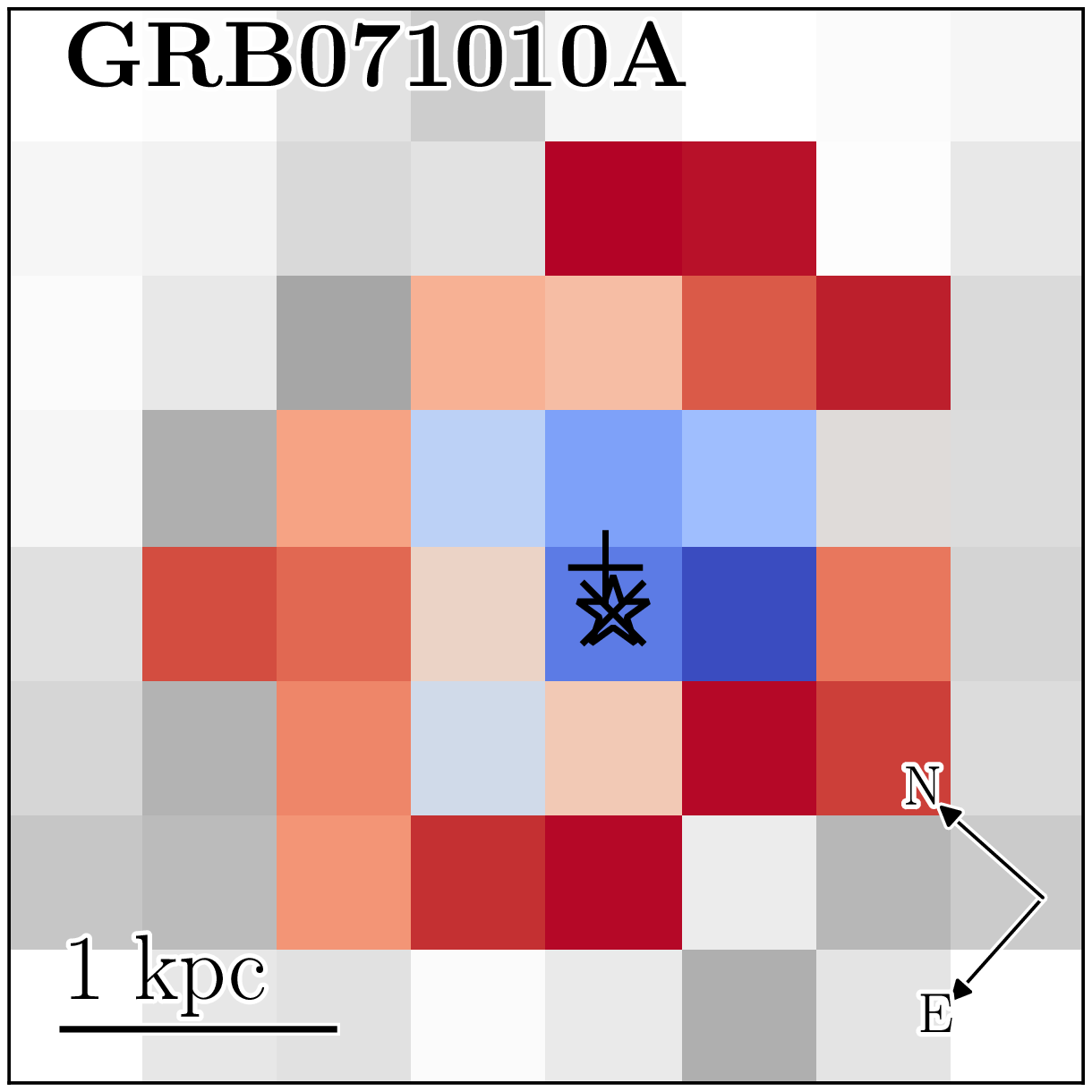}}\hspace*{-0.185cm}
\subfloat{\includegraphics[width=0.25\linewidth]{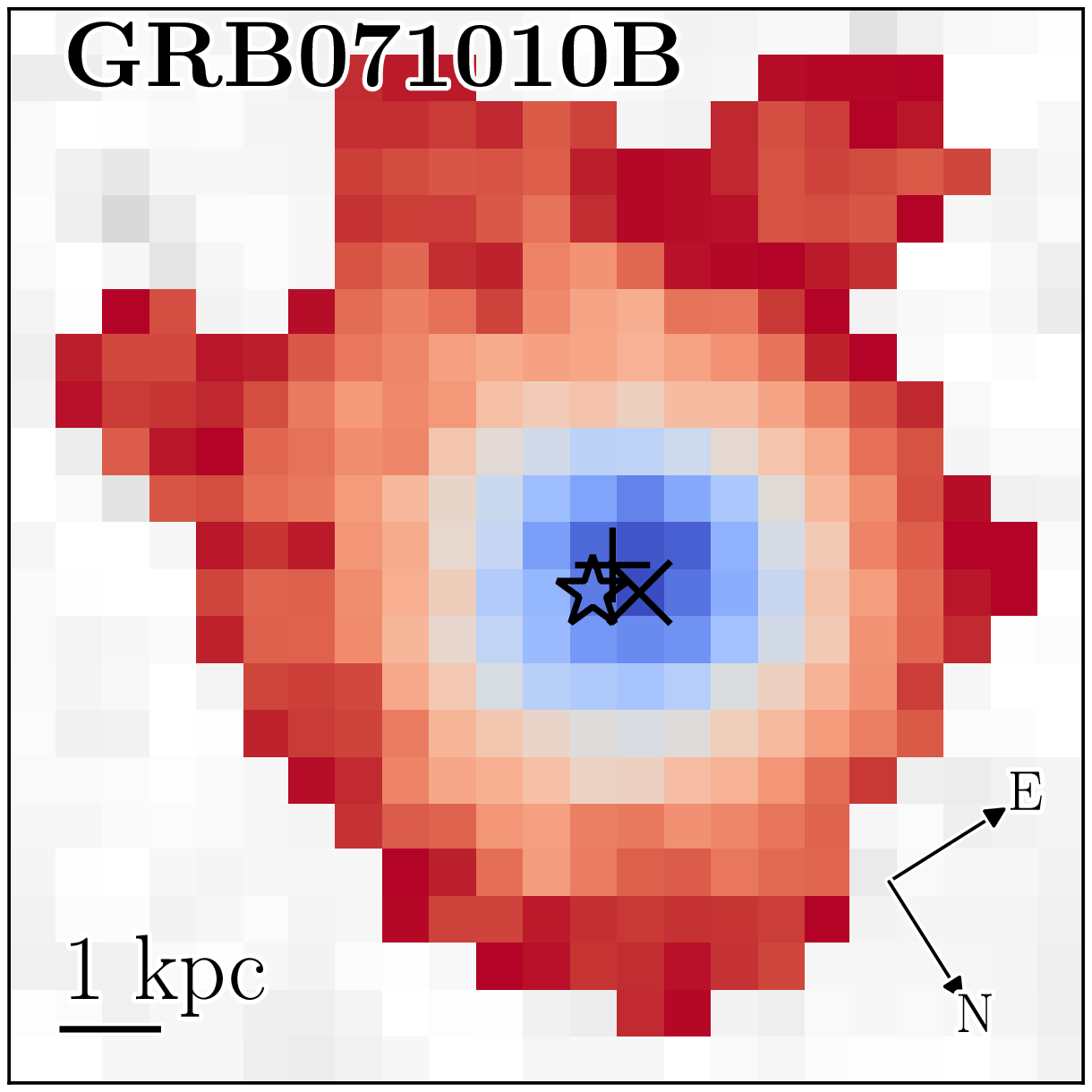}}\\
\vspace{-0.4cm}
\subfloat{\includegraphics[width=0.5\linewidth]{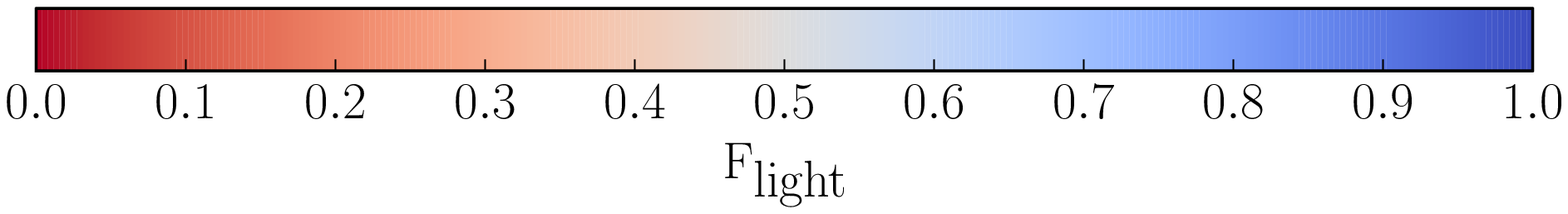}}
 \caption{Visual representation of the offset and \fl{} analysis for each GRB host. The pixels selected in the \sex{} segmentation map for each host are colour-coded by their \fl{} value and superposed onto an inverted stamp of the \hst{} image. The pixel used to give the \fl{} value of the GRB is indicated on each host with a black star, with the barycentre of the host as determined by \sex{} shown by the black plus symbol. The brightest pixel location is indicated with an $\times$ -- note this is the brightest pixel in the \sex{} {\em filtered} image (see text), and thus is not synonymous with the location of the \fl{}~$= 1$ pixel. Fainter regions of the hosts are denoted by red pixels with the brightest regions shown with blue pixels. Orientation and linear scale at the distance of the GRB/host are indicated on each stamp. Pixels are 0.065~arcsec on a side.} 
 \label{fig:segmaps}
\end{figure*}
\addtocounter{figure}{-1}
\begin{figure*}
\centering
\subfloat{\includegraphics[width=0.25\linewidth]{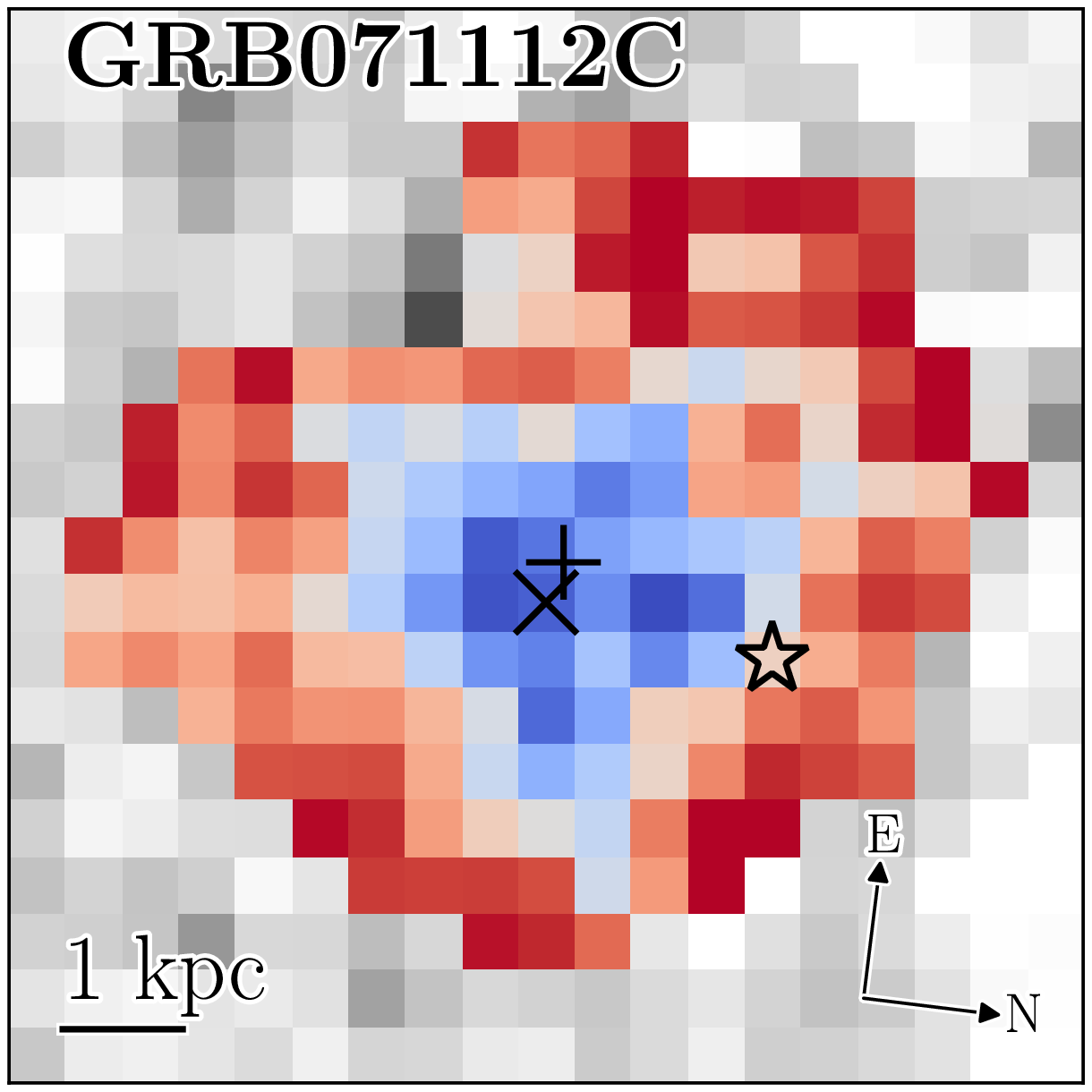}}\hspace*{-0.185cm}
\subfloat{\includegraphics[width=0.25\linewidth]{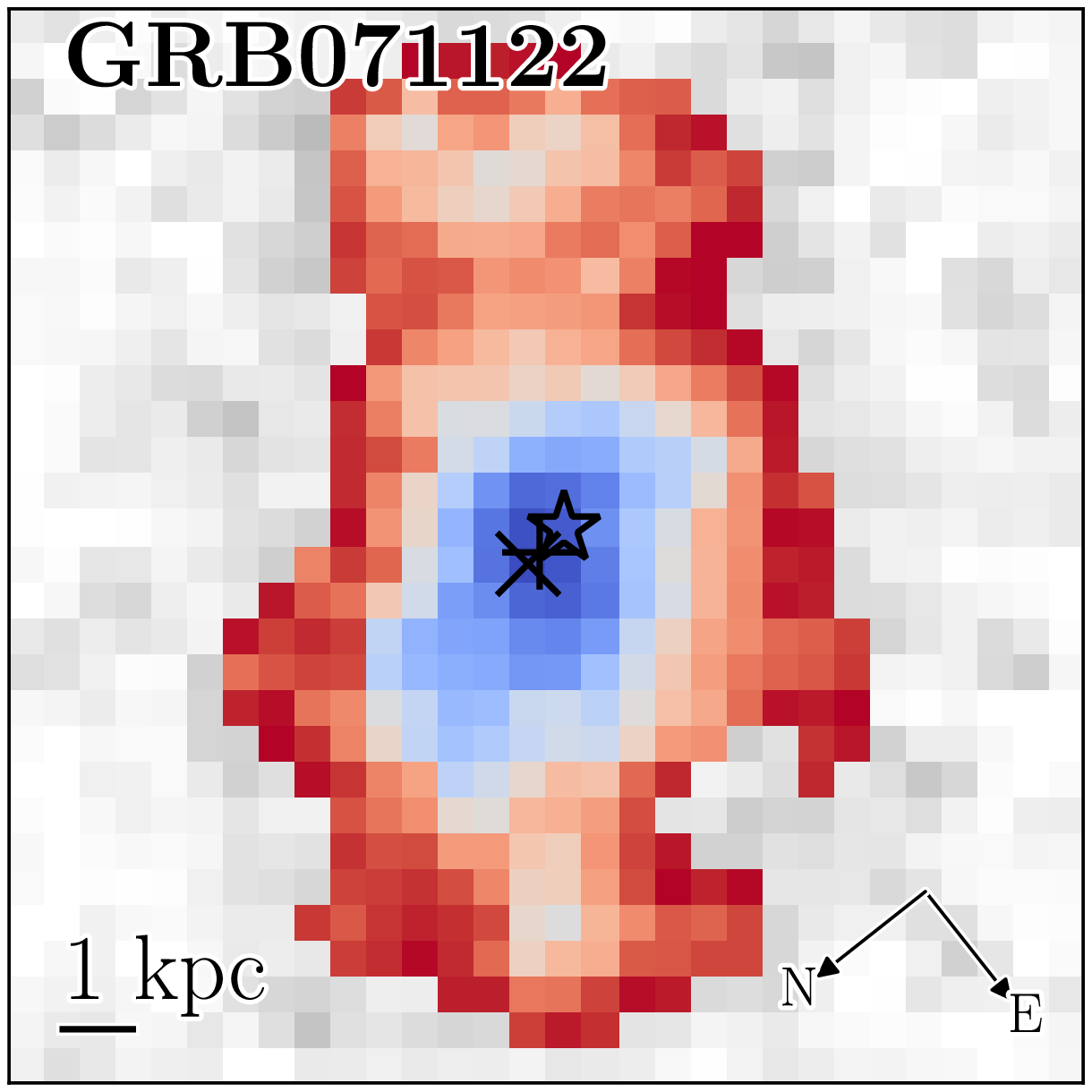}}\hspace*{-0.185cm}
\subfloat{\includegraphics[width=0.25\linewidth]{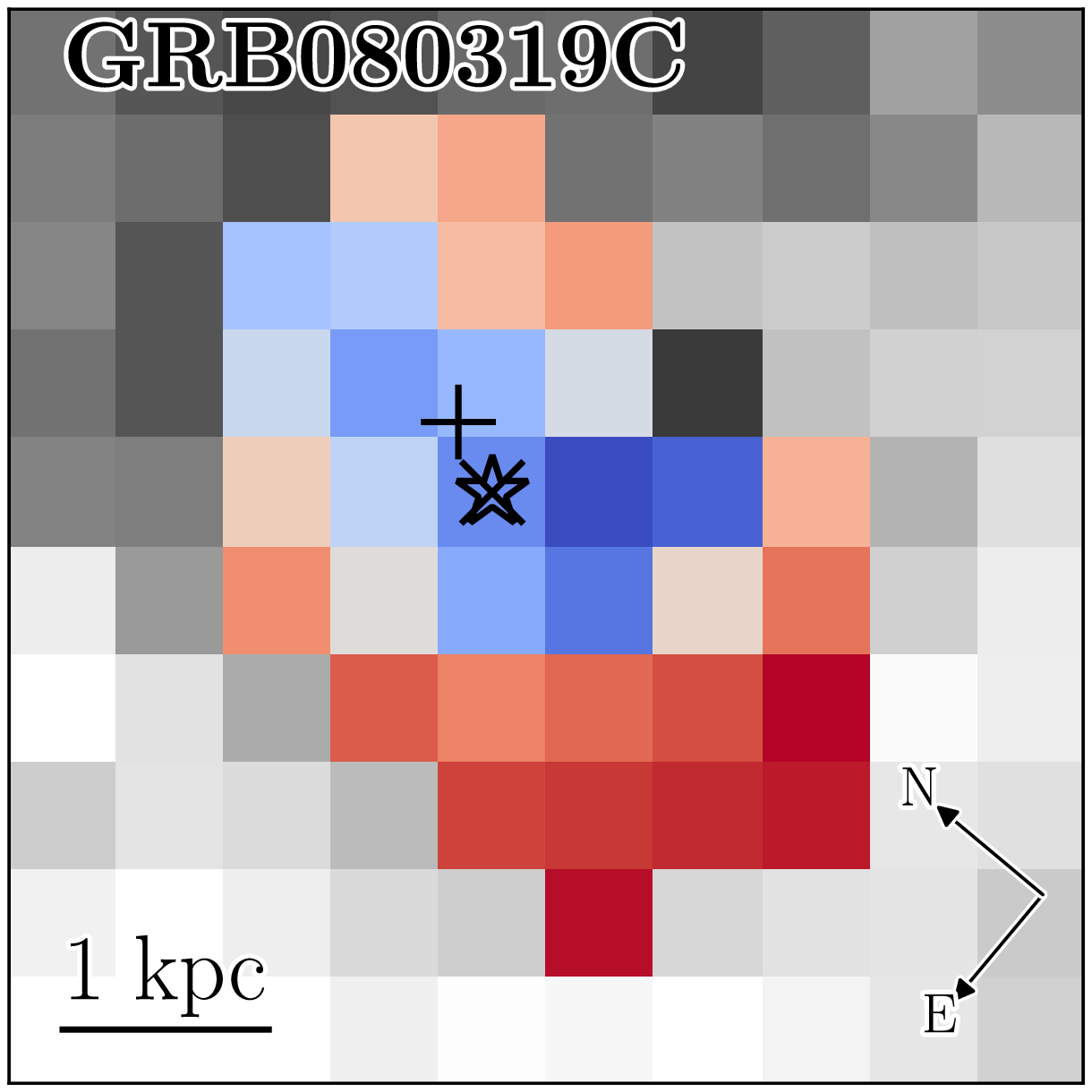}}\hspace*{-0.185cm}
\subfloat{\includegraphics[width=0.25\linewidth]{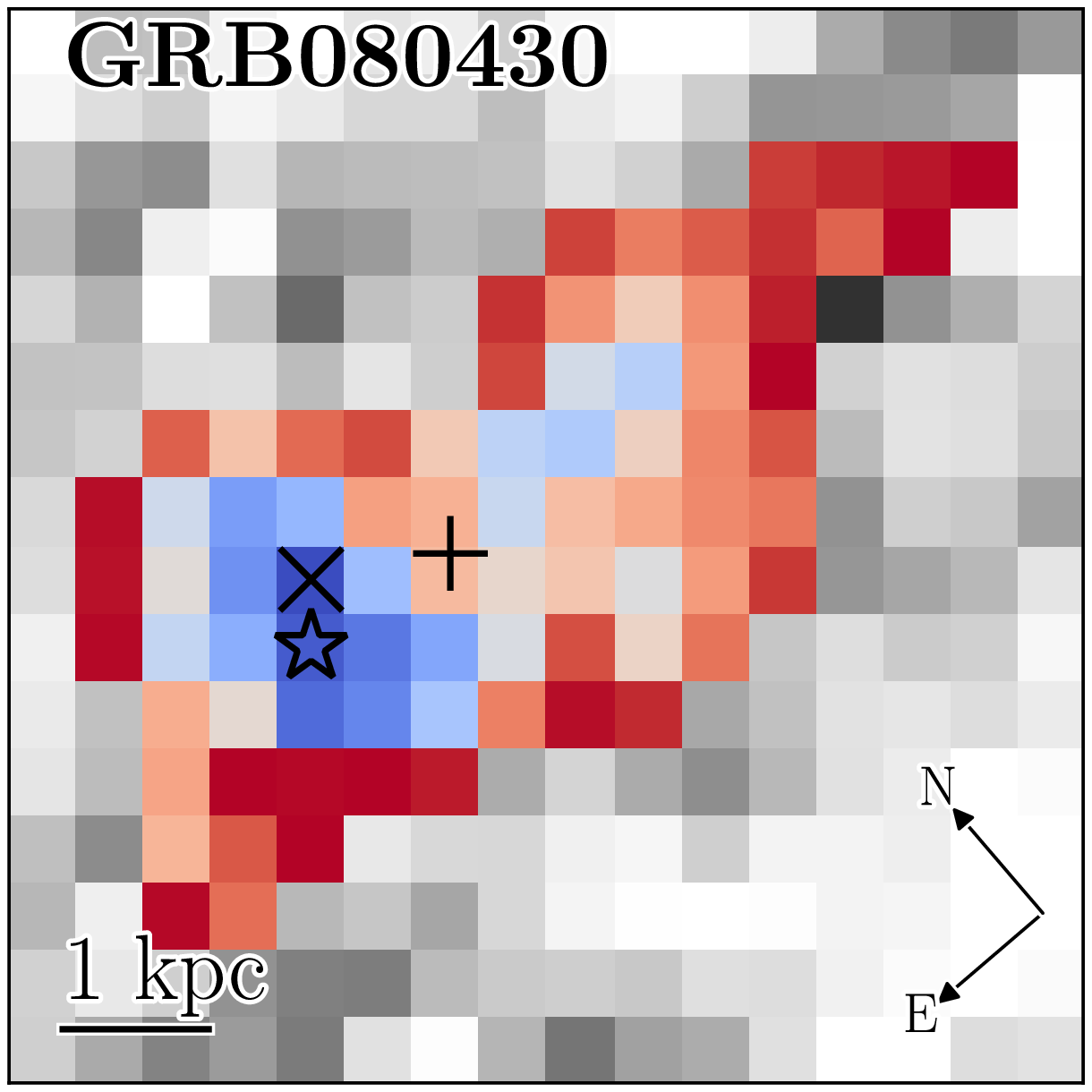}}\\
\vspace{-0.4cm}
\subfloat{\includegraphics[width=0.25\linewidth]{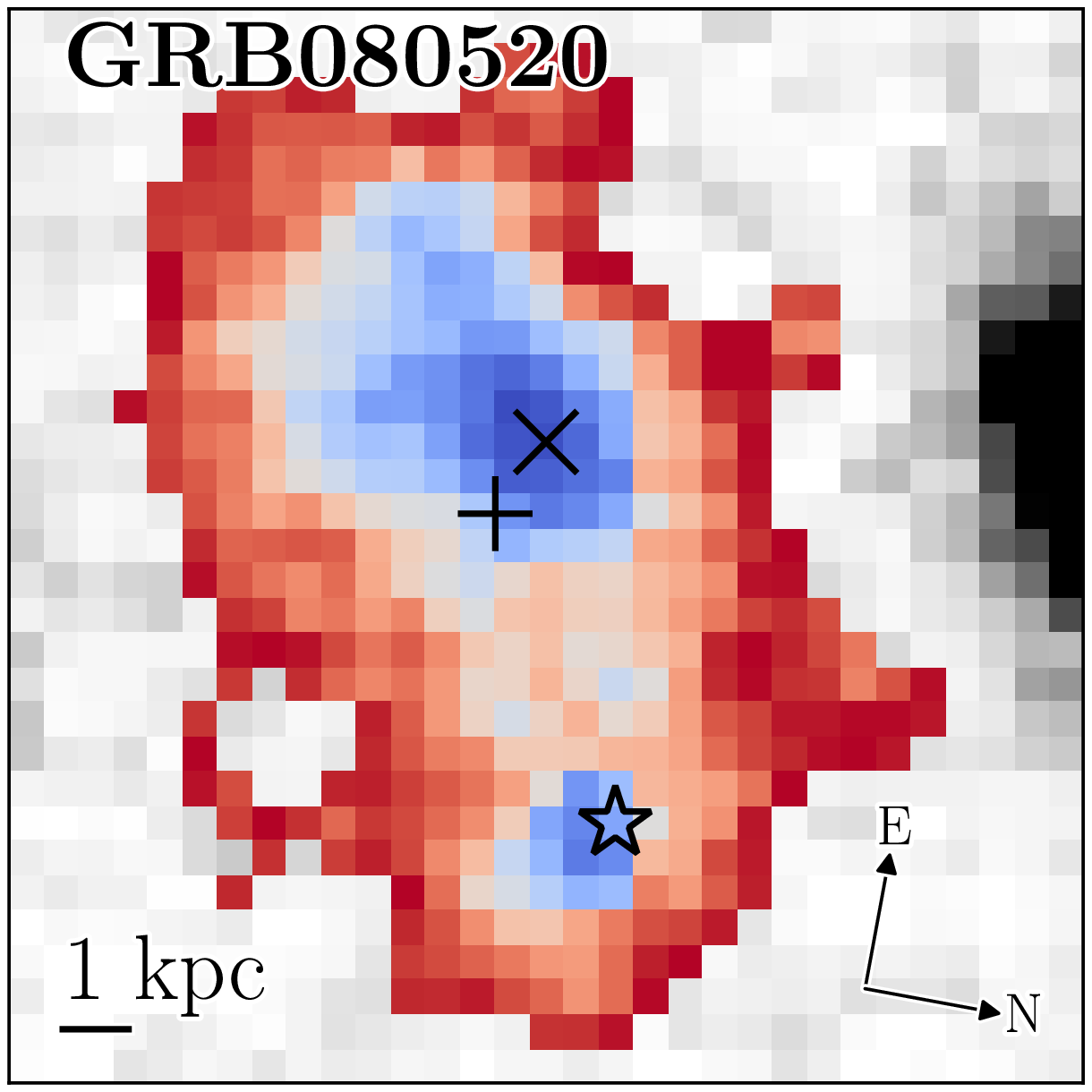}}\hspace*{-0.185cm}
\subfloat{\includegraphics[width=0.25\linewidth]{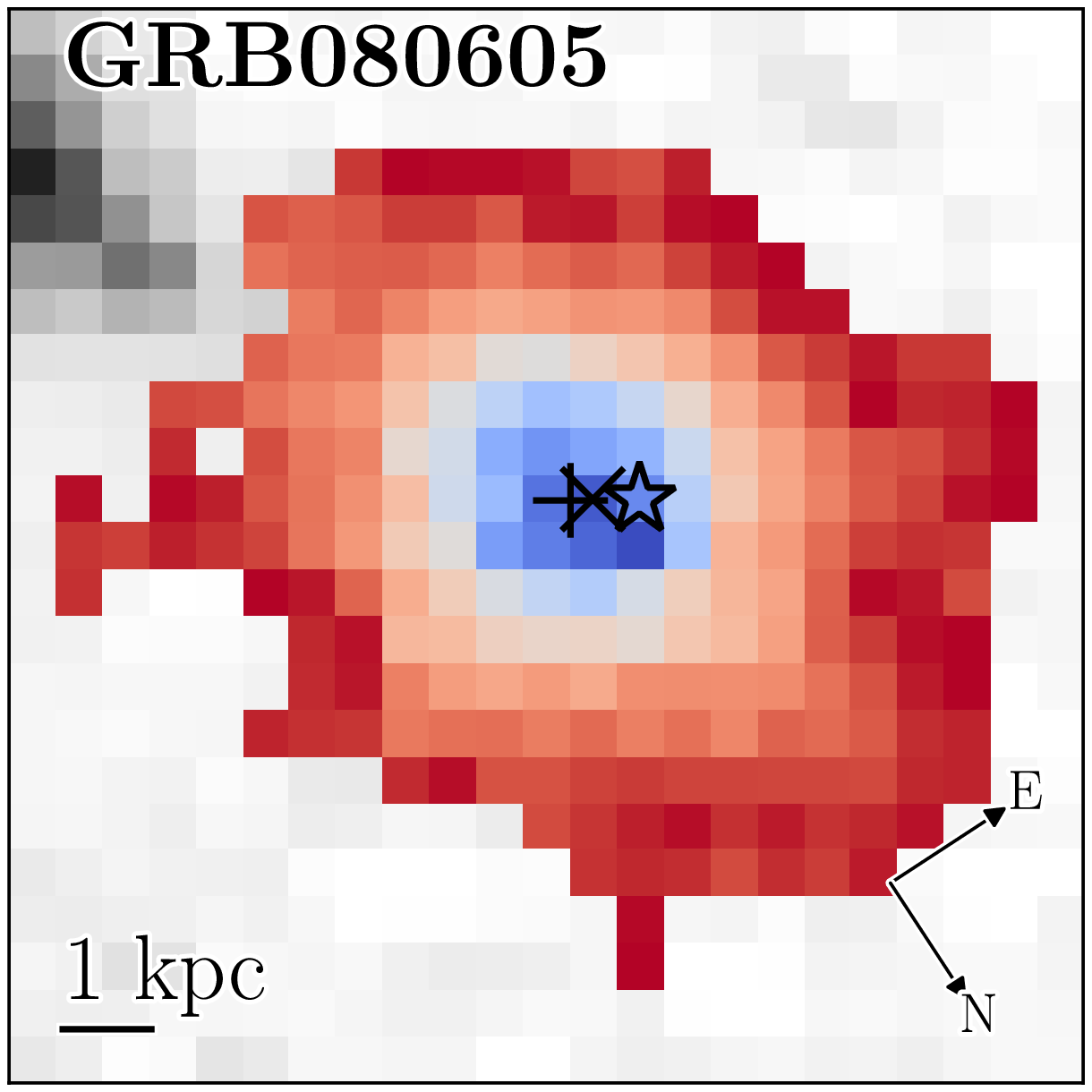}}\hspace*{-0.185cm}
\subfloat{\includegraphics[width=0.25\linewidth]{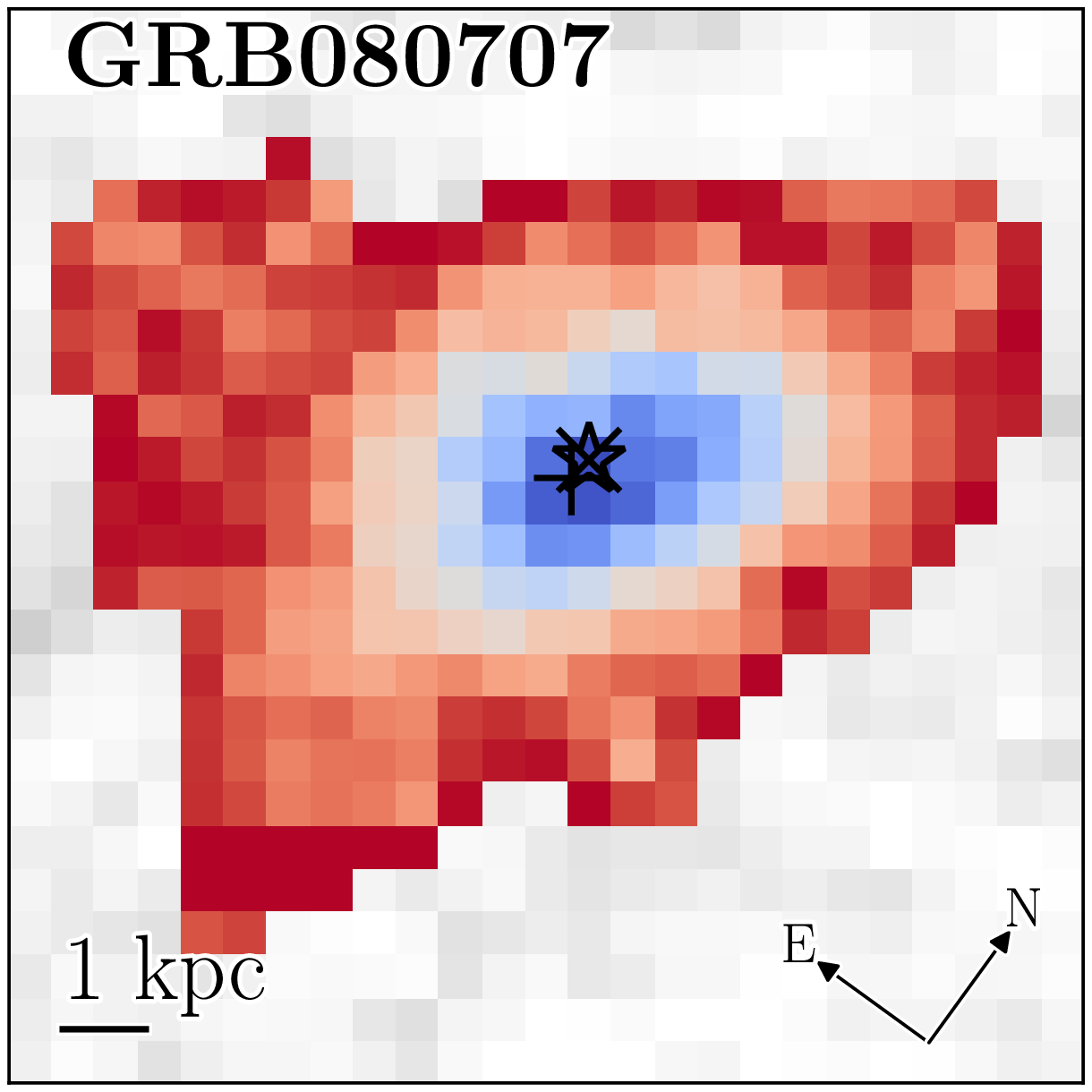}}\hspace*{-0.185cm}
\subfloat{\includegraphics[width=0.25\linewidth]{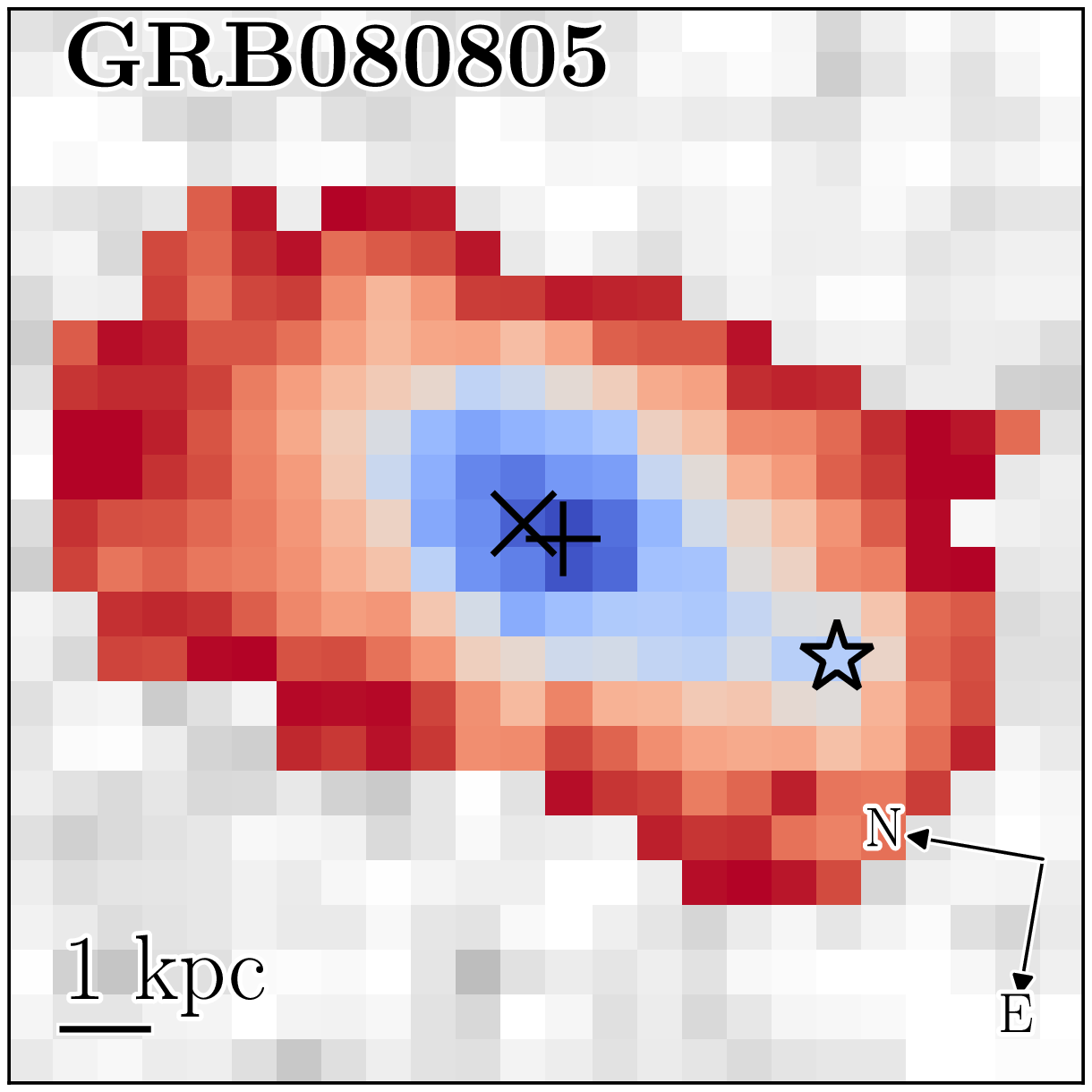}}\\
\vspace{-0.4cm}
\subfloat{\includegraphics[width=0.25\linewidth]{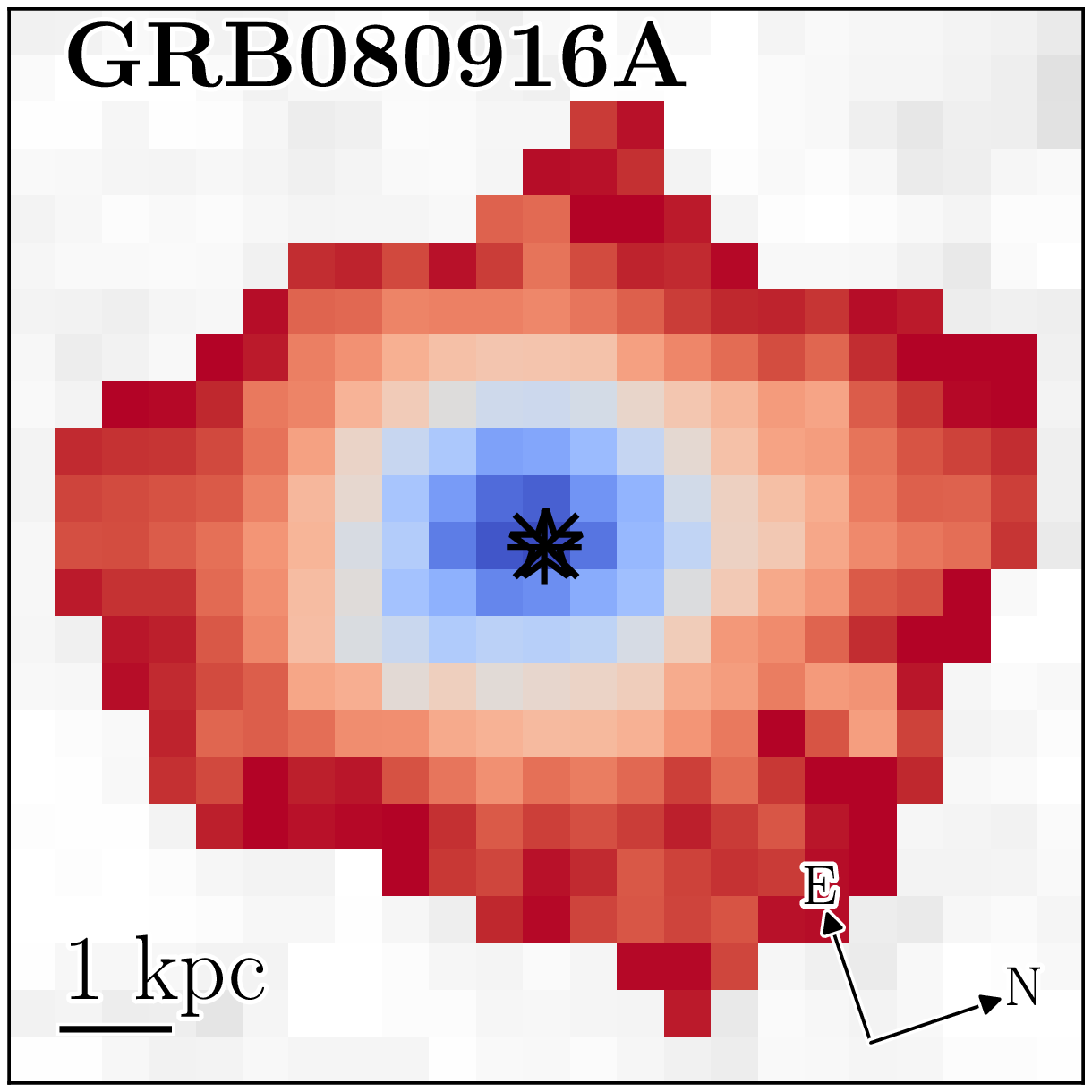}}\hspace*{-0.185cm}
\subfloat{\includegraphics[width=0.25\linewidth]{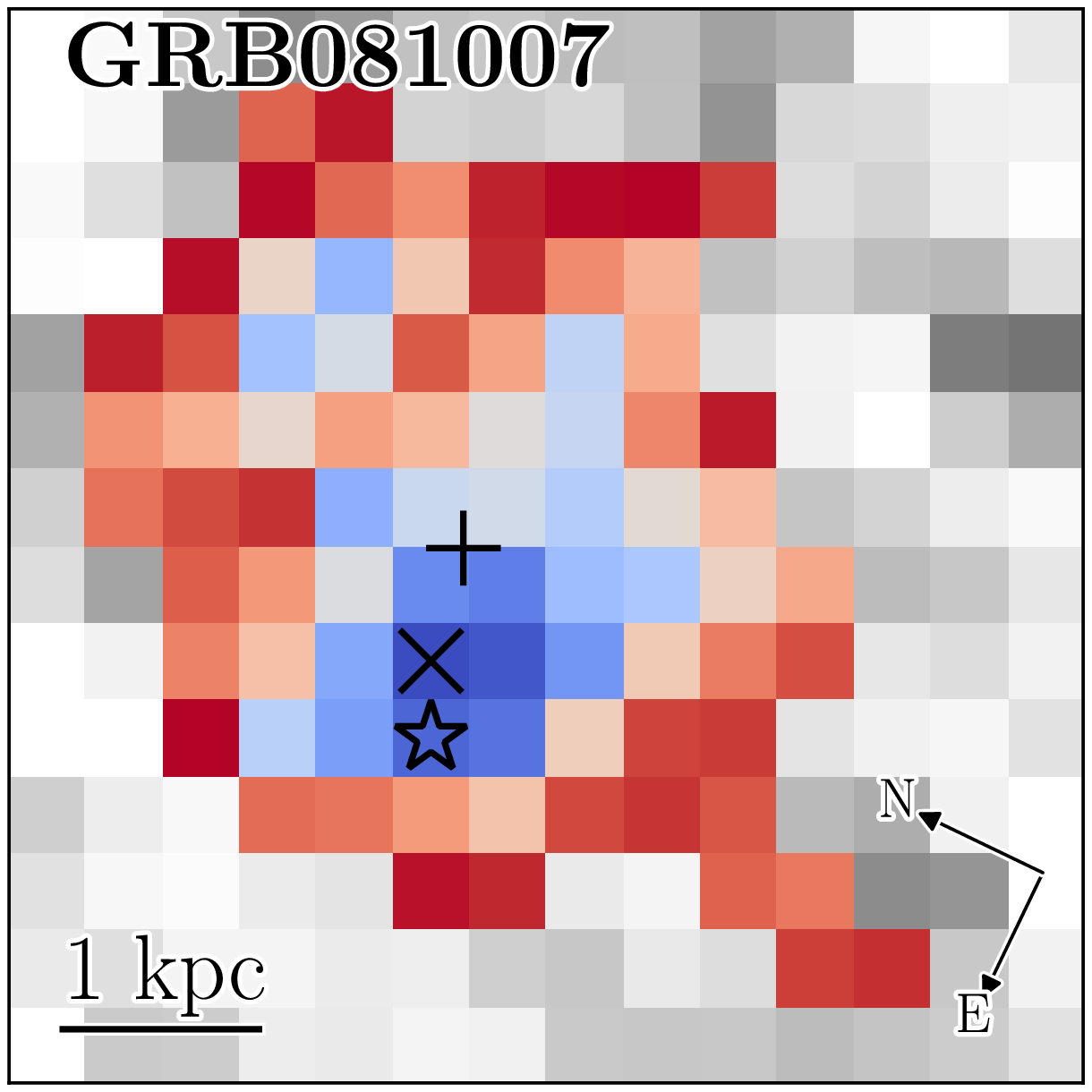}}\hspace*{-0.185cm}
\subfloat{\includegraphics[width=0.25\linewidth]{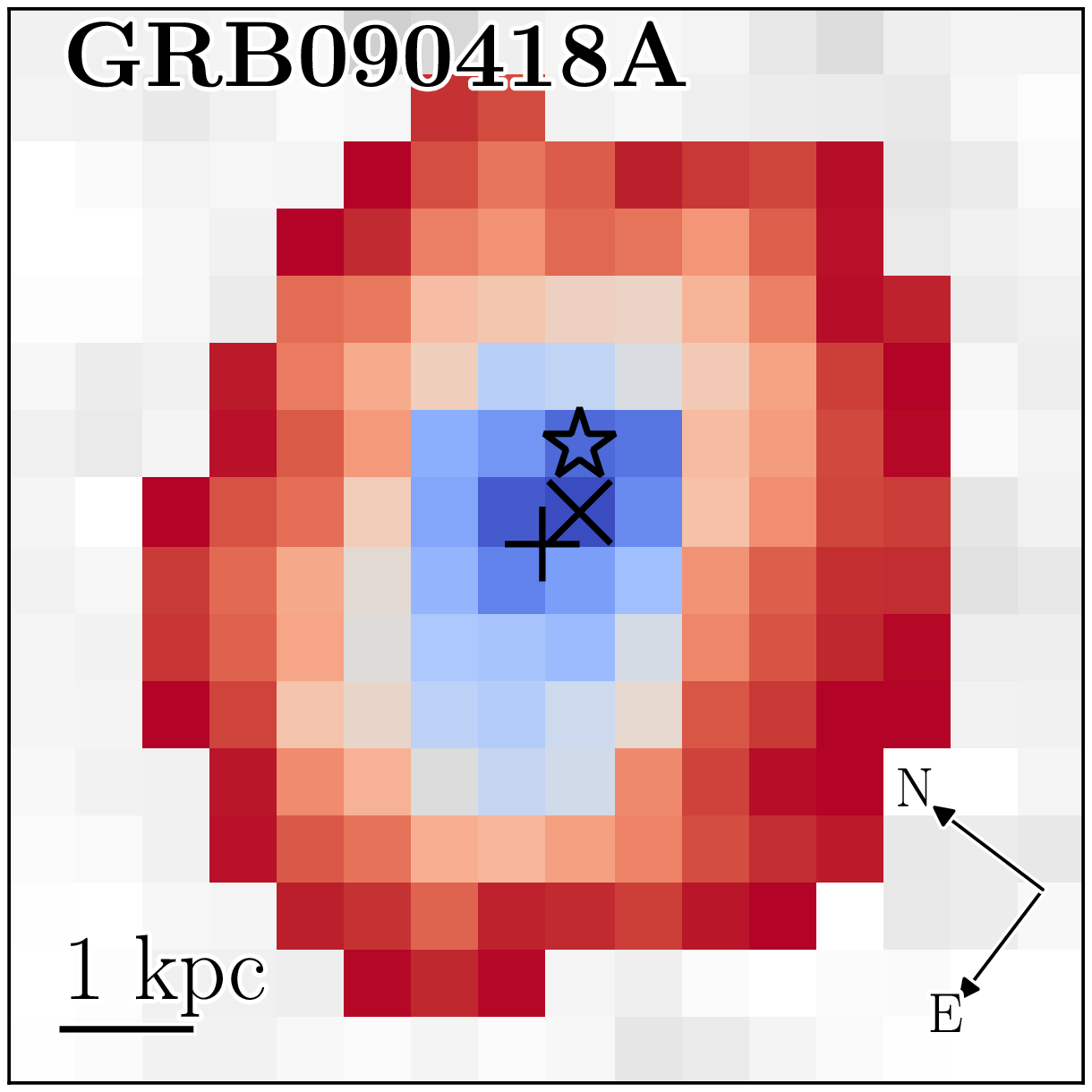}}\hspace*{-0.185cm}
\subfloat{\includegraphics[width=0.25\linewidth]{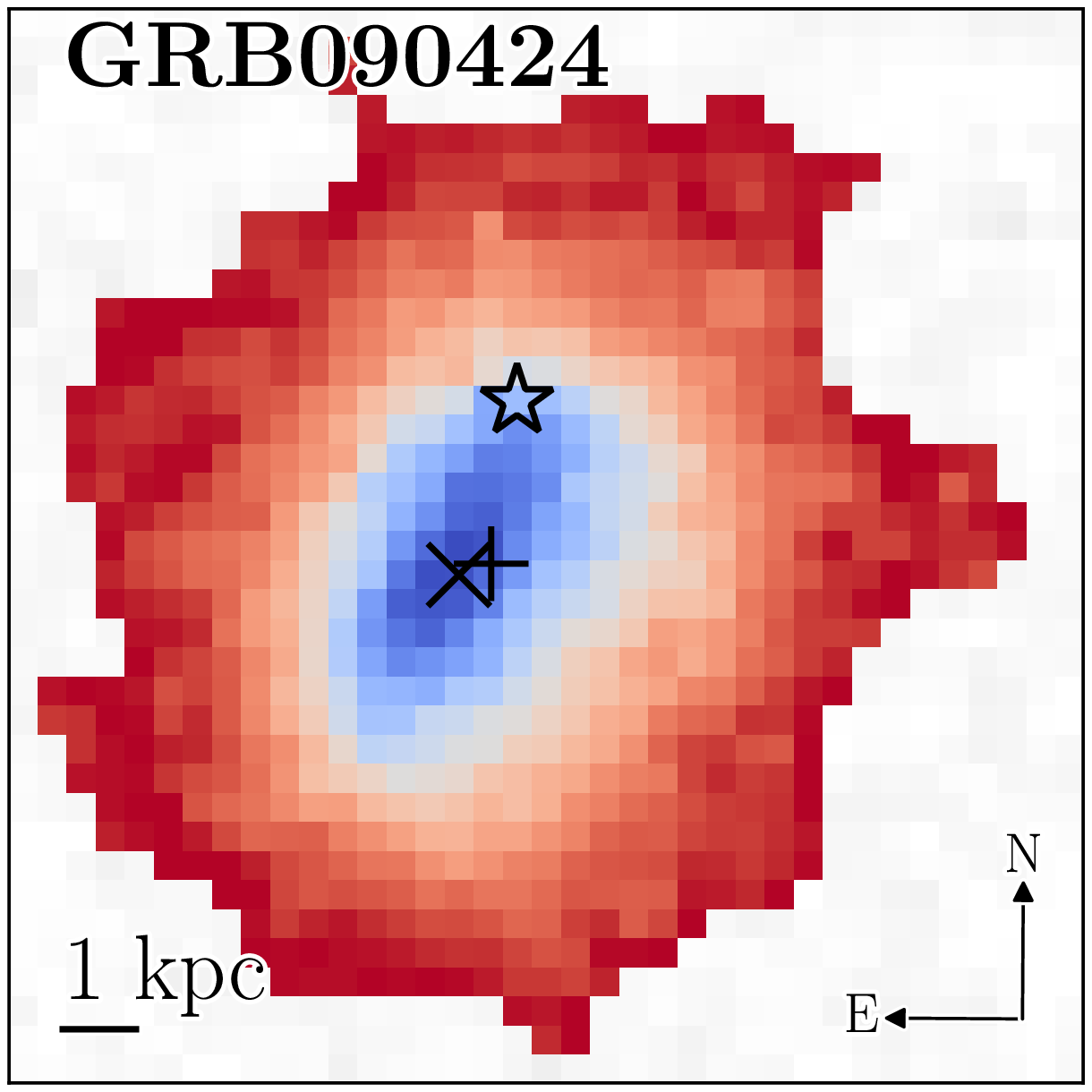}}\\
\vspace{-0.4cm}
\subfloat{\includegraphics[width=0.25\linewidth]{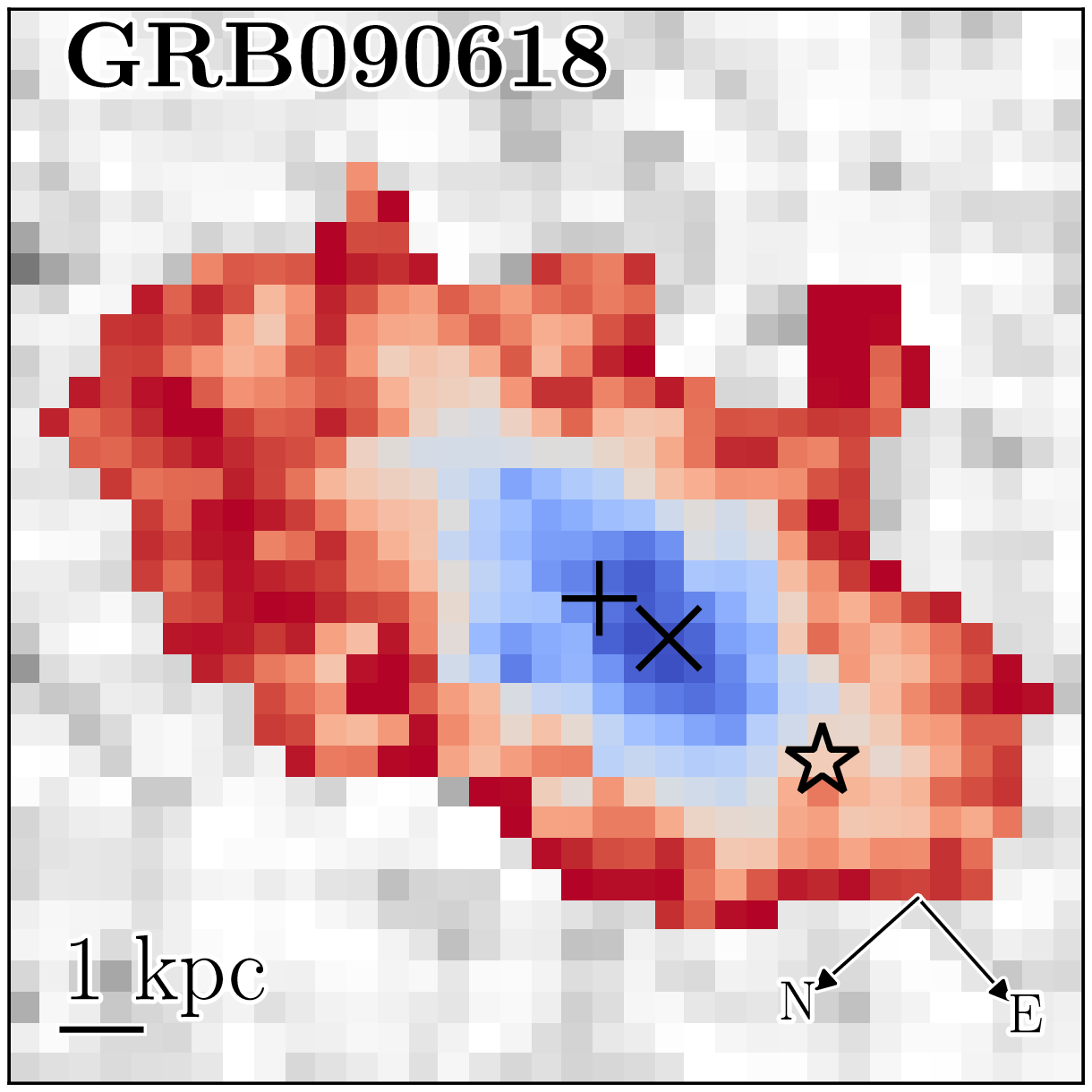}}\hspace*{-0.185cm}
\subfloat{\includegraphics[width=0.25\linewidth]{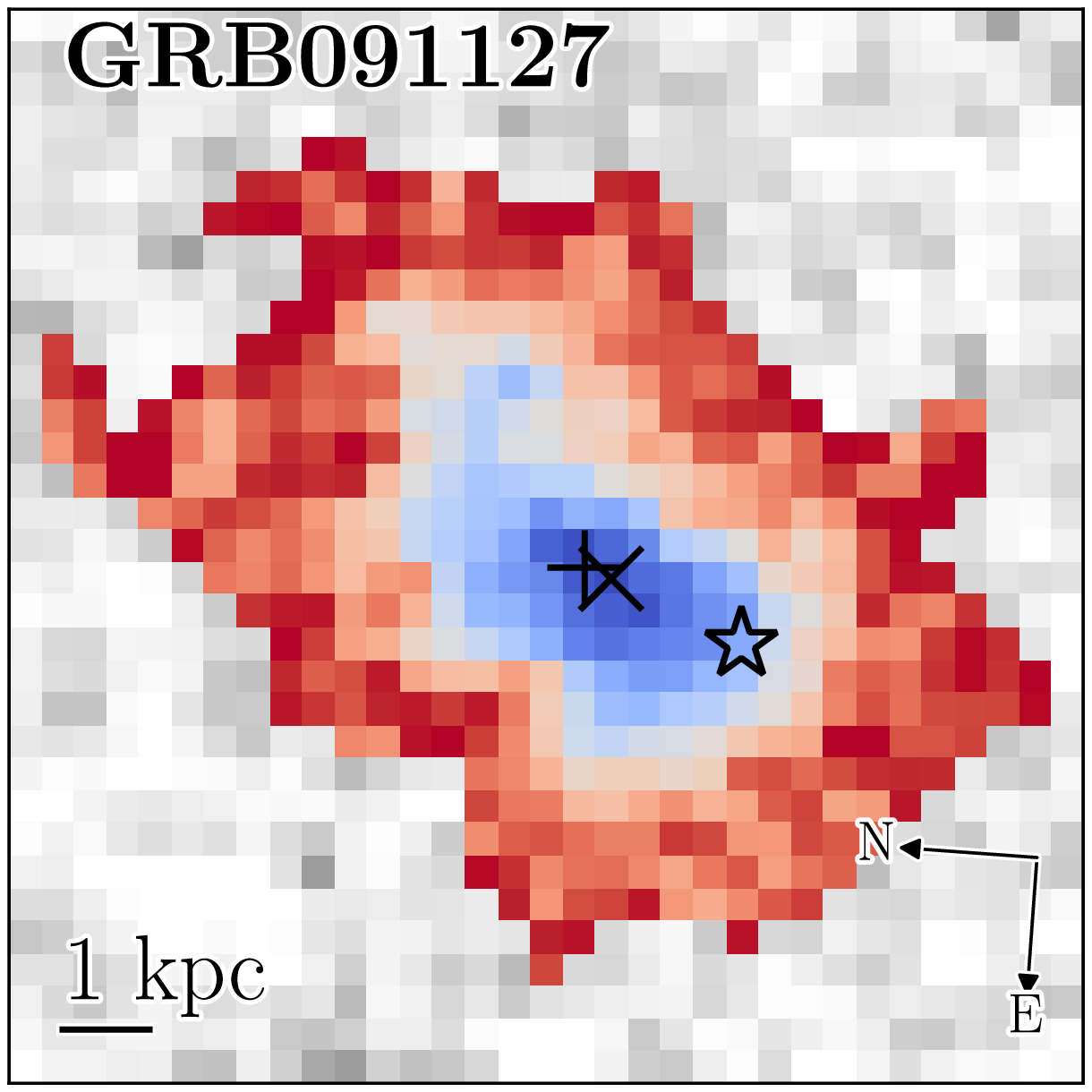}}\hspace*{-0.185cm}
\subfloat{\includegraphics[width=0.25\linewidth]{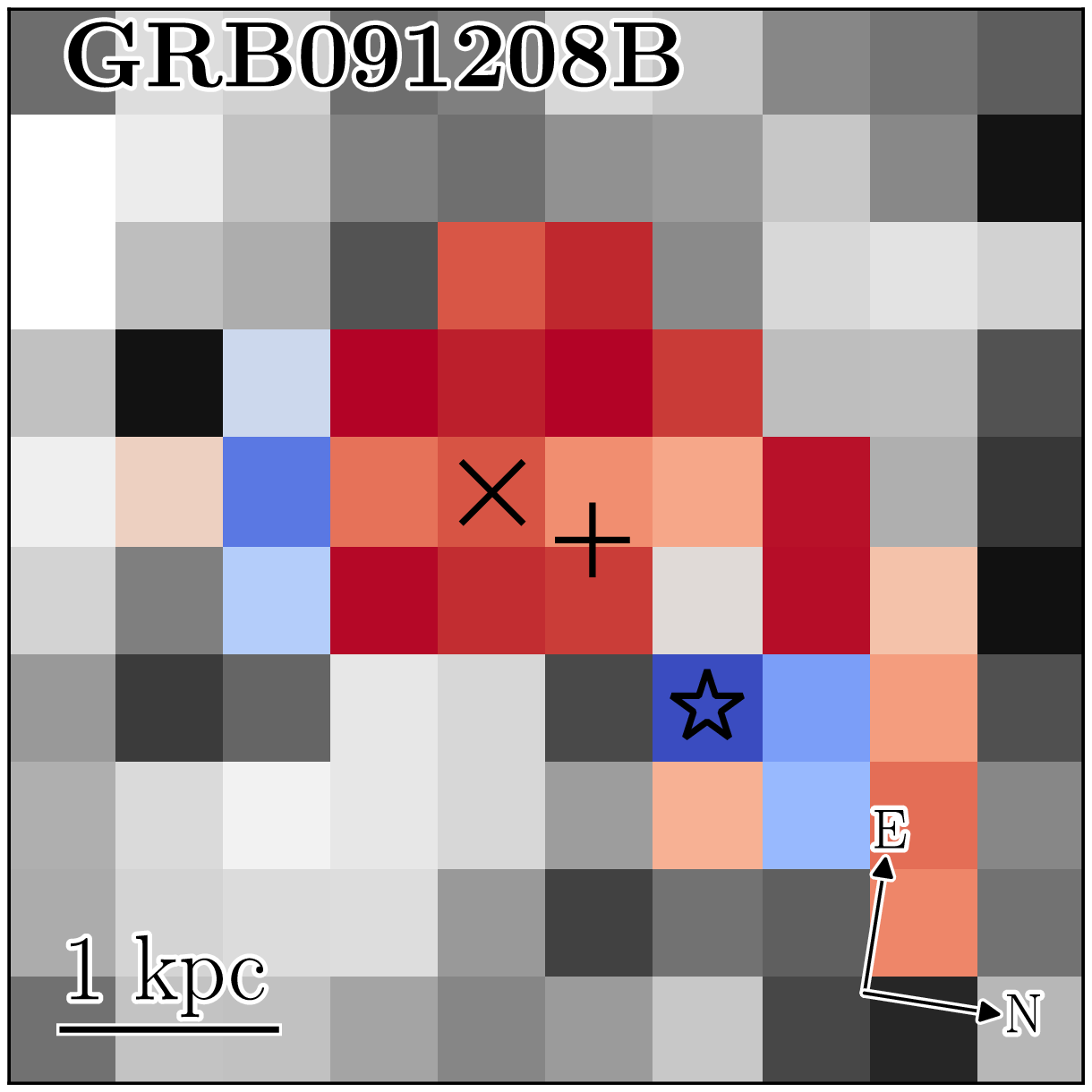}}\hspace*{-0.185cm}
\hbox to 0.27\linewidth{}\\
\vspace{-0.4cm}
\subfloat{\includegraphics[width=0.5\linewidth]{colourbar.eps}}
 \caption{{\em continued.}} 
 \label{fig:segmaps2}
\end{figure*}

\begin{figure*}
\centering
\subfloat{\includegraphics[width=0.25\linewidth]{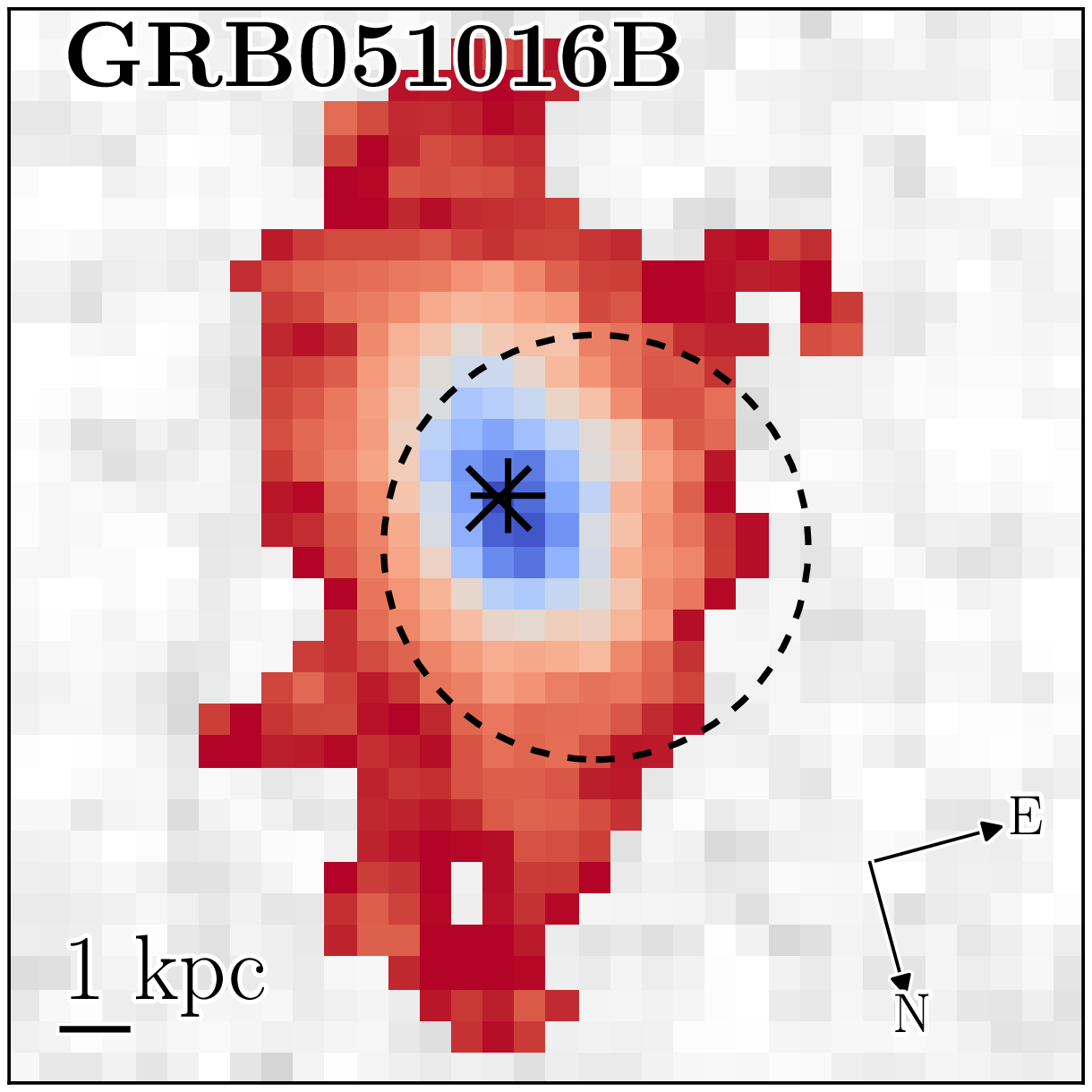}}\hspace*{-0.185cm}
\subfloat{\includegraphics[width=0.25\linewidth]{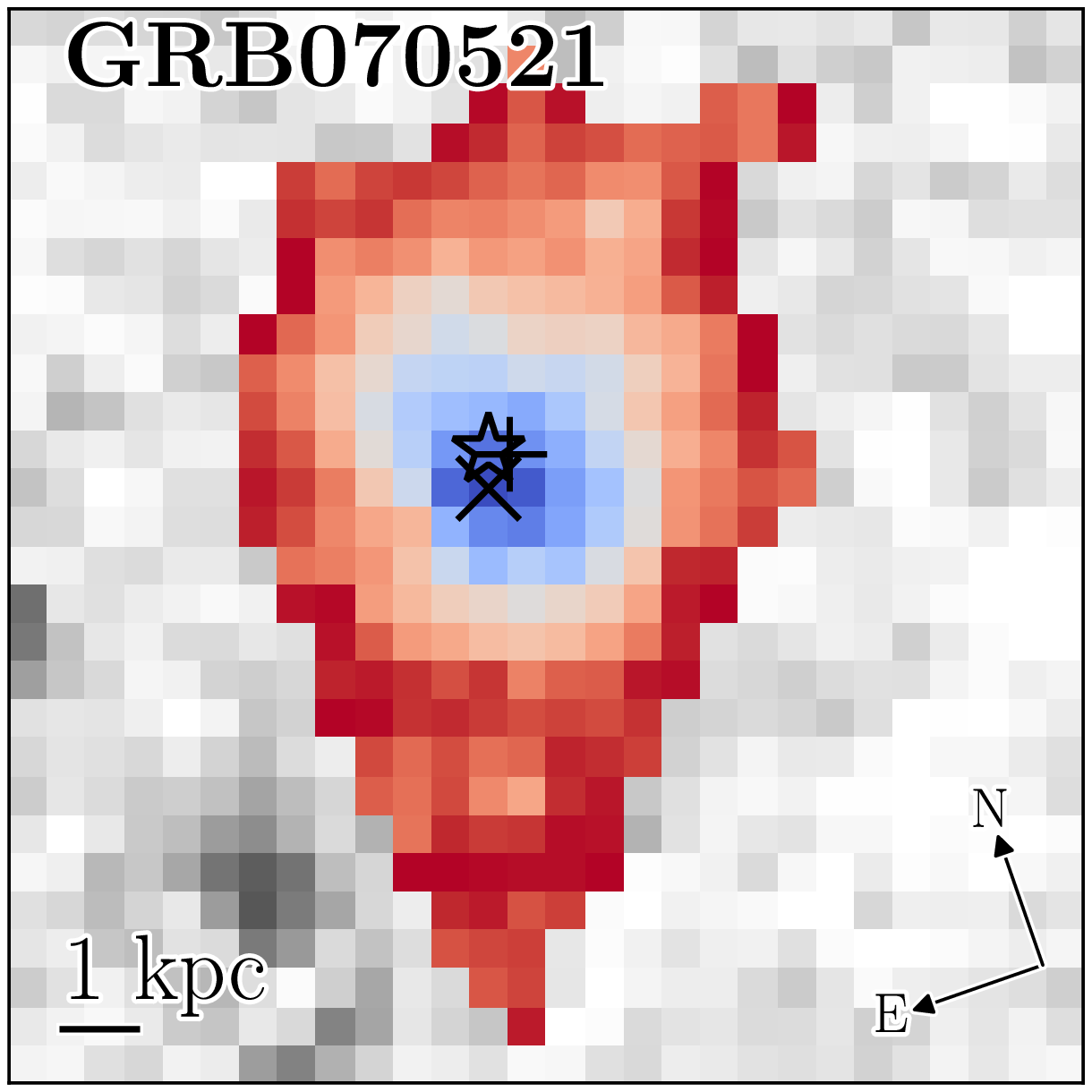}}\hspace*{-0.185cm} 
\vspace{-0.4cm}
\subfloat{\includegraphics[width=0.25\linewidth]{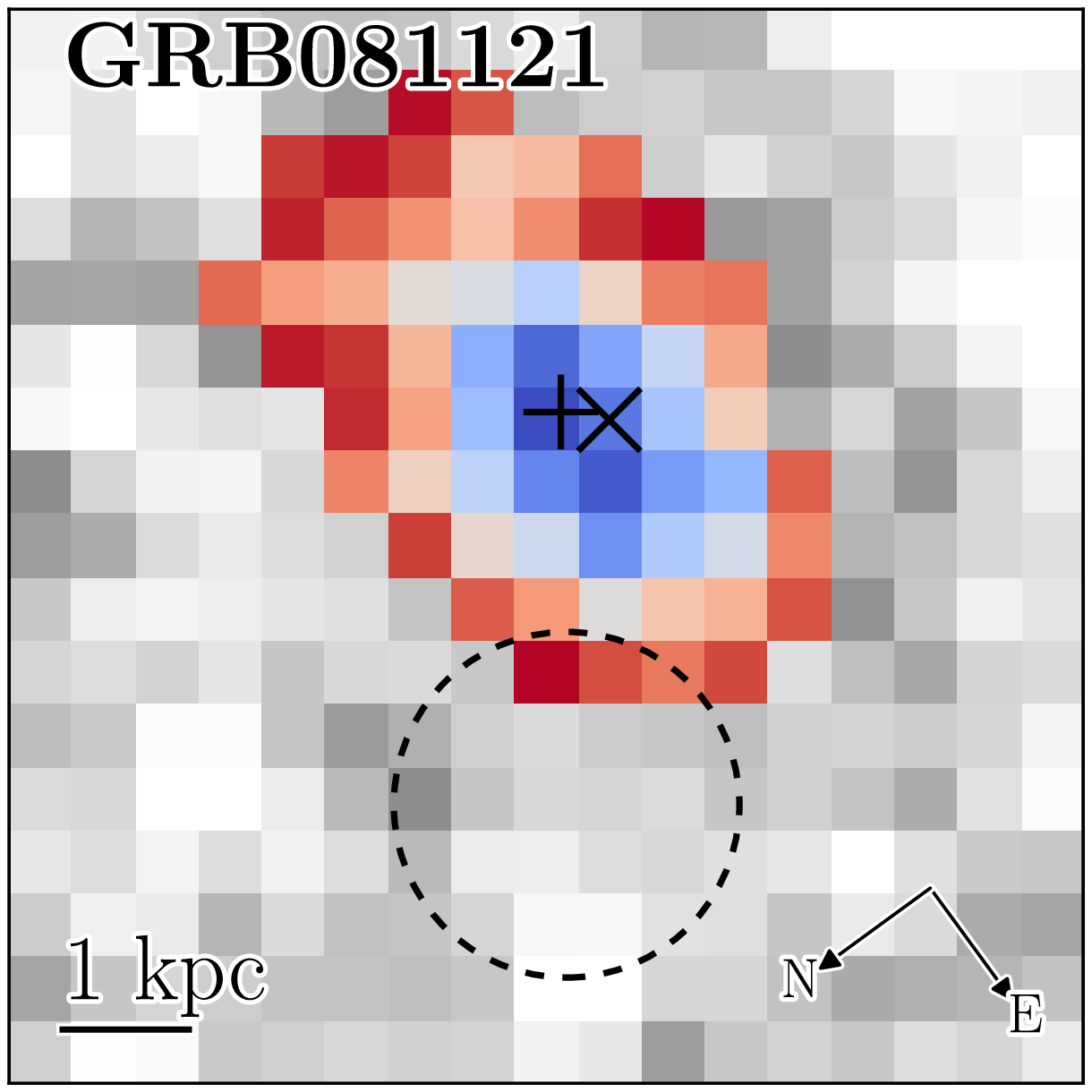}}\hspace*{-0.185cm}\\
\subfloat{\includegraphics[width=0.5\linewidth]{colourbar.eps}}
\caption{Same as \cref{fig:segmaps} but for the GRBs where the location of the burst could not be determined to sufficient accuracy to perform the \fl{} analysis. The \sex{} segmentation map for the chosen host (see \cref{sect:astrometry}) is again colour-coded by the \fl{} statistic for comparison to the others, but an \fl{} value is not calculated for these bursts. The 1$\sigma$ locations determined from alignment with afterglow imaging is shown by the dashed circles, except for GRB~070521 where we simply follow the host assignment of \citet{perley09}.} 
 \label{fig:segmaps_noalign}
\end{figure*}

\subsection{Comparison data}
\begin{figure}
 \includegraphics[width=\columnwidth]{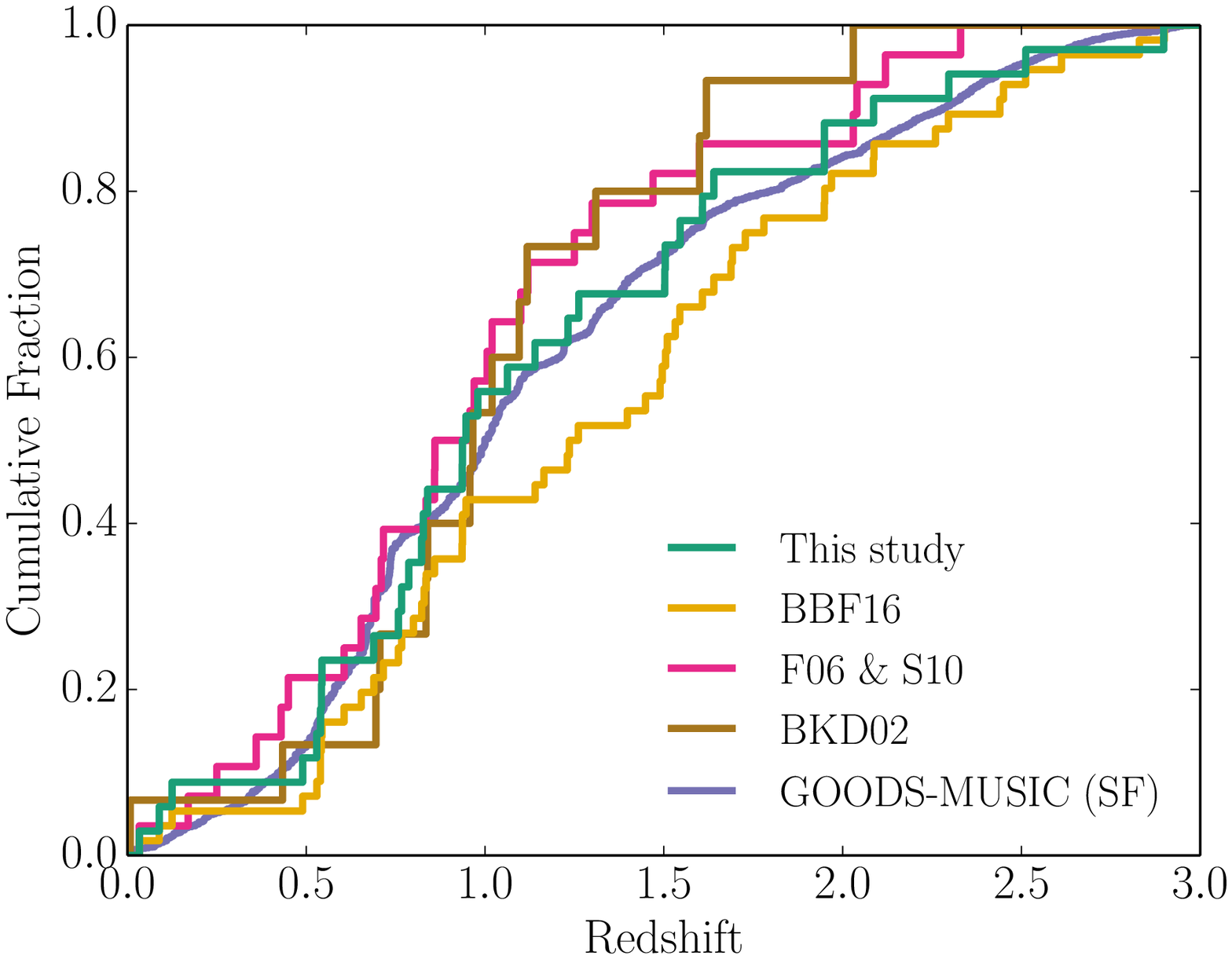}
 \caption{The redshift distributions of the LGRBs of this study, the comparison LGRB host samples, and the subset of the GOODS-MUSIC star-forming galaxy catalogue used.}
 \label{fig:redshift}
\end{figure}

In order to investigate the nature of our sample of LGRB hosts, we utilise a deep survey of field galaxies and results from other LGRB host studies. The comparison samples used are introduced below and their redshift distributions are shown in \cref{fig:redshift}.

\subsubsection{GOODS-MUSIC}
\label{sect:goodssample}
The GOODS-MUSIC \citep[GOODS-MUltiwavelength Southern Infrared Catalogue,][updated in \citealt{santini09}]{grazian06} is a mixture of space- and ground-based literature data of near-infrared (NIR) selected sources. The catalogue has deep NIR coverage, \citep[90~per~cent complete to $K_s \sim 23.8$~mag, with accurately determined spectroscopic and photometric redshifts][]{grazian06}. On all non-stellar, non-AGN objects in the GOODS-MUSIC catalogue we require SNR~$> 5$ in the $H$ band (which the F160W filter closely resembles) and $z < 3$. This catalogue was then cross-correlated with the star formation rates (SFRs) determined by \citet{santini09}. These SFRs were determined using two estimators, spectral energy distribution (SED) fitting and mid IR fluxes. Our results are not sensitive to the choice of estimator (\cref{sect:hostlum}).

\subsubsection{Bloom, Kulkarni \& Djorgovski}
\label{sect:bkd02sample}
The host offset distribution of LGRBs was investigated by \citet[][hereafter \citetalias{bloom02}]{bloom02}. Ground-based observations of the afterglows were astrometrically aligned to HST STIS/CLEAR ($\lambda_\text{cen} \sim 5850$~\AA{}) imaging of the hosts, in order to determine the projected offset of the burst from the host centre. Where physical (linear) distances and offsets are presented here, we have recalculated them from the angular separations given in \citetalias{bloom02} and using the cosmology defined in \cref{sect:intro}. We only include those bursts which have a redshift determined to be $< 3$ (i.e. we do not include GRBs 971214, 980326, 980329, 980519, 981226 and 990308).

\subsubsection{Fruchter et al. \& Svensson et al.}
\label{sect:f06s10sample}
\citetalias{fruchter06} present an analysis of the host galaxies of LGRBs and the location of the bursts within their hosts. Additional data analysed in the same manner is provided in \citetalias{svensson10}. \hst{} imaging of the hosts were primarily STIS/CLEAR observations, supplemented with ACS/F606W and WFPC2/F555W observations. Measurements of pixel statistics and host galaxy sizes from these studies provide comparison samples in a different wavelength regime. We restrict the comparison sample to $z < 3$ to match our selection criterion. As with the \citetalias{bloom02} sample, any physical distances are recalculated using the cosmology of the paper.

\subsubsection{Blanchard, Berger \& Fong}
\label{sect:bbf15sample}
A recent collation sample of a large number of LGRBs observed with \hst{} is presented in \citetalias{blanchard16}. Following previous work they investigate host galaxy properties as well as explosion site diagnostics such as offsets and pixel statistics. Data are used from a wide variety of instrument setups and filters, mainly optical and NIR observed wavelengths (corresponding to rest-frame UV/visual over the redshift range investigated here). As with the other comparison samples, we restrict comparison only to those with a measured redshift of $z < 3$, and recalculate any physical sizes to the cosmology of this paper. This sample is slightly weighted to higher redshifts compared to other samples (median $z \sim 1.25$, compared to $\sim 1$ for other samples). If the sample was heavily weighted to the extremes of our redshift cut, galaxy evolution issues may arise, however the tails of the distribution are in good agreement and thus this does not compromise a comparison to our results. We note that this sample is not independent of the sample presented here. The bursts of our sample are included in the total sample of 105 in \citetalias{blanchard16}, with 37 where their analysis was performed on the same imaging as analysed here.

\section{Methods}
\label{sect:methods}
\subsection{Optical afterglow astrometry}
\label{sect:astrometry}

In order to securely identify the host and accurately pinpoint the location of the GRB on its host, optical imaging of the afterglow is required. Using imaging from a variety of telescopes and instruments (see \cref{tab:sample}), a geometric alignment was found between the follow-up afterglow imaging and the \hst{} host imaging using common sources in the field. After selecting common sources, the transformation solution was computed by the {\sc iraf}\footnote{{\sc IRAF} is distributed by the National Optical Astronomy Observatory, which is operated by the Association of Universities for Research in Astronomy (AURA) under cooperative agreement with the National Science Foundation.} task {\sc geomap}. The rms of the fit gives the contribution to uncertainty in the GRB location on the \hst{} frame due to this alignment. In addition to this uncertainty, an estimate of the uncertainty in the centroid of the afterglow is given by FWHM/(2.35$\times$SNR), where FWHM is that of the afterglow image and SNR is the signal-to-noise ratio of the afterglow detection. Uncertainties on the location from both image axes and the centroiding were added in quadrature. The coordinates of the afterglow centroid were transformed with {\sc geoxytran} to give coordinates in the \hst{} frames. Note for GRB 060729 the \hst{} WFC/IR imaging was aligned directly to \hst{} imaging of the afterglow (GO 10909, PI: Bersier), this was repeated for another epoch where the optical transient was bright to give an estimate of the uncertainty on the alignment. 

For three bursts, an accurate alignment with the \hst{} images could not be made, these are highlighted in \cref{tab:sample}. This may be due to the afterglow not being detected on the follow up imaging, the coarse pixel resolution/PSF of the afterglow image, or too few sources in common between the two images to accurately tie an alignment solution. For these bursts we do not calculate offsets or the pixel statistic detailed in \cref{sect:pixstat}, however, where we can securely identify the host through visual inspection or using absolute WCS coordinates of the host given in GCN circulars of the afterglow, we include the overall properties of the host in our subsequent analysis. In the following, we briefly detail the host assignment method for these bursts. 

For GRB 051016B, the afterglow was barely detected with UVOT in individual filters, and available ground-based imaging was of poor quality. To increase the depth of the UVOT imaging for the purposes of astrometric alignment we summed all exposures taken in all filters ($uvw2\rightarrow V$, including the white filter) using {\sc uvotimsum} within the {\sc ftools}\footnote{\url{http://heasarc.gsfc.nasa.gov/ftools/}} package. From this stacked image we had 6 sources that we could use as astrometric tie points between the \hst{} and UVOT image. Given the low number of tie points and the faintness of the source even in the stacked UVOT image, our alignment uncertainty is considerably larger than the seeing of the \hst{} image (and likely to be underestimated, since we have a low number of sources from which to determine the rms). Nevertheless, the 1-sigma uncertainty is directly overlying a source, which we consider the host galaxy. The coordinates are in agreement with those used by \citet{soderberg05}, who took a late time spectrum of the GRB's position and found the host to be at $z=0.9364$.

GRB 070521 was a dark burst with no robust optical/nIR afterglow detection. We made the host assignment according to that in \citet{perley09}, where the XRT error circle is used. Similar to their deep $K$ band data, we find a single source (the putative host) within the XRT error circle, on the eastern side. 

For GRB 081121, we used UVOT imaging of the afterglow. Unfortunately there are only a low number of common sources between the UVOT and WFC3 imaging. Following the procedure employed for GRB 051016B, we stacked the early UVOT exposures in all filters to achieve a greater depth. Our uncertainty for the astrometric position of the burst in the \hst{} image (which, as above, is likely to be an underestimate since we have a low number of tie points) overlaps with a single source, which we consider to be the host galaxy. The next nearest source is $\sim$~2.2~arcsec away.

\subsection{Host radii and photometry}
\label{sect:radphot}
The extraction software \sex{} (\citealt{bertin96}; {\sc v2.8.6}) was used on the final drizzled, CTE-corrected \hst{} images to perform morphological and photometric measurements of the GRB hosts. After applying a Gaussian filter with a FWHM of 3 pixels, a S/N cut of 1 across a minimum area of 5 pixels was used to identify objects in the images. Visual inspection of the output segmentation maps showed extraneous, spatially distinct, islands of pixels were being associated with the host galaxies. A higher value of the cleaning parameter was used to either remove these from the final segmentation maps or assign them as a separate object. Note for GRBs 060912A and 080605 the presence of a bright nearby galaxy and diffraction spikes due to nearby bright stars respectively warranted aggressive cleaning and de-blending by \sex{}.\footnote{For GRB~060912A the segmentation map still contains multiple islands of pixels that appear distinct from the host, despite employing aggressive cleaning and deblending -- therefore the determined flux radii, and thus photometry, are likely to be affected for this host.} For GRB~080319C, a brighter foreground galaxy is superimposed, however the \sex{} segmentation map distinguishes this as a separate object from the source (the adopted host) that is underlying our chosen GRB position (\cref{sect:pixstat}). The output \sex{} source catalogue for each host image was also used to determine $r_{20}$, $r_{50}$ and $r_{80}$ for each host -- the radii at which 20, 50 and 80~per~cent of the host flux is enclosed, respectively. The uncertainties in these quantities were estimated by repeating measurements on different realisations of the drizzled image: individual {\sc \_flt.fits} images were resampled based on the original pixel values and their uncertainty (given by the {\sc [ERR]} extension) before drizzling (as in \cref{sect:sample}). This was repeated $\sim$few hundred times per burst. Radii values are given as those based on the original drizzled images, with the uncertainty given by the 1$\sigma$ limits of the resampled distribution of values. Where the radius value from the original drizzled image is outside these 1$\sigma$ limits, we quote that as the lower or upper value, as appropriate.

Observed absolute magnitudes of the hosts were obtained using the apparent magnitude measurements {\sc mag\_auto} and the redshift of the burst (\cref{tab:sample}) -- results are presented in AB magnitudes for the observed wavelength (i.e. $\sim 15400/[1+z]$~\AA{}) of the hosts using the STScI-provided zeropoints in the image headers. Milky Way extinction was accounted for using the dust maps of \citet{schlafly11}. 

\subsection{GRB location offsets and pixel statistics}
\label{sect:pixstat}
Knowledge of the location of a transient within the host can be used to provide additional constraints on the progenitor population. Here we determine three additional values for those GRBs where we can localise the transient to sufficient accuracy. 

Firstly, we determine offsets of the GRBs by calculating the separation between the afterglow centroid and the host galaxy's barycentre at the redshift of the GRB. The host galaxy's centre in this case is determined by \sex{} as the barycentre of the pixel distribution (the parameters {\sc x\_image} and {\sc y\_image}). Secondly, we calculate the offset of the GRB from the brightest pixel in the \sex{} {\em filtered} extraction image (i.e. after applying a $3\times3$ Gaussian convolution, given by parameters {\sc xpeak\_image} and {\sc ypeak\_image}). In the case of a smooth elliptical or even a well-formed spiral, the position of the barycentre and brightest pixel of the galaxy should agree very well. However, in the case of multiple cores, potentially merging systems and generally asymmetrical profiles as is the case for a large fraction of the LGRB hosts (\cref{fig:segmaps,fig:segmaps_noalign}), these two measures can be significantly offset from each other. Calculation of the GRB offset from its host's barycentre and brightest pixel gives an estimate of how much the morphology of the hosts impacts our resulting offset distributions.

The third method is a measure of the GRBs association to the light distribution of their host, based on the flux of the explosion site. Two independent means of measuring this quantity were developed, the `fractional light' method \citep[\fl{} hereafter,][]{fruchter06} and the normalised cumulative rank method \citep{james06}. The \fl{} method will be used here, which was developed and presented in \citet{fruchter06}, but is repeated here in the interest of completeness. All pixels in the segmentation map output from running \sex{} (as in \cref{sect:radphot}) that are associated to the GRB host were sorted and cumulatively summed. This cumulative sum was then normalised by the total sum of the pixels associated with the host. As such, each pixel in the host now has a value between 0 and 1, where its value is equivalent to the fraction of host galaxy flux below the level at that pixel. The pixel containing the GRB location (determined as in \cref{sect:astrometry}) was found and its value in the normalised cumulative sum gives the \fl{} value for that GRB. In each case where we attempted an accurate star-matched alignment, the uncertainties in the afterglow position due to alignment and centroiding of the afterglow (see \cref{sect:astrometry}) were added in quadrature to give the total positional uncertainty (\cref{tab:siteprops}). These were found to be less than the seeing of the \hst{} images (FWHM $\sim$ 0.17 arcsec) -- thus, given the observations are subject to a seeing-induced smoothing above our determined positional uncertainties, no additional smoothing was required (cf. \citealt{fruchter06}). Discussion of location uncertainties and the \fl{} statistic is given in \cref{sect:flight_compare}. Briefly, we consider an alternative measure of \fl{} incorporating directly the uncertainties in the location. Since our alignment uncertainties are relatively small for the bursts where we determine \fl{}, we find similar results for the \fl{} distribution of our sample based on the single pixel choice detailed above. For this study we use the method described above in order to facilitate comparison to previous works.

We do not consider GRBs 051016B, 070521 and 081121 when calculating offsets and \fl{} values since we could not determine their positions to sufficient accuracy so as to make results meaningful, although they are included when looking at overall host galaxy properties.

\section{Results}
\label{sect:results}
\cref{tab:hostprops,tab:siteprops} contain the results determined from our investigations following the methods above, which are presented in the following sections.

\subsection{Host assignment}
\label{sect:hostassignment}

Inferring results on the hosts and explosion locations of LGRBs is naturally contingent on making the correct host assignment for each burst. The increasing rapid response of follow up observations and improved localisation of high energy alerts over recent years has worked to minimise the contamination in host galaxy studies due to unassociated line-of-sight galaxies by accurately pin-pointing the location.

Our host assignments where accurate astrometry between the afterglow and \hst{} images could not be performed are detailed in \cref{sect:astrometry}. Here we use the formalism of \citetalias{bloom02} (see also discussion in \citealt{perley09}) to quantify the potential that we have spurious host associations. Specifically, $P_{i\text{,ch}}$, the probability that in a random sky circle with an effective radius $r_i$ there will be $\geq$1 galaxies brighter than magnitude $m_i$ is given by
\begin{equation}
 P_{i\text{,ch}} = 1 - \exp(\pi {r_i}^2 \sigma_{\leq m_i})\text{,}
\end{equation}
where $\sigma_{\leq m_i}$ is the average surface density of galaxies brighter than $m_i$. In contrast to \citetalias{bloom02}, where this value is parameterised based on $R$-band galaxy counts, we estimate this based on galaxy counts in a similar band to our observations. 
We use the results of \citet{metcalfe06} who provide galaxy counts in $H$-band, which extend to faint magnitudes in F160W using Hubble Deep Field data.

$P_\text{ch}$ values for our hosts are given in \cref{tab:hostprops}. For most hosts where we perform an accurate alignment, we can be confident in our host assignments since the probability of observing an unrelated underlying galaxy for these localisations is generally of order 0.1--1 percent.
Again following \citetalias{bloom02}, we use a Poisson binomial distribution to estimate the contamination of spurious host assignments in our sample. The probabilities are 0.754, 0.218, 0.026, 0.002 for zero to three spurious assignments, respectively. This indicates we have $<$~3 spurious assignments, with $\sim 0-1$ most likely, thus not undermining our statistical results for the host sample.\footnote{We note this approach treats the potential contaminant galaxies as point sources, whereas in reality they are extended and thus the $r_i$ search radius should include some measure of the extent of the galaxies at the magnitude of interest. This is unlikely to affect our results greatly since when searching for bright, extended hosts, where the correction is greatest, they have a very low surface number density and when searching for the higher surface number density fainter hosts, they appear more point-like.}

\subsection{Undetected hosts}
\label{sect:undetectedhosts}

\begin{figure*}
\centering
\subfloat{\includegraphics[width=0.2\linewidth]{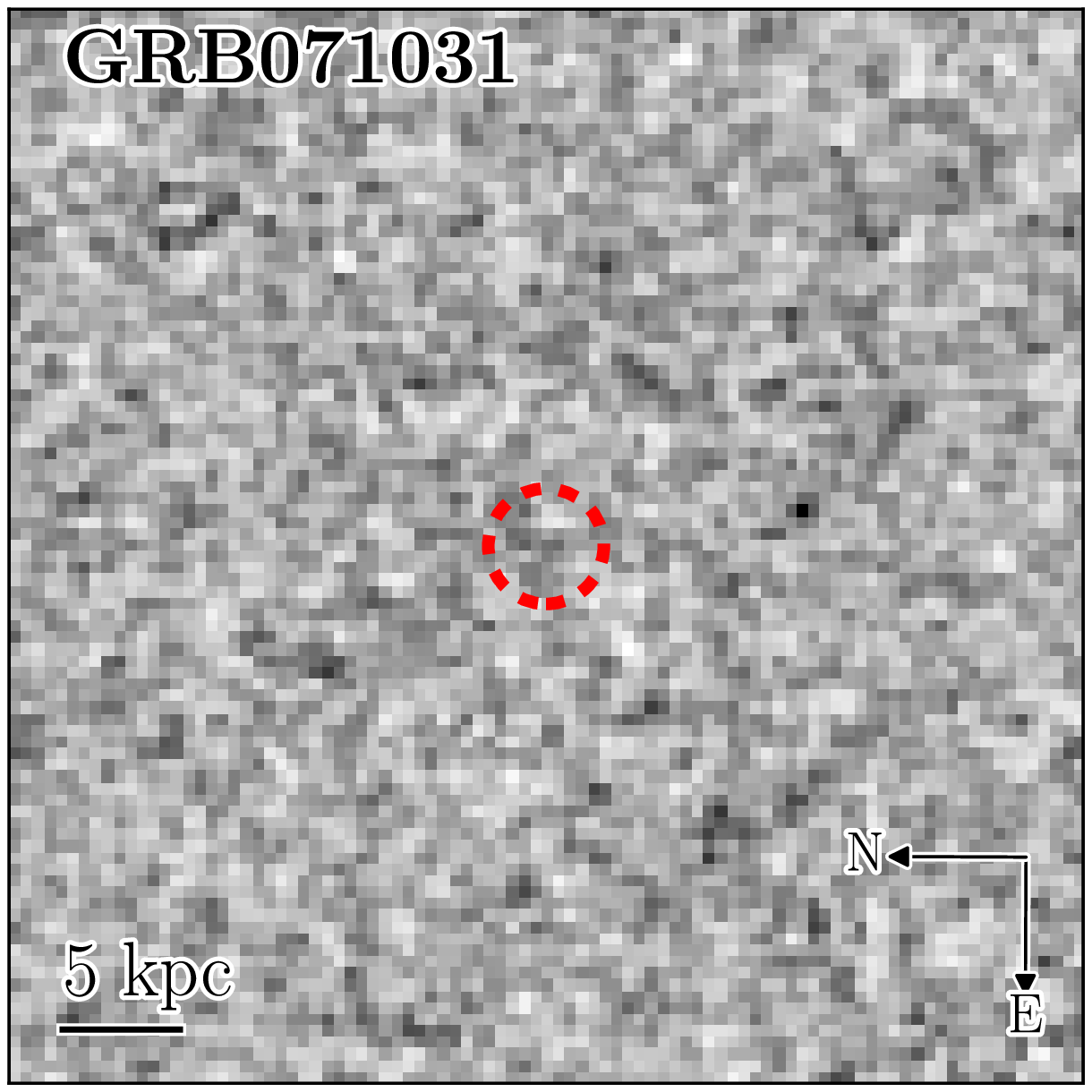}}
\subfloat{\includegraphics[width=0.2\linewidth]{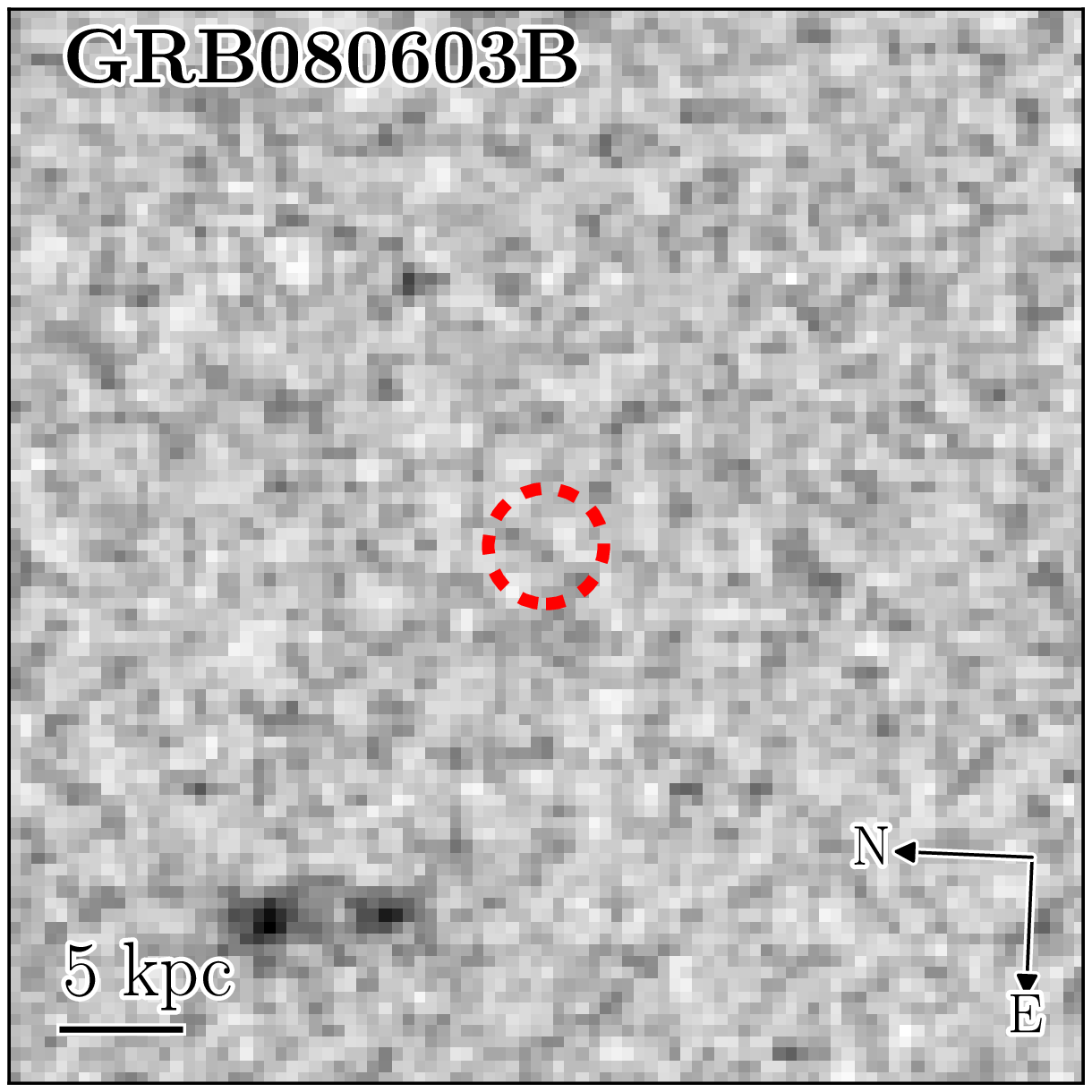}}
\subfloat{\includegraphics[width=0.2\linewidth]{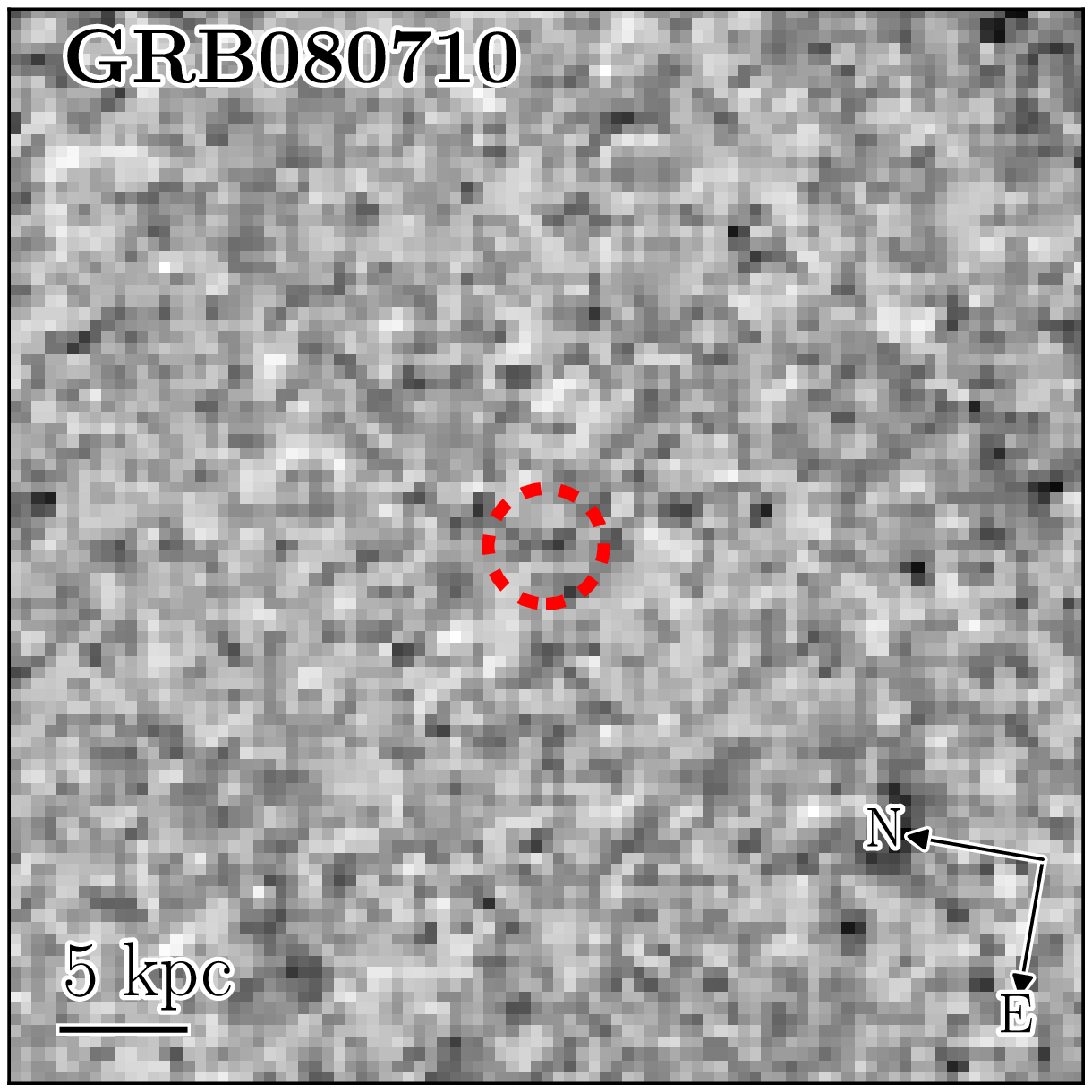}}
\subfloat{\includegraphics[width=0.2\linewidth]{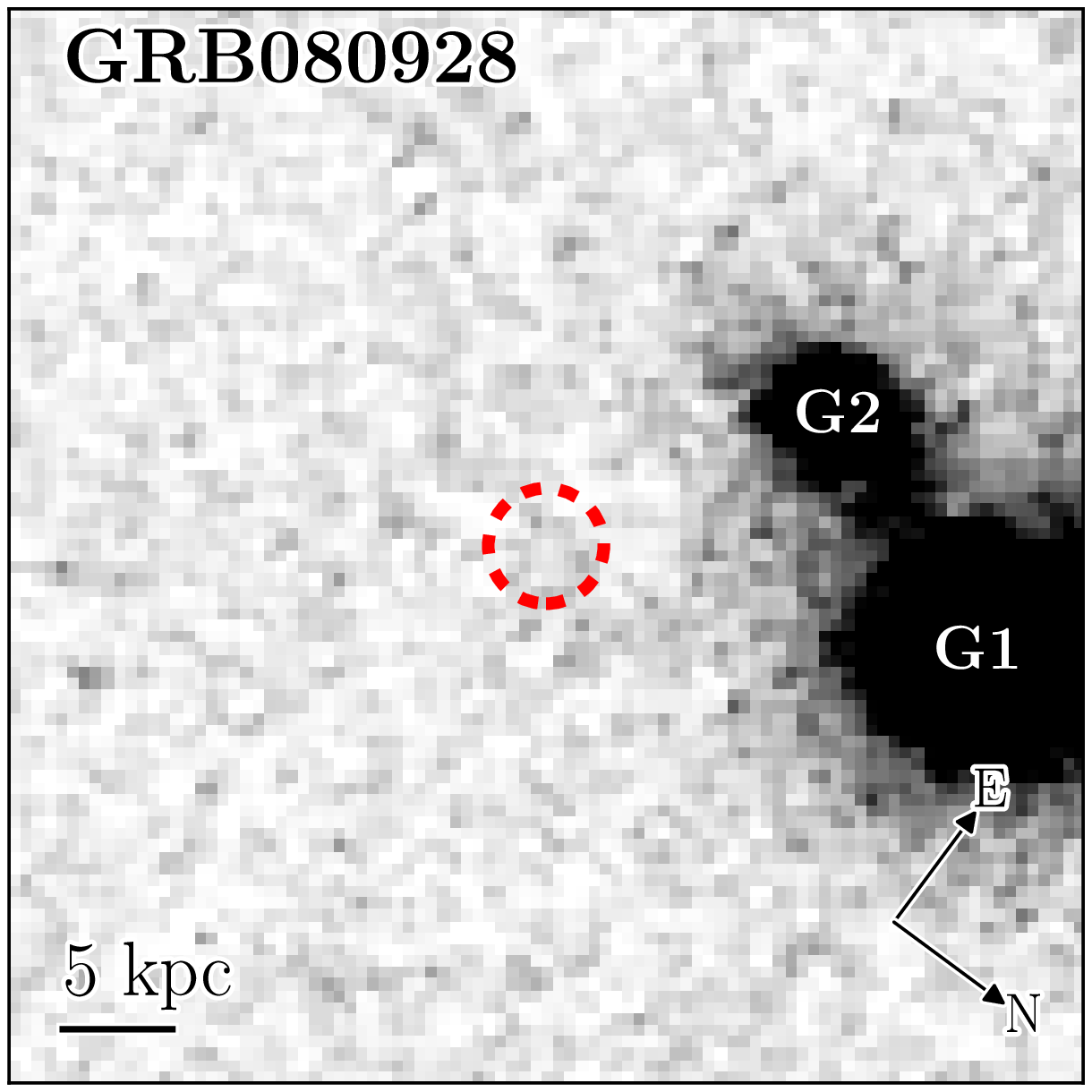}}
\subfloat{\includegraphics[width=0.2\linewidth]{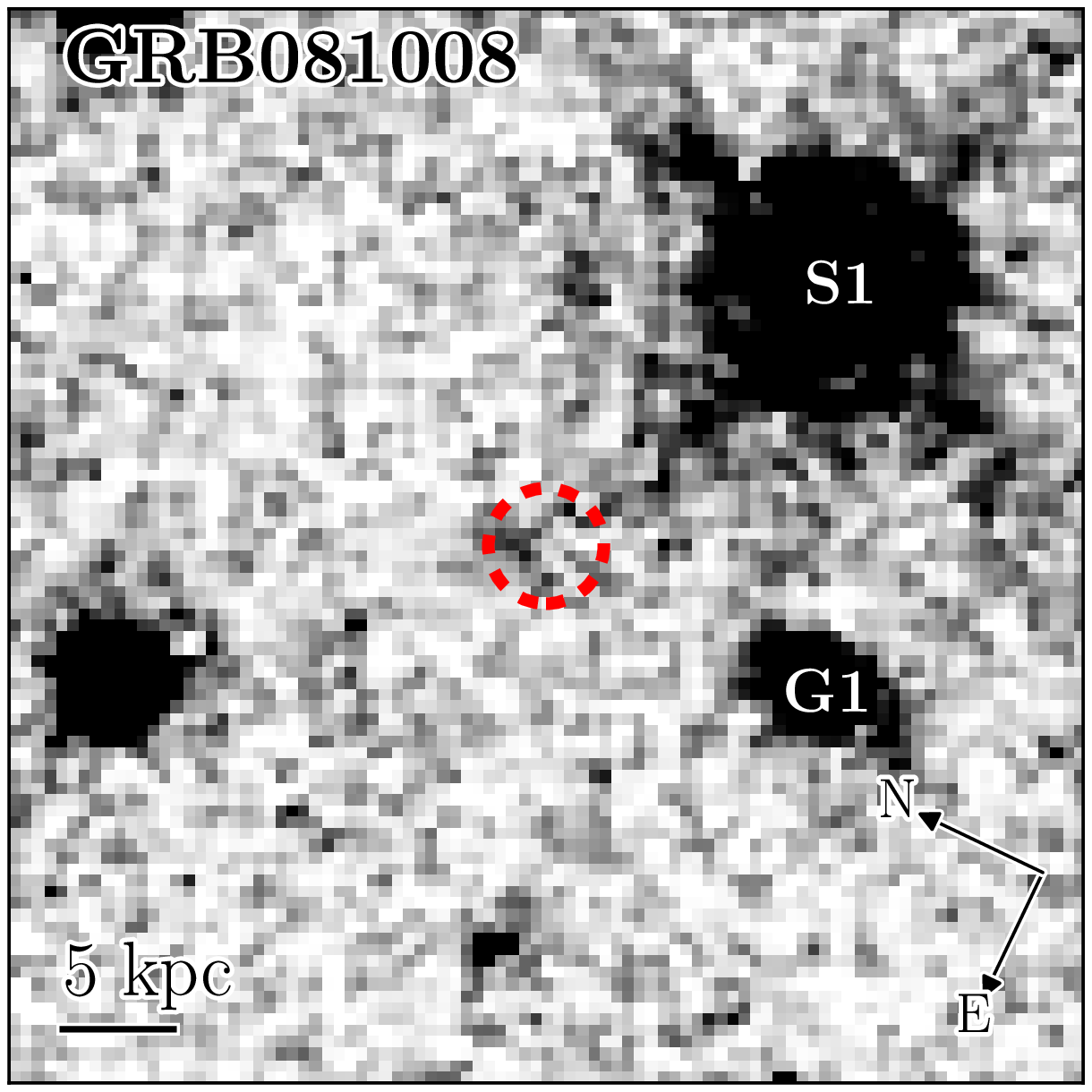}}
 \caption{Inverted stamps at the location of the GRBs in our sample for which no host was detected by our \sex{} detection criteria at the location of the afterglow. We note that a $\sim4 \sigma$ detection of flux was made underlying GRB 080710. Each stamp is 5~arcsec on a side with the orientation and linear scale at the redshift of the GRB indicated. The dashed red circle on each plot is centred on the location of the afterglow.} 
 \label{fig:nohosts}
\end{figure*}

For five bursts we do not detect any underlying host with our original \sex{}-based source selection criteria in the WFC3 imaging. The sky locations of these bursts are shown in \cref{fig:nohosts} and we here discuss their immediate environments.

GRBs 071031 and 080710 appear on a blank regions of sky with no obvious potential hosts nearby. For GRB 080603B a pair of (possibly interacting) galaxies are located NE of the GRB. The separation between this system and the GRB is $\sim$~2.2~arcsec, much larger than our estimated astrometic uncertainty on the GRB's position of $\sim$~80~mas. The distance of this system is $\sim$~16.7~kpc at the redshift of the burst -- this offset is much larger than the offset determined for any other burst in this sample or previously \citepalias[e.g.][]{bloom02,fruchter06,svensson10}, thus discrediting this system as the host of GRB 080603B.

GRB 080928 appears reasonably close to two bright, overlapping galaxies. \citet{rossi11} presented an investigation into the potential for either of the two nearby galaxies to be the host of the GRB 080928, with G2 here being clearly detected as a separate galaxy as oppose to a diffuse feature of G1, which, as the authors state, wasn't obvious from their ground-based observations. \citet{fynbo09} give an absorption-line redshift of $z = 1.6919$ for the burst. At this redshift, we find the centres of G1 and G2 are $\sim$21.5 and 16.8~kpc away from the GRB, respectively -- in agreement with \citet{rossi11}. Similar to the case of GRB 080603B, these offsets are much larger than for any other LGRB, very strongly disfavouring either as the host. \citet{rossi11} also have difficulty in modelling the SED of either galaxy at $z = 1.6919$. G2 is well fit by an irregular galaxy at $z = 0.736$, the redshift of an intervening absorption system seen in the spectra of \citet{fynbo09}, indicating this could be the system responsible.

For GRB 081008, photometry is complicated by the diffraction spike of a nearby bright star. The sources of flux near our astrometric afterglow location cannot be confidently distinguished as coming from a putative host galaxy or the bright star, and do not constitute significant detections in any case, due to the higher background noise.

We performed simple aperture photometry at the adopted location of the GRBs using 0.4~arcsec radius apertures, with sky determination made using an annulus aperture, and the appropriate F160W zeropoint\footnote{\url{http://www.stsci.edu/hst/wfc3/phot_zp_lbn}}. This resulted in the $\sim4 \sigma$ detection at the location of GRB 080710; although we cannot perform our morphological and explosion site analysis for the host, we include this measurement in our discussion of GRB host luminosities. For the other four, the following 3$\sigma$ magnitude limits for the GRB hosts were found\footnote{These limits were not affected significantly when calculations were repeated over our typical alignment uncertainty.}:

\begin{center}
\begin{tabular}[h]{lcc}
 \hline
 GRB & \multicolumn{2}{c}{Host magnitude (AB mag)} \\
     & $m_\text{F160W}$ & M$_{\text{F160W}/(1+z)}$ \\
 \hline
 071031  & $>$27.08 & $>-18.15$\\
 080603B & $>$27.13 & $>-18.09$\\
 080710  & 26.02$\pm0.26$ & $-16.87\pm0.26$\\
 080928  & $>$26.85 & $>-17.49$\\
 081008  & $>$26.88 & $>-17.75$\\
 \hline
\end{tabular}
\end{center}

Absolute magnitudes are given for the observed wavelength ($\lambda_\textrm{cen} = 15400/(1+z)$~\AA{}). These limits are within the observed magnitude distribution, albeit at the faint end (as expect given our deep imaging and redshift cut), and thus cannot be considered a separate population of extremely faint hosts with our current observations. 

Despite a non-detection with \hst{}\footnote{A non-detection underlying the GRB explosion site was independently made by \citetalias{blanchard16} using the same observations.}, within the literature we find a detection of the host of GRB~071031 in Ly$\alpha$ ($F$(Ly$\alpha$) $= 23.6\pm2.7 \times 10^{-18}$~erg~cm$^{-2}$~s$^{-1}$) by \citet{milvang12} based on the spectrum of \citet{fynbo09}, with \citet{kann10} finding that the afterglow did not suffer any significant extinction due to dust. We redrizzled this image with a coarser resolution of 0.26~arcsec/pixel to attempt to detect any lower surface brightness sources, however our detection routine did not find a nearby source in this case either. Using the relation between $L$(Ly$\alpha$) and SFR \citep{milvang12}, the detection implies a SFR of 1.3~\msun{}~yr$^{-1}$ (neglecting any impact from dust attenuation). The location of GRB071031 has been observed by Spitzer-IRAC at 3.6$\mu$m. Following the procedure described in \citet{perley16b} we find $m_{3.6\mu\textrm{m}} >$ 24.63 mag (AB) and estimate a corresponding host galaxy mass limit of $\log(\textrm{M}_\textrm{star}) < 9.6$~\msun{}. The limit on the specific SFR is therefore $> 0.33$~Gyr$^{-1}$, this is not stringent enough to mark the host as exceptional amongst LGRB hosts \citepalias[e.g.][]{svensson10}. Using the mean $\textrm{UV} - B = 1.22$~mag colour for star-forming galaxies at $z = 2-3$ \citep[][see \citealt{chen09}]{shapley05} to make an estimate of the continuum level, our $\sim$B-band rest-frame observation with \hst{} implies a limit of $>67.8$~\AA{} for the equivalent width of Ly$\alpha$. We note this is of course quite uncertain due to the unknown true spectral shape.

The host of GRB~081008 was suggested to have been detected in a Gemini/GMOS-S $r$-band acquisition image \citep{cucchiara08c}, located 2.1~arcsec from the optical afterglow and at $m_r = 20.8\pm0.1$. This object was within the slit used for spectroscopic observations of the optical afterglow and was determined to contain multiple metal absorption features at a common redshift of $z = 1.967$, the same as that of the GRB. The authors note that, if this was the host, it would be one of the brightest LGRB hosts seen. Furthermore, the associated projected offset of 16.8~kpc would be double that seen for any other LGRB (e.g. \citetalias{bloom02}, \cref{fig:offsets}). After astrometrically aligning the acquisition image with our HST image, we determine the suggested host, labelled `S1' in \cref{fig:nohosts}. This object is not extended and probably stellar in nature. We attempted to reveal any potential extended component by rotating the \hst{} image about this source, and then subtracting the rotated image from the original, but found no evidence for any extended component. The galaxy located to the south of the afterglow position (labelled `G1') is at a similar angular distance but is not detected in the acquisition image. Nevertheless, there is no detected stellar population component from either of these sources extending to the location of the afterglow and the source-afterglow offsets are much larger than seen for any other LGRB, discrediting either as the host, notwithstanding the probable stellar nature of one of them.

\subsection{LGRB Host properties}

\subsubsection{Luminosities}
\label{sect:hostlum}

\begin{figure*}
 \includegraphics[width=\linewidth]{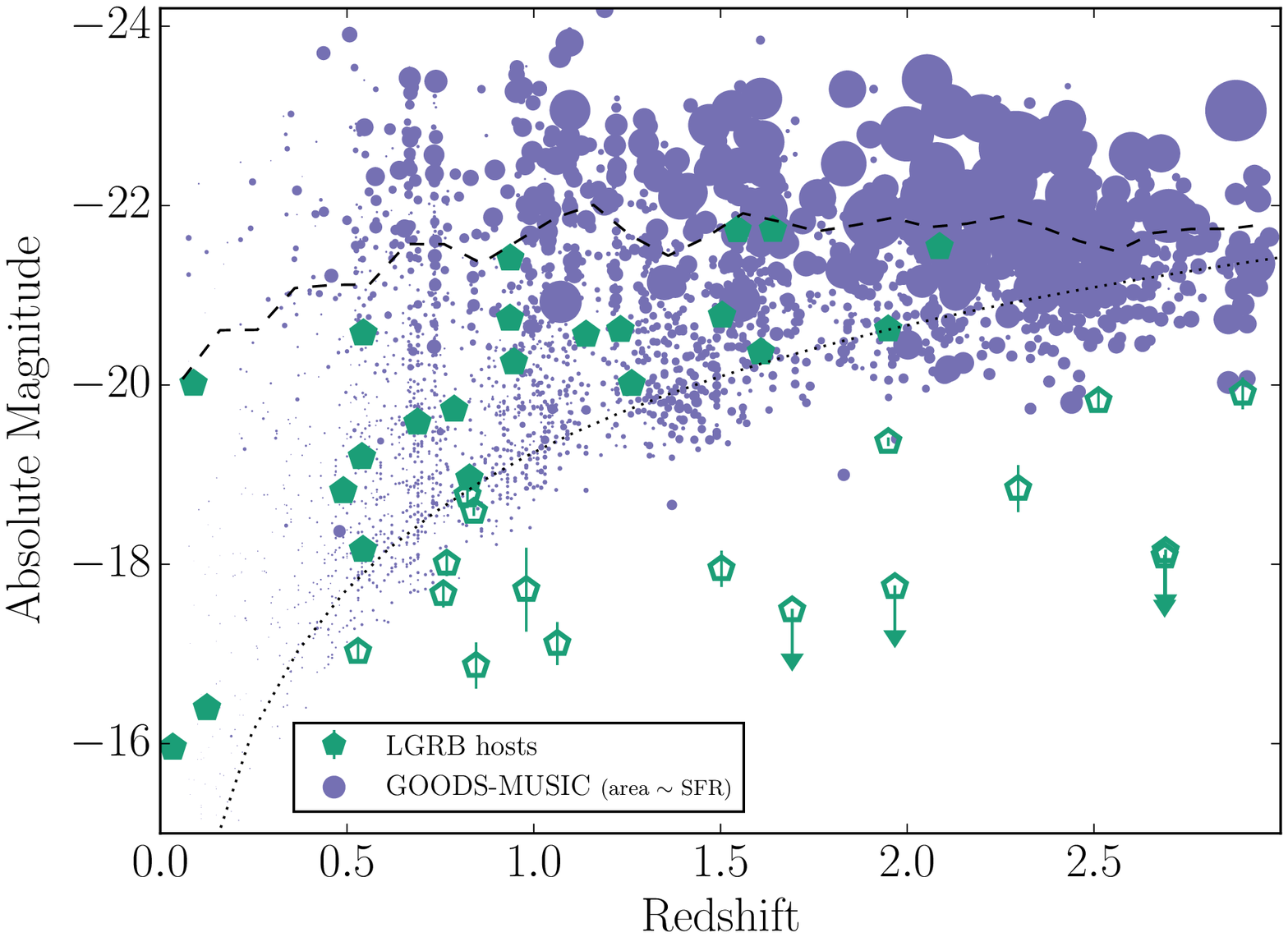}
 \caption{The redshift distribution of F160W ($\sim$H-band) observed absolute magnitudes for the LGRB hosts in this study and a sample of galaxies from the GOODS-MUSIC survey \citep[][see text]{santini09}. The black dotted line indicates $m = 24$~mag, roughly the 90~per~cent completeness limit of GOODS-MUSIC. LGRB detections fainter than this depth are show as open symbols, with arrows indicating non-detections. GOODS-MUSIC marker sizes are weighted by their SFR. A moving SFR-weighted absolute magnitude average for the GOODS-MUSIC sample is shown by the dashed line.}
 \label{fig:absmag}
\end{figure*}

The observed luminosities of the LGRB sample are shown in \cref{fig:absmag} alongside those of the GOODS-MUSIC comparison sample of field galaxies over the same redshift range. In order to facilitate a comparison to the GOODS-MUSIC sample we also remove those host detections fainter than the limit of the GOODS-MUSIC survey ($m_\textrm{H} = 24$~mag, assuming a conservative $H-K = 0.2$~mag,  \citealt{glass84}). As the progenitors of LGRBs are expected to be massive stars, the chance of a galaxy to host a LGRB can be expected to be proportional to its SFR. We thus plot the GOODS-MUSIC sample with markers corresponding to their SFRs determined in \citet{santini09}.\footnote{SFRs were determined using two indicators: SED fitting, and the mid IR flux. We present results based on comparison to the SED fitting technique but note the choice of SFR indicator has minimal impact on quoted significances.} From \cref{fig:absmag} it is clear the LGRB host population is not representative of the field galaxy population weighted by their SFR, and therefore that our sample of LGRBs are not unbiasedly following the distribution of star formation. We find that all 22 LGRB hosts within the survey limits of the GOODS-MUSIC sample are below the average SFR-weighted luminosity of the field galaxy population at that redshift (although we note that for the lowest redshift events, at $z < 0.5$, the small surveyed volume is hampering field galaxy numbers). Luminous galaxies ($M \simlt -22$~mag) are not present in our LGRB host population, although such galaxies can be host to dark-GRBs (\citealt{perley16b}, \cref{sect:discusshosts}).

\subsubsection{Sizes}
\label{sect:hostsize}

Resolved imaging allows us to probe the physical size of the host galaxies. 
Median values of enclosed-flux radii for the hosts (and GRB host-offsets, presented in \cref{sect:grboffsets}) are given in \cref{tab:medianprops}, alongside those of our comparison samples.\footnote{Uncertainties on median quantities were determined through bootstrapping of each sample. Within the large number of realisations, we remove a fraction of entries for each value given by $P_{i,\text{ch}}$ to account for the confidence of each host association.}

\begin{figure}
\centering
 \includegraphics[width=\columnwidth]{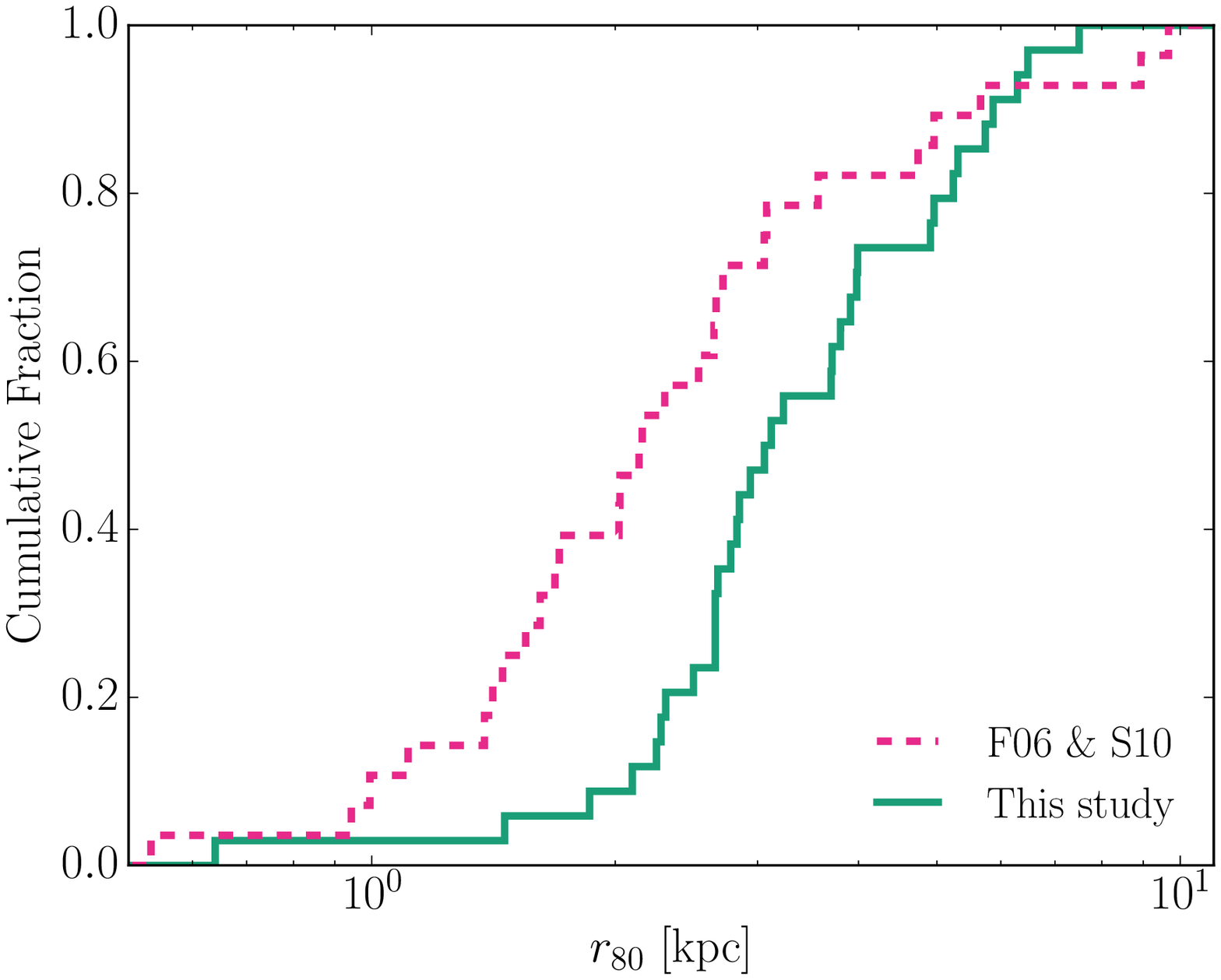}
 \caption{The cumulative distribution of $r_{80}$ for GRB hosts in this study and the combined sample of \citetalias{fruchter06} and \citetalias{svensson10} ($z < 3$). The data of \citetalias{fruchter06} and \citetalias{svensson10} were taken predominantly using STIS/CLEAR or ACS observations and observed wavelengths in the blue--visual. The rest-frame wavelengths of these observations are sensitive to much shorter wavelengths than the F160W observations presented in this study, which is a potential explanation for the difference observed (see text).}
 \label{fig:r80}
\end{figure}

Our $r_{50}$ distribution was found to be in good agreement with ones based on imaging in rest-frame UV \citepalias{bloom02} and a UV-optical-NIR amalgamation \citepalias{blanchard16}. However, we find our $r_{80}$ distribution to be shifted to larger values in comparison to the sample of \citetalias{fruchter06} and \citetalias{svensson10} (\cref{fig:r80}). $r_{80}$ values were determined in \citetalias{fruchter06} and \citetalias{svensson10} using the same \sex{} detection method as employed here. The medians are discrepant at the 3$\sigma$ level, with an Anderson-Darling (AD) test giving $p = 0.003$ that the two samples are drawn from the same parent population.
Despite the samples being over the same $z < 3$ redshift range, we unfortunately only have one host in common between the samples, GRB~060218. This burst was in an extremely low redshift host and atypical of the samples. For this host we find $r_{80} = 0.65$~kpc compared to the value from \citetalias{svensson10} (corrected to our cosmology) of 0.53~kpc. This is not a particularly striking difference, albeit in the same direction as our overall discrepancy between the samples. Measurements in the \citetalias{fruchter06} and \citetalias{svensson10} sample were predominantly made on STIS/CLEAR, supplemented with ACS observations, therefore most of the flux is from observed blue-visual light -- i.e. rest-frame UV. In contrast, our WFC3/F160W observations probe down only to visual wavelengths even for the most distant members of the sample, with more typical rest-frame wavelengths being around red-optical to NIR. 
A more extended distribution of $r_{80}$ values is similar to the findings of \citetalias{blanchard16}, who noted their $r_{80}$ distribution was shifted to higher values in general, compared to that of \citetalias{svensson10}. Our distribution of $r_{80}$ is in good agreement with that of \citetalias{blanchard16}, who used an amalgamation of optical and NIR observations. We again note significant overlap with the sample of \citetalias{blanchard16}, and as such this agreement is expected given the similar analysis used to determine the host sizes, but also serves to verify these findings on the sizes of LGRB hosts.

 \begin{table*}
 \begin{threeparttable}
  \caption{Median properties of the hosts and offsets of LGRBs}
  \begin{tabular}{lcccc}
 \hline
 Quantity                 & This Study      & \multicolumn{3}{c}{Literature} \\
                          &                 & \citetalias{bloom02} & \citetalias{svensson10} & \citetalias{blanchard16} \\
 \hline
 $r_{50}$                 & $1.7\pm0.2$~kpc & $1.5\pm0.5$~kpc      & ---                     & $1.8\pm0.1$~kpc          \\
 $r_{80}$                 & $3.1\pm0.4$~kpc & ---                  & $2.2\pm0.3$~kpc         & $\sim 3$~kpc             \\
 Offset (centre)          & $1.0\pm0.2$~kpc & $1.4\pm0.8$~kpc      & ---                     & $1.3\pm0.2$~kpc          \\
 Offset (centre)/$r_{50}$ & $0.6\pm0.1$     & $0.8\pm0.3$          & ---                     & $0.7\pm0.2$              \\
 Offset (brightest pixel) & $0.8\pm0.2$~kpc & ---                  & ---                     &  ---                     \\
 \hline
 \end{tabular}
 \label{tab:medianprops}
 \end{threeparttable}
 \end{table*}

\subsubsection{Morphologies}
\label{sect:hostmorph}

For those bursts where we detect a host, a wide variety of morphologies are seen -- \sex{} segmentation maps of the hosts are plotted in \cref{fig:segmaps,fig:segmaps_noalign}. Some exhibit smooth disk profiles (e.g. GRBs 060218, 090418A) with many showing lower surface brightness, extended features that may be the signatures of recent disturbance of the host (e.g. GRBs 060614, 080707). A number display very asymmetric or even multiple cores within the light distribution ($\sim$~11/34). However, a caveat to consider for morphologies determined from imaging alone is the prospect of chance alignment being responsible for producing multiple cores/clumps \citep[e.g. as was the case for GRB 060814,][]{jakobsson12}. Our observations are in good agreement with the findings of both \citet{wainwright07}, who found 30~per~cent of their GRB host sample exhibit signatures of irregular or asymmetric structure, and \citet{conselice05}, who found a diverse spread of morphological types for GRB hosts over similar redshift ranges.

\begin{figure*}
\centering
 \includegraphics[width=\linewidth]{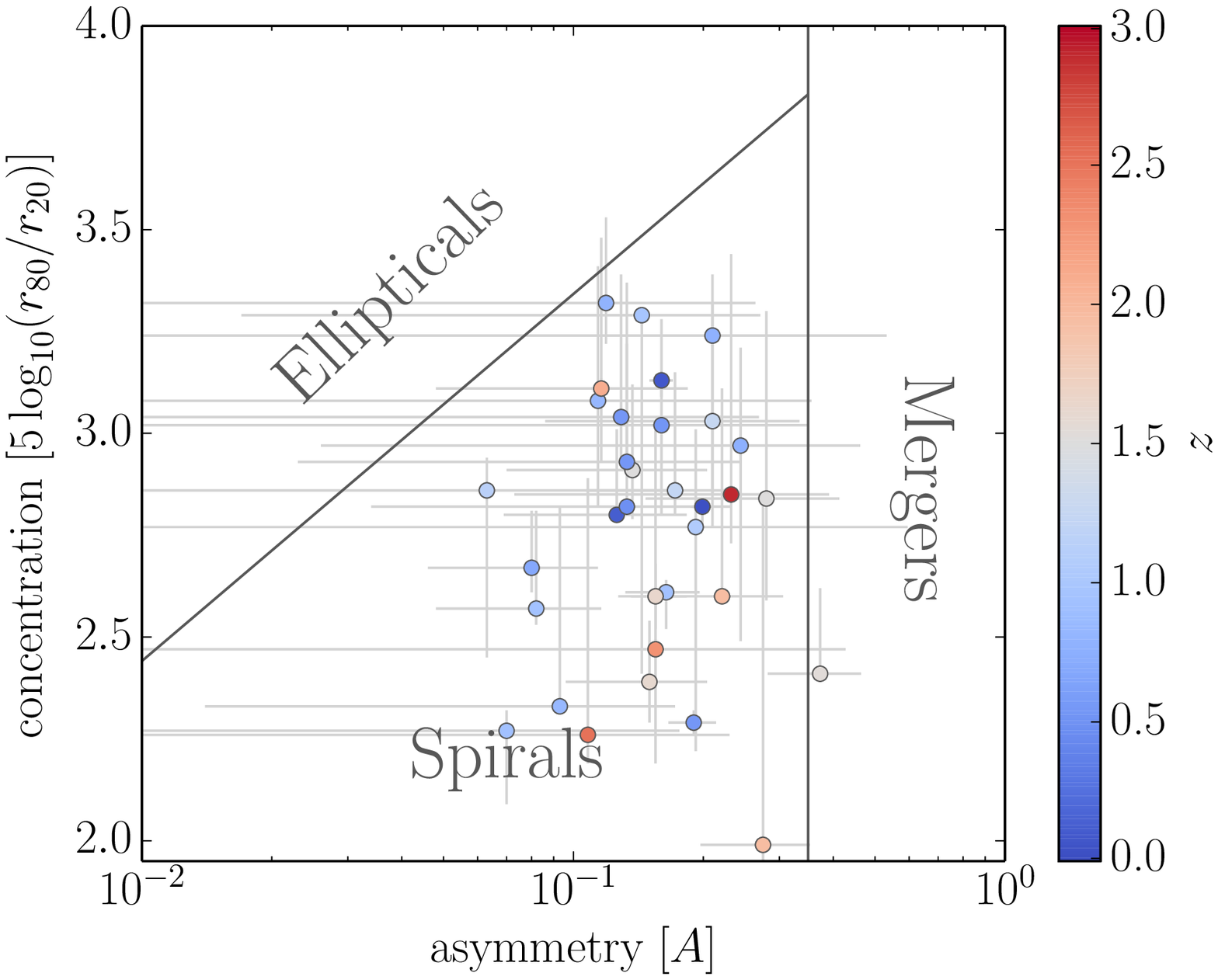}
 \caption{The morphology of the LGRB hosts parameterised by their concentration and asymmetry \citep[see][and references therein]{conselice05}. Regions that separate different galaxy classes are plotted using the definitions in \citet{conselice05} -- these are based on visual inspection of low-redshift galaxies. The majority of hosts appear similar to spiral-like galaxies in this parameter space, with a small number appearing in the region of merger-like and elliptical-like. However, these strictly defined borders are in reality a simplified representation of the parameter space even for low-redshift samples, and the $CA$ parameter space does not fully encapsulate the morphology of a galaxy. As such individual hosts cannot be confidently assigned a Hubble type morphology, which are in any case more appropriate for low-redshift galaxies, based on this analysis alone. Markers are colour coded by redshift.}
 \label{fig:CA}
\end{figure*}

In order to study the host morphologies, we measure the concentration and asymmetry of the hosts following the work of \citet{kent85, bershady00, conselice00}. Concentration, $C$, uses the flux radii determined as above and is given by $C = 5\log_{10}(r_{80}/r_{20})$. Larger values of $C$ mean the light of the host is dominated by a compact central region, such as in elliptical galaxies, whereas spiral galaxies tend to have a larger fraction of their flux in their extended profile. Asymmetry, $A$, is determined by rotating the image about the host's centre, subtracting this rotated image from the original and analysing the absolute size of the residual. $A$ is parametrised in the range 0 to 1 and it's calculation is extensively detailed in \citet{conselice00, conselice05}. In the idealised case, a smooth, axisymmetric galaxy profile would return $A = 0$. For spiral galaxies, with extended patchy emission, and merging or disturbed systems, $A$ increases. These broad morphological distinctions between galaxy types can be used as a guide for the underlying population of our LGRB sample. We follow the methods of previous works in the calculation of these quantities with 2 small alterations: we use Monte Carlo to determine the location of the centre for the asymmetry rotation and use resampled realisations of the \hst{} images to determine the uncertainty in $C$ in the same manner as for the enclosed flux radii (see \cref{sect:radphot}). Concentration and asymmetry values for individual hosts are given in \cref{tab:hostprops} and we plot them in \cref{fig:CA} along with the morphological grouping borders of \citet{conselice05} based on low redshift populations with visual classifications, with the `Spirals' region also being where irregular galaxies are typically located \citep{conselice03,conselice05}.

The population is dominated by galaxies in similar region of this parameter space to spiral galaxies, with a small number being close to or just inside the regions defined by ellipticals and mergers. However, the assignment of morphological types based on the $CA$ parameter space is complicated. Morphological definitions and the inferred properties of the hosts based on these morphologies (e.g. elliptical galaxies dominated by old, passive stellar populations) are appropriate for field galaxy populations at low redshift. At higher redshifts ($z \simgt 1$) there is a drop in normal Hubble types of galaxies, in favour of peculiar and irregular types \citep{abraham96,brinchmann98}, which, for example, can be found at a wide range of asymmetries \citep[albeit more typically at higher values][]{conselice05b}. Furthermore, even for low-redshift galaxies where a visual morphology can be assigned, different morphologies are not cleanly distinguished in the $CA$ parameter space. Regions, particular near borders, are comprised of composite populations when looking at large numbers of field galaxies \citep{conselice05}. These issues make it difficult to confidently assign a morphology for individual hosts based on the $CA$ parameter alone. As such, these regions are shown as indicators of the distribution of more well-known galaxy types and we mainly use the $CA$ parameter space to investigate changes within the LGRB host population in this parameters with redshift.

The markers in \cref{fig:CA} are colour coded by redshift, which upon visual inspection indicates higher redshift events ($z > 1.5$) appear less concentrated than the lower redshift events. A number of effects are to be considered here, however: the shift in rest-frame wavelength of the observations in the high- and low-redshift regimes, the changing resolution and brightness of the detections with distance, and, thirdly, evolution of the underlying galaxy population over cosmic time. This is further discussed in \cref{sect:discusshosts}.

\subsection{GRB explosion sites}

\subsubsection{Offsets}
\label{sect:grboffsets}

\begin{figure}
\centering
 \includegraphics[width=\columnwidth]{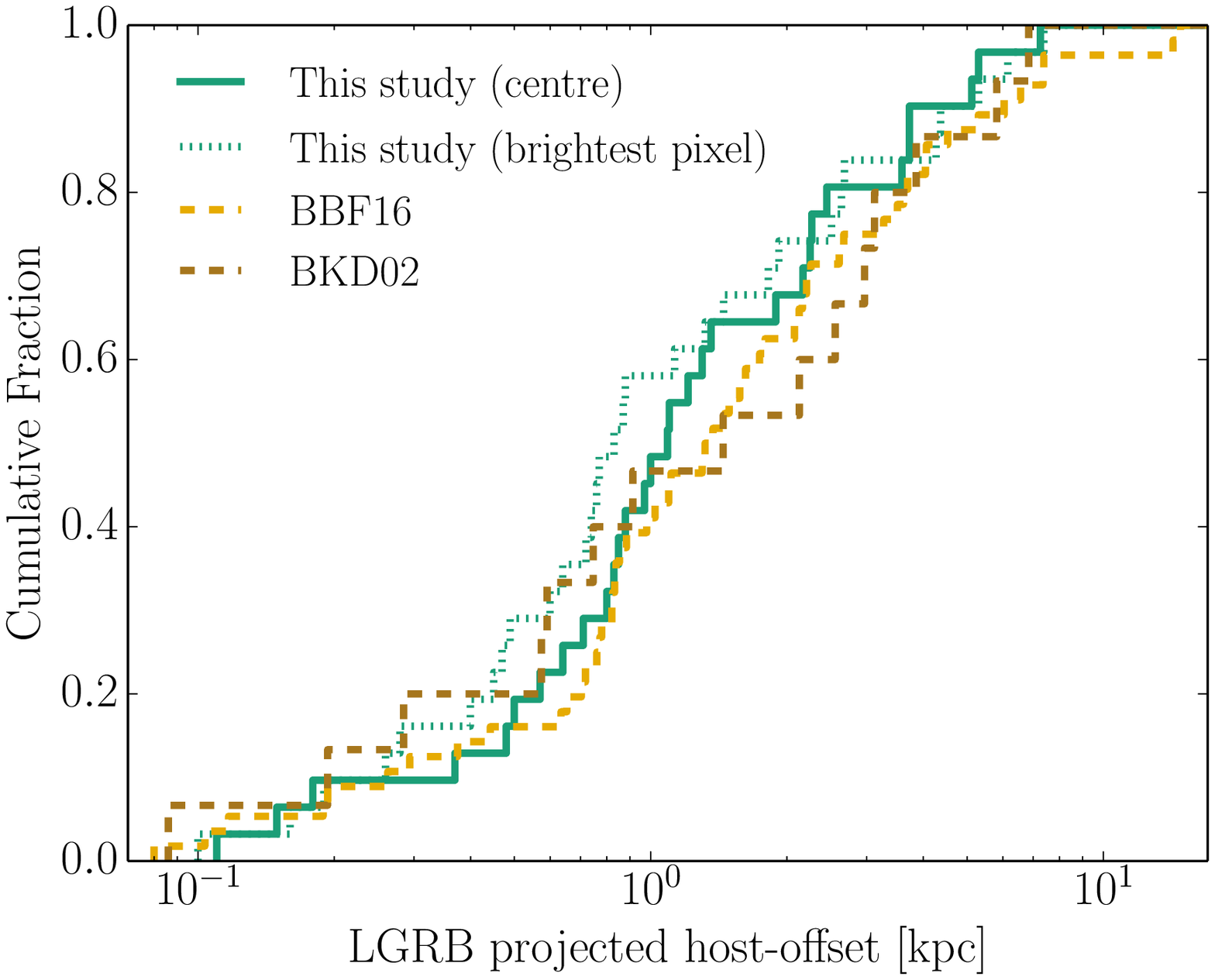}
 \caption{The cumulative distribution of host galaxy barycentre offsets for GRBs in this study and those of \citetalias{bloom02} and \citetalias{blanchard16} ($z < 3$), which were recalculated using the cosmology adopted in this paper. We also plot the distribution of GRB offsets from their host's brightest pixel for our sample.}
 \label{fig:offsets}
\end{figure}

The host-offset distribution for transients can be a powerful diagnostic to their origin. The median LGRB offset of the \citetalias{bloom02} sample used here is $1.4\pm0.8$~kpc (recalculated using the cosmology of this paper), with $1.3\pm0.2$~kpc found for the sample of \citetalias{blanchard16}. These values are in good agreement with $1.0\pm0.2$~kpc determined for our sample (\cref{tab:medianprops}). Furthermore, when considering the distributions of the offsets, we find excellent agreement with those of \citetalias{bloom02} and \citetalias{blanchard16}, as shown in \cref{fig:offsets}. Although values for the different samples were found with different filter observations, since the barycentre of a galaxy remains reasonably stable across the UV to IR, this good agreement is expected. The brightest pixel offset distribution for our GRB sample is in very good agreement with our host barycentre offsets, indicating the choice of host centering method, and typical offset uncertainties, has little impact on our results. 

\begin{figure}
\centering
 \includegraphics[width=\columnwidth]{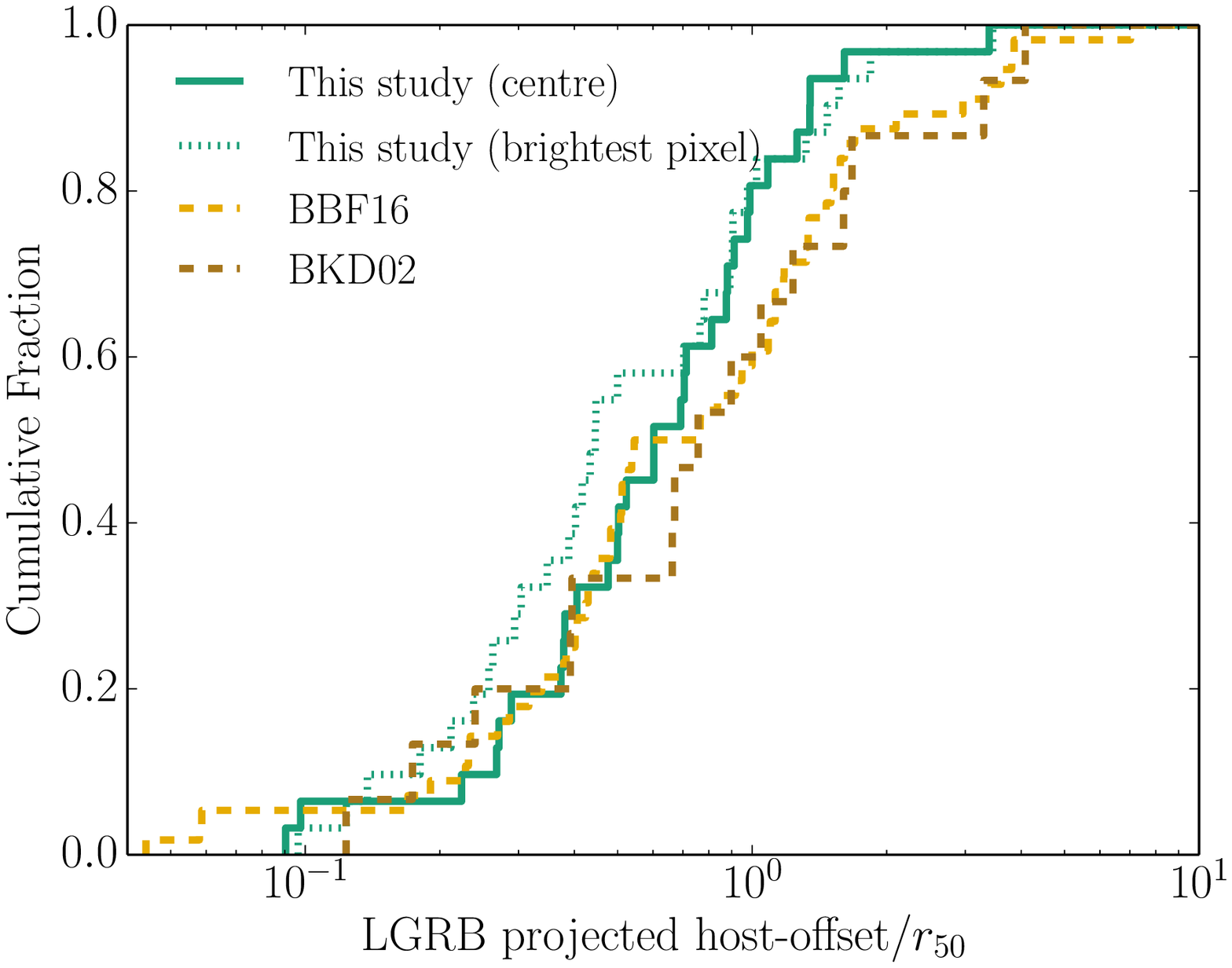}
 \caption{The cumulative distribution of host-normalised offsets for GRBs in this study and those of \citetalias{bloom02} and \citetalias{blanchard16} ($z < 3$). We also plot the distribution of normalised GRB offsets from their host's brightest pixel for our sample.}
 \label{fig:offsetsnorm}
\end{figure}

The GRB offsets can also be expressed normalised to measurements of their host galaxy. The GRB offsets normalised by $r_{50}$ are shown in \cref{fig:offsetsnorm}, and we again note similarity with the corresponding distributions of the other studies. There appears to be some indication that our values do not extend to the higher values found by these studies but the distributions as a whole are statistically indistinguishable. We find the median $r_{50}$-normalised offset is $0.6\pm0.1$, again in good agreement with $0.8\pm0.3$ and $0.7\pm0.2$ found by \citetalias{bloom02} and \citetalias{blanchard16}, respectively.

\subsubsection{\fl{} analysis}
\label{sect:grbflight}

\begin{figure}
\centering
 \includegraphics[width=\columnwidth]{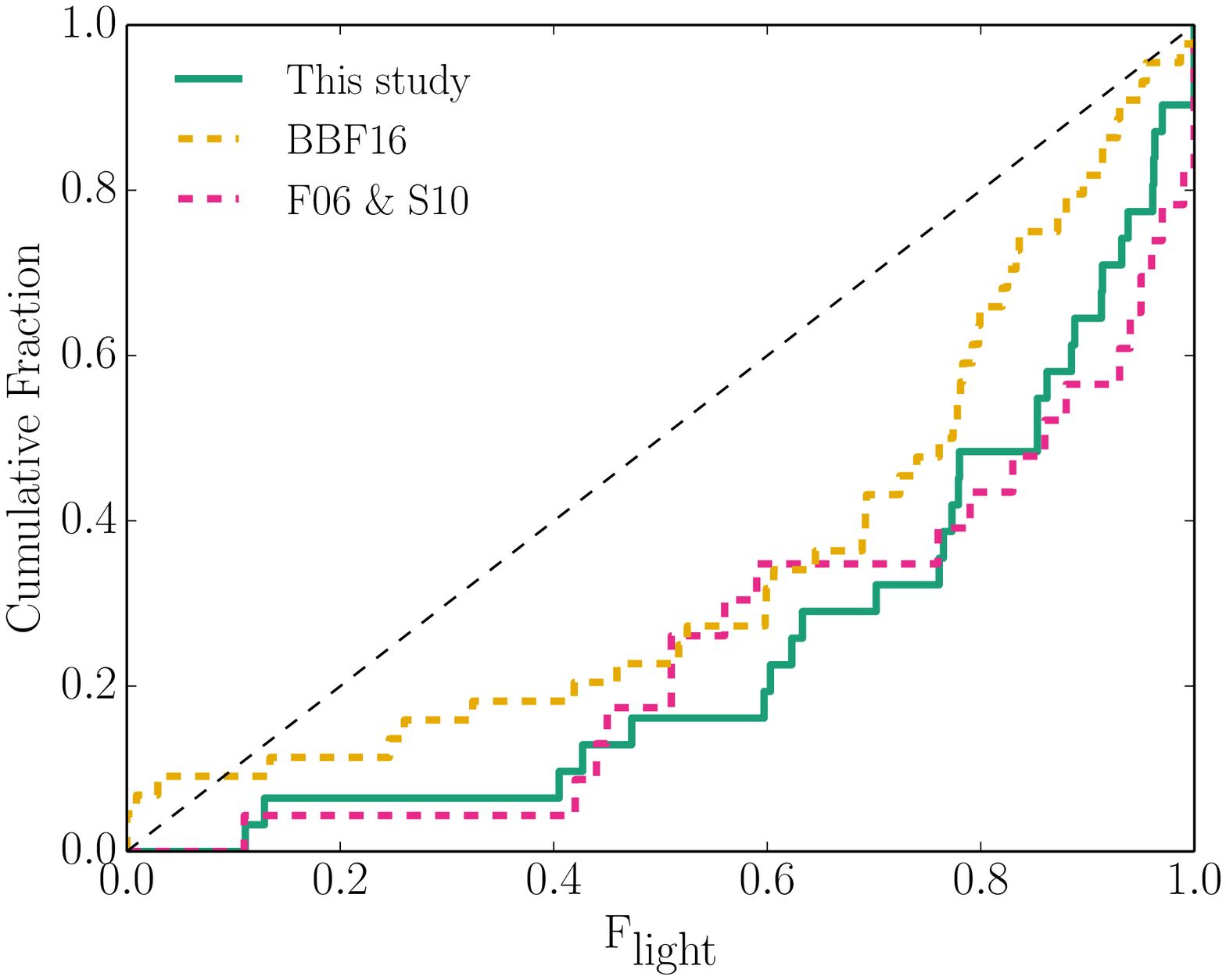}
 \caption{The cumulative distribution of \fl{} values for GRB hosts in this study and the combined sample of \citetalias{fruchter06} and \citetalias{svensson10}, and that of \citetalias{blanchard16} ($z < 3$). Observations for this study are exclusively in F160W. Those of \citetalias{fruchter06} and \citetalias{svensson10} were made in optical filters, with \citetalias{blanchard16} comprised of both optical and NIR.}
 \label{fig:flight}
\end{figure}

A visual representation of the \fl{} analysis for the GRBs where the analysis was performed is given in \cref{fig:segmaps}. Pixels used for the \fl{} calculation (those deemed by \sex{} to belong to the host) are highlighted on a colour scale showing each pixel's \fl{} value, with the pixel used for the \fl{} of each GRB indicated by a black star. Visual inspection shows a bias for the GRB locations being coincident with high \fl{} values. Even when the GRB has a noticeable offset from the bulk of the light of its host (i.e.\ where high value \fl{} pixels are located), it is often seen that the GRB's location is overlaid on or next to a bright knot of galaxy light -- this is particularly evident for GRBs 050824, 060912A and 080520. In \cref{fig:flight}, the cumulative distribution of the \fl{} values for the GRBs is plotted, confirming that LGRBs preferentially explode on brighter regions of their hosts. In comparison, we plot the \fl{} analysis of \citetalias{svensson10} where the rest-frame wavelength of observations is UV and the degree of association found is similar. Conversely, the analysis of \citetalias{blanchard16} found a somewhat lower degree of association between LGRBs and the brightest regions of their hosts.\footnote{Note, we only include \fl{} values from \citetalias{blanchard16} where the location was determined to sufficient accuracy, following the authors' criterion -- i.e. that the area of the location uncertainty circle is $\leq 0.1 \times$ galaxy area.} Our distribution is formally consistent with both the \citetalias{svensson10} and \citetalias{blanchard16} distributions. Further discussion of \fl{} can be found in \cref{sect:discusssites} and we directly compare our \fl{} results for overlapping events with \citetalias{blanchard16} in \cref{sect:flight_compare}.

\begin{figure}
\centering
 \includegraphics[width=\columnwidth]{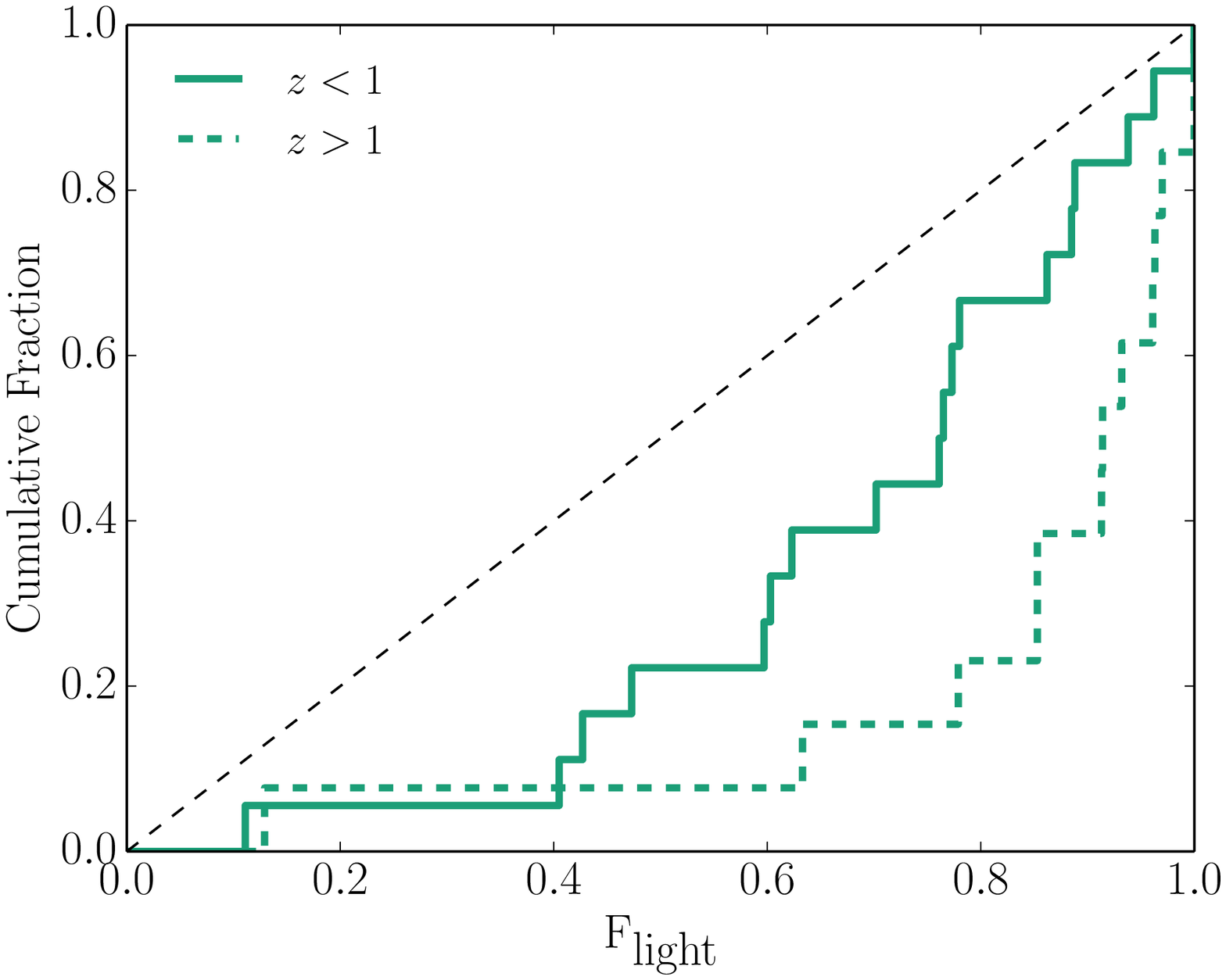}
 \caption{The cumulative distribution of \fl{} values for GRBs, split by redshift. The redshift split corresponds approximately to splitting the observations at rest-frame wavelength $\sim$7800\AA{}.}
 \label{fig:flight_zsplit}
\end{figure}

Given the large redshift range of the GRBs in the sample, observations in a single wavelength band will detect various regimes of rest-frame light from the hosts. Using a cut at $z = 1$ to form low- and high-redshift samples of roughly equal size, the events are observed in rest-frame red--NIR wavelengths ($>7800$~\AA{}) and visual ($\sim 4000$--$7500$~\AA{}) respectively. When plotting the cumulative \fl{} distributions for each sample (\cref{fig:flight_zsplit}) the high-redshift events have, in general, higher \fl{} values than the low-redshift events, with the AD test giving $p = 0.03$ that the two samples are drawn from the same parent population.

\section{Discussion}
\label{sect:discussion}

We have presented a sample of 39 LGRB locations imaged using WFC/IR on \hst{} in F160W. A search for the hosts was performed using astrometric ties from ground-based data of the GRB afterglows. Photometric, morphological and explosion site diagnostics have been determined in the case of a host detection. Five of the LGRBs have undetected hosts in our imaging, with no convincing nearby detections. Our detection limits, however, appear consistent with the detected luminosity distribution, without having to invoke a population of very low-luminosity hosts to explain these non-detections.

\subsection{LGRB Host Galaxies}
\label{sect:discusshosts}

Our sample of LGRB host galaxies is fainter than a distribution expected from a field galaxy population if the LGRBs are taken to unbiasedly trace star formation. These results are in agreement with the findings of \citet{vergani15}, where LGRB hosts at $z < 1$ were observed in the K-band. Although our sample extends to more distant sources than the sample of \citet{vergani15}, the rest-frame wavelengths of our observations remain \simgt{} 4000~\AA{}. We can then similarly use the luminosities as proxies for mass, inferring that the stellar masses of LGRB hosts are inconsistent with the SFR-weighted masses of field galaxies up to $z \sim 2$. The comparison at $z = 2-3$ is hampered by the completeness of the completeness limit of the comparison sample (\cref{fig:absmag}). Nevertheless, our detections and limits for LGRB hosts in this redshift range are of comparable luminosities to our lower redshift sample and we find no hosts with $M < -22$~mag in our full sample. The spectroscopic redshift restriction on our sample, however, disfavours the inclusion of optically-dark LGRBs. For these dark bursts the afterglow is subject to significant dust attenuation and is thus not detected, whereas the majority of redshifts for this sample were determined from afterglow features. Using the definition of \citet{jakobsson04}, our sample comprises only $\sim 10$~per~cent dark bursts. In contrast, it is estimated that overall $25-40$~per~cent of LGRBs are dark bursts \citep{fynbo09, greiner11}. 
The host galaxies of dark bursts appear distinct from their optically bright counterparts at similar redshifts \citep[e.g][]{kruhler11, perley13}. 
There appears to be a significant preference for LGRBs to explode in fainter galaxies up to $z \sim 1.5$, while the inclusion of dark LGRB hosts at higher redshifts improves the consistency between the LGRB host population and the SFR-weighted field galaxy population \citep{perley13, perley16b}.

The small physical sizes of LGRB hosts has been suggested in the literature \citepalias[e.g.][]{svensson10}, indicating they are comparable to the intriguing hosts of super-luminous SNe \citep{lunnan15,angus16}. This previous determination of the $r_{80}$ distribution was made using rest-frame UV imaging. Although we find good agreement with other works in the $r_{50}$ distribution of LGRB hosts \citepalias[from][]{bloom02, blanchard16}, our $r_{80}$ distribution is larger than that of \citetalias{svensson10} (\cref{fig:r80}). The combination of bluer filters and smaller pixel sizes is likely to mean that the surface brightness limits of the rest-frame UV data are somewhat bright, potentially accounting for some of this discrepancy.
Conversely, a visual inspection of our $r_{80}$ distribution and that of \citetalias{blanchard16} shows we are in good agreement, thus supporting their argument that LGRB hosts are in fact intermediate between SLSN hosts and CCSNe hosts, and not as compact as previously thought based on smaller samples of solely rest-frame UV observations. For comparison to the median LGRB host $r_{50}$ values in \cref{tab:medianprops}, \citet{paulino16} looked at a sample of low mass ($9 <\log_{10}M_\star/M_\odot < 10$) \Ha{}-selected galaxies and found the median $r_{50}$ values of to be $2-4$~kpc for $z = 0.4-2.23$, increasing up to $3-5$~kpc at $z \sim 0$.
Although we find that LGRB hosts are somewhat larger in comparison to other transient hosts than previously thought, we confirm the centrally concentrated distribution of galactocentric offsets of LGRBs \citepalias{bloom02, blanchard16}, being smaller than that of CCSNe \citep[e.g.][]{anderson09, prieto08,svensson10} and comparable to super-luminous SNe \citep{lunnan15}. 

The morphological diversity of LGRB hosts, specifically the abundance of irregular or disturbed hosts is well established \citep[e.g.][]{conselice05,wainwright07}. Around a third of the sample presented here appear to show some degree of asymmetry or multiple cores, and most display extended, low surface-brightness features. NIR imaging allows us to probe redder wavelengths of light, which trace more of the stellar mass of the hosts and are less skewed by ongoing star-formation regions. Such features as we see here therefore better approximate true morphological features of the mass distribution in the hosts. 
When considering asymmetry ($A$) and concentration ($C$) measures of the hosts in order to compare to morphological types, the population is shown to be comprised mainly of spiral-like and irregular-like galaxies but with some fraction of elliptical-like and merging systems. Only a small number appear on the border of merging systems, despite a considerable fraction of them being visually disturbed (\cref{fig:segmaps,fig:segmaps_noalign}). As noted in \citet{conselice05}, the $CA$ parametrisation does not capture all phases of merger activity, and these may be in the post-merger stage. We also find a number of hosts on the border region between ellipticals and spirals. The distribution of GRB hosts in the $CA$ parameter space is broadly similar to that found by \citet{conselice05}, where observations of GRB hosts over a similar redshift range were made in optical wavelengths (i.e. rest-frame blue-optical and UV). However, these authors found the population of GRB hosts extended to high concentration (around a quarter having $C \simgt 3.5$), whereas we find no such highly concentrated hosts from this sample. It was suggested that nuclear starbursts, likely to be prevalent amongst the strongly star-forming hosts of GRBs, could be acting to mimic the highly concentrated nature of elliptical galaxies. Our lack of highly concentrated hosts may then be attributed to the fact we are observing the hosts at longer rest-frame wavelengths, such that the contribution of a nuclear star-burst to the overall light profile is relatively less (as opposed to rest-frame UV imaging). The results of our $CA$ investigation here confirm the varied nature of GRB host morphologies found by \citet{conselice05}.

\begin{figure}
\centering
 \includegraphics[width=\columnwidth]{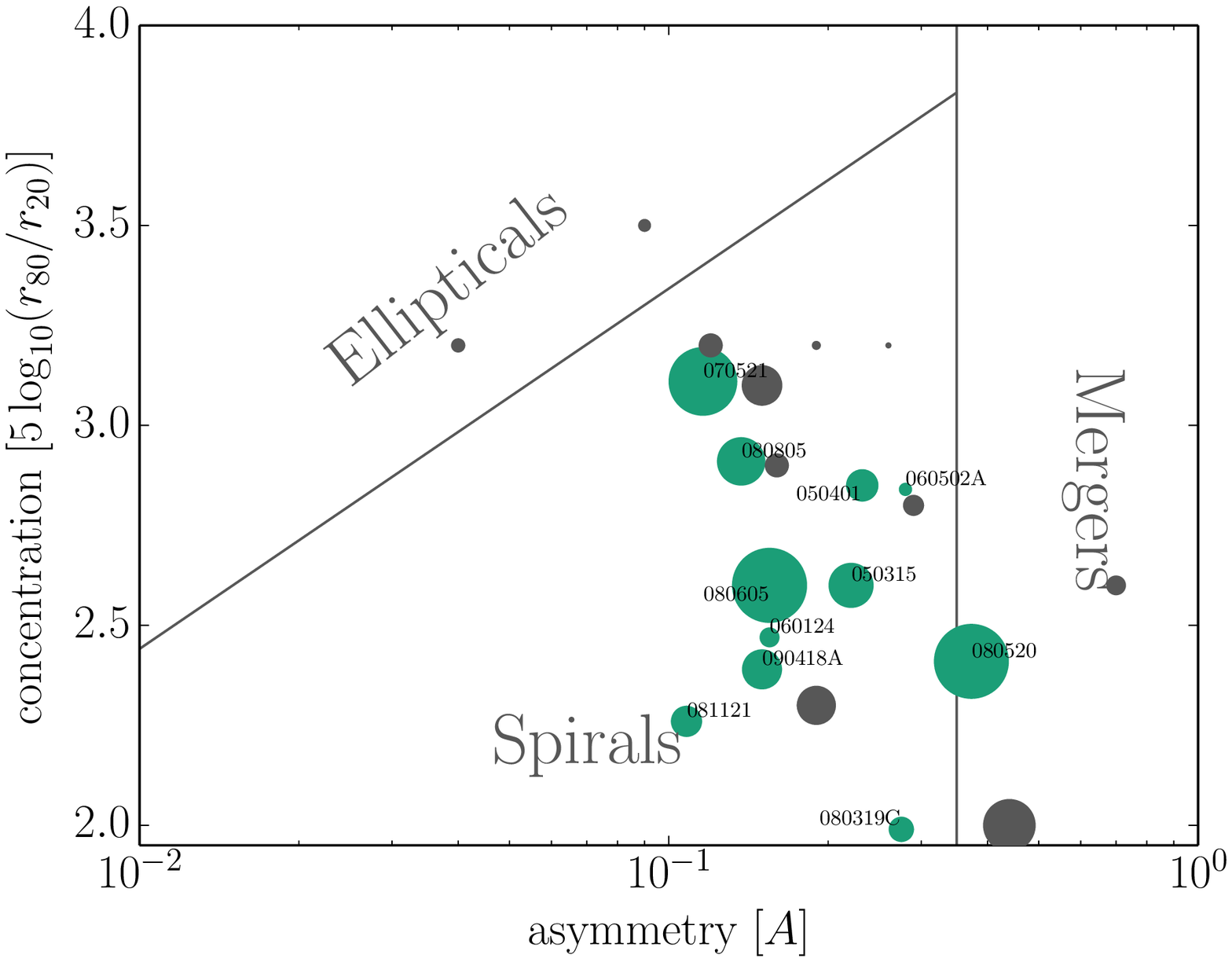}
 \caption{As for \cref{fig:CA}, but here we only plot events at $1.5 < z < 3.0$ from our sample. A $z < 1$ sub-sample from \citet{conselice05} is shown by grey markers. Due to the different observed wavelength filters used in each sample, both correspond to roughly optical rest-frame observations. Marker areas indicate the relative optical luminosities of the galaxies.}
 \label{fig:CA_conselice}
\end{figure}

As noted in \cref{sect:hostmorph}, a visual inspection of the LGRB hosts in the $CA$ parameters space suggests higher redshift events appear distinct from the lower redshift sample in concentration of their hosts, perhaps indicative of galaxy evolution (see \cref{fig:CA}). However, this observation is complicated by the rest-frame wavelength of the observations also changing significantly. Although galaxies are expected to have relatively stable morphology from the optical to NIR \citep{taylor07,conselice11}, we can test the evolution of $C$ independent of band-shifting effects by using the results of \citet{conselice05}. These authors calculated $C$ and $A$ values for a sample of GRBs based on \hst{} imaging in optical filters, with the typical observed wavelength centred at $\sim6000$~\AA{}. We make a cut at $z = 1$ of their sample to obtain $C$ and $A$ values for LGRB hosts in the optical regime. By taking a $z > 1.5$ sub-sample from this study, we then also have optical ($\sim 4000-6000$~\AA{}) measures of $C$ and $A$ for a sample LGRB hosts but at higher redshift. These two samples are plotted in \cref{fig:CA_conselice}. The distribution of asymmetry values are similar between the two samples. There appears evidence of a separation between high- and low-redshift events in concentration ($p = 0.03$); the median concentration of the low redshift sample of \citet{conselice05} is comparable to the largest of our $z > 1.5$ sample. The marker areas are proportional to the optical luminosities of the hosts, showing evolution of the typical host from a relatively luminous, diffuse spiral-like at higher redshifts to less luminous, compact knots at lower redshifts. This is consistent with the cosmic downsizing of star formation \citep{cowie96}, particularly low-metallicity star-formation if LGRBs are subject to a metallicity bias or cut-off \citep[see also][and references therein]{perley16b}.

\subsection{LGRB explosion sites}
\label{sect:discusssites}

When investigating the locations of the LGRBs within their hosts F160W light distribution, we find a strong association to the brightest pixels, with $\sim50$~per~cent of LGRBs exploding on the brightest 20~per~cent of their hosts (\cref{fig:flight}), in agreement with the findings of \citetalias{fruchter06} and \citetalias{svensson10}. 
\citetalias{blanchard16}, however, found a somewhat lower degree of association and, in particular, investigated this association in comparison to the offset of the bursts. When considering those bursts that have a galactocentric offset $> 0.5 \times r_{50}$, it was found that the \fl{} distribution lies close to a uniform distribution -- i.e. that these larger offset bursts have no preference for brighter regions of their hosts. For the low offset sample they found these have exclusively large \fl{} values.

However, the imposition of a threshold on the LGRB offset means that the \fl{} values cannot be used in their raw form, since there are now only certain (i.e. outer or inner) regions of the host where an LGRB can explode to be a member of the respective samples, by definition. Furthermore, since the barycentre of the host is likely to be spatially coincident with the brightest regions of the host, the exclusion of LGRBs close to this location naturally restricts the presence of large \fl{} values in their distribution, artificially weighting them towards lower values. Conversely, when considering only small offset bursts, almost any location within $0.5 \times r_{50}$ of the host centre has a large \fl{} value, weighting these bursts to higher values. This is unsurprisingly what was found by \citetalias{blanchard16} when using the \fl{} values in their raw form. When imposing cuts on the locations of LGRBs, one must also make the same cuts on the light distribution of the host. This is in order to correctly determine the \fl{} value for the LGRB considering only the flux from regions where the LGRB is permitted to explode in order to make it into the sample. 

\begin{figure}
\centering
 \includegraphics[width=0.45\columnwidth]{GRB090618_drz_0.065_0.8_flheat.eps}
 \includegraphics[width=0.45\columnwidth]{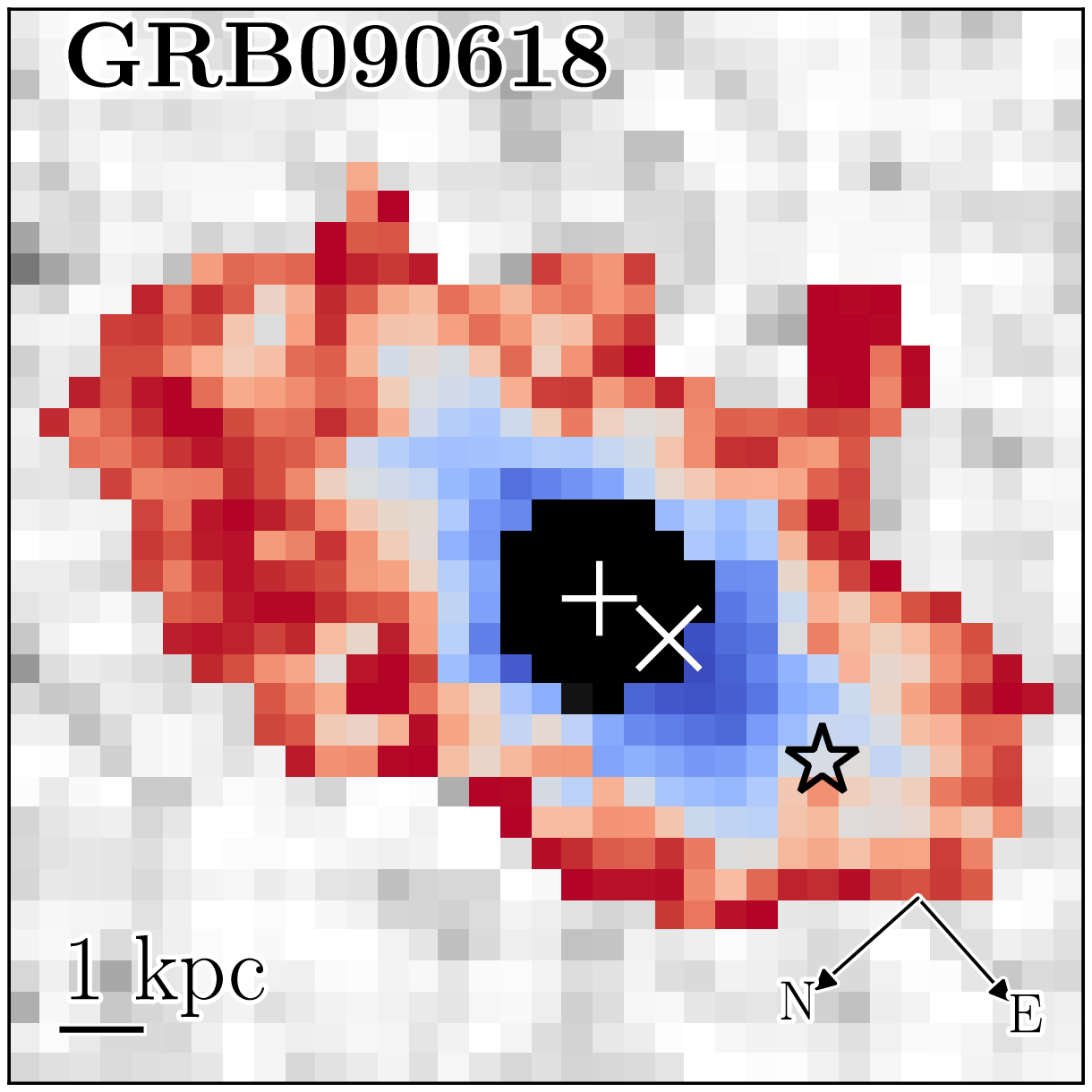}
 \includegraphics[width=\columnwidth]{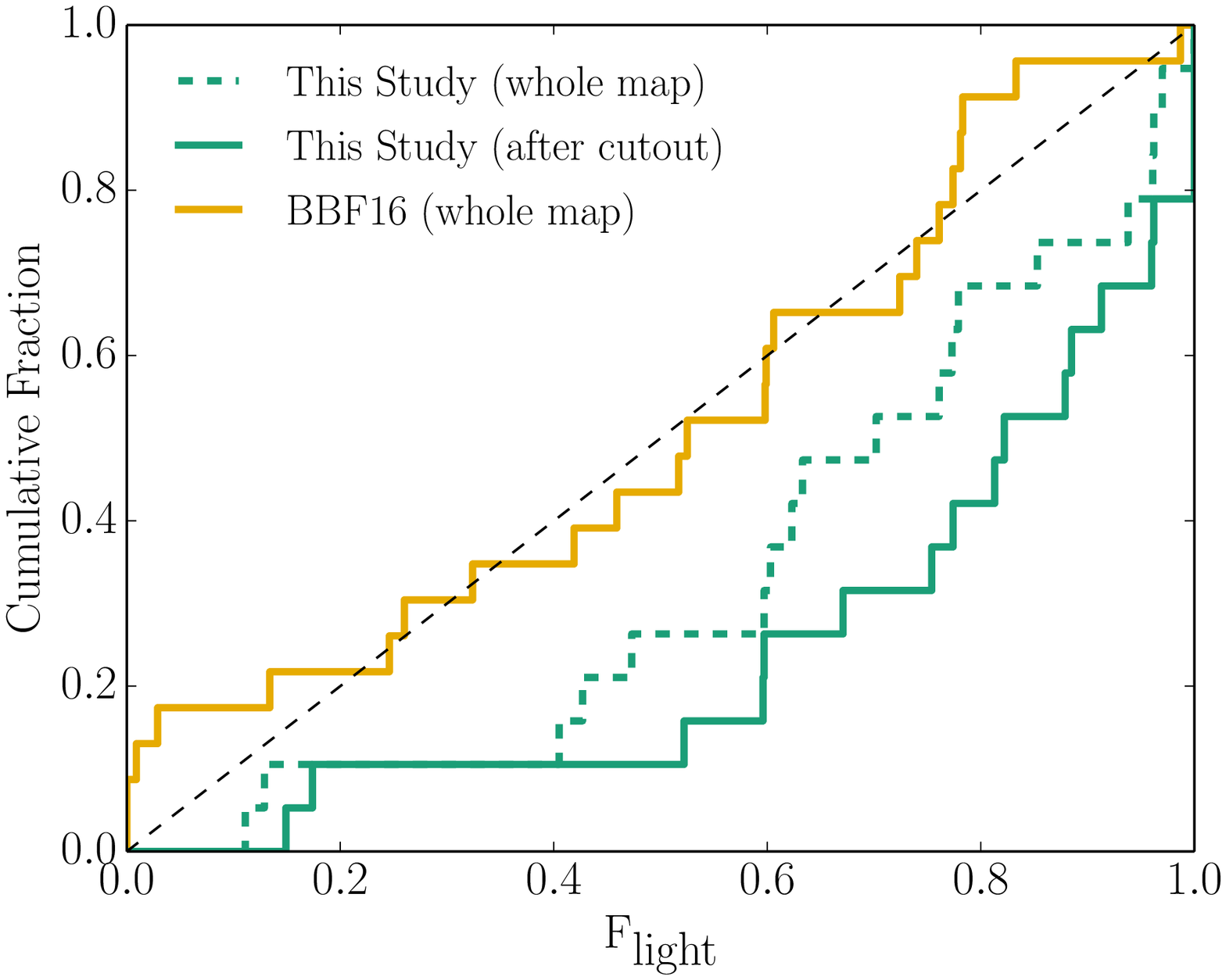}
 \caption{{\em Top:} An example \fl{} heatmap from \cref{fig:segmaps} shown including {\em (left)} and excluding {\em (right)} a cut-out of radius $0.5 \times r_{50}$ (appearing as black) centred on the host. Markers have the same meaning as \cref{fig:segmaps}. {\em Bottom:} Cumulative distribution of \fl{} calculated only for bursts with offsets $> 0.5 \times r_{50}$ when including the whole map and after cutting out the central regions, also shown is the distribution of \citetalias{blanchard16}. We find LGRBs at large offsets trace the brighter regions of their hosts, following the behaviour seen for the entire sample (\cref{fig:flight}), this effect becomes more pronounced when accounting for the offset threshold of the LGRBs by applying a cut-out to the heatmap.}
 \label{fig:flight_offset}
\end{figure}

\begin{figure}
\centering
 \includegraphics[width=0.45\columnwidth]{GRB090618_drz_0.065_0.8_flheat.eps}
 \includegraphics[width=0.45\columnwidth]{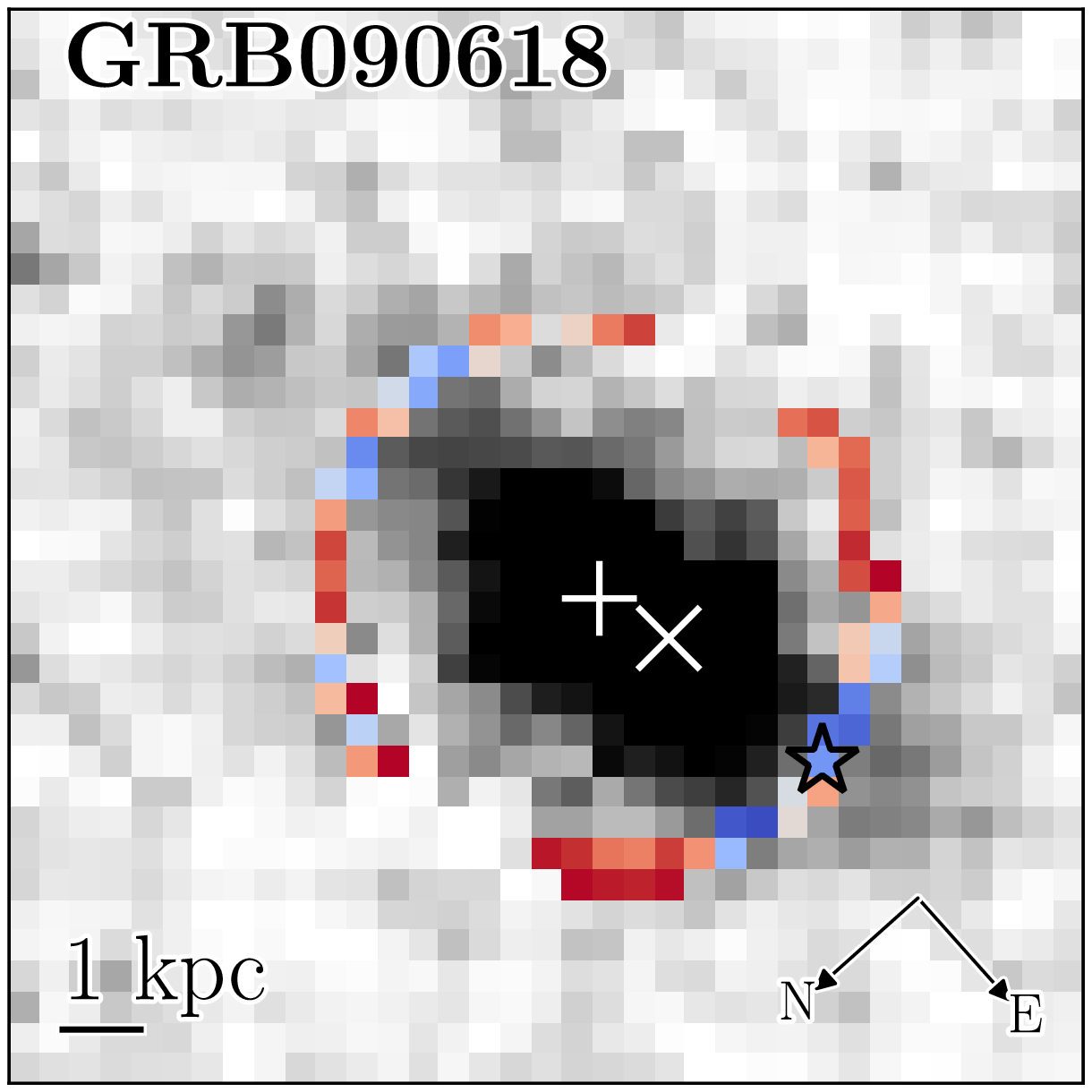}
 \includegraphics[width=\columnwidth]{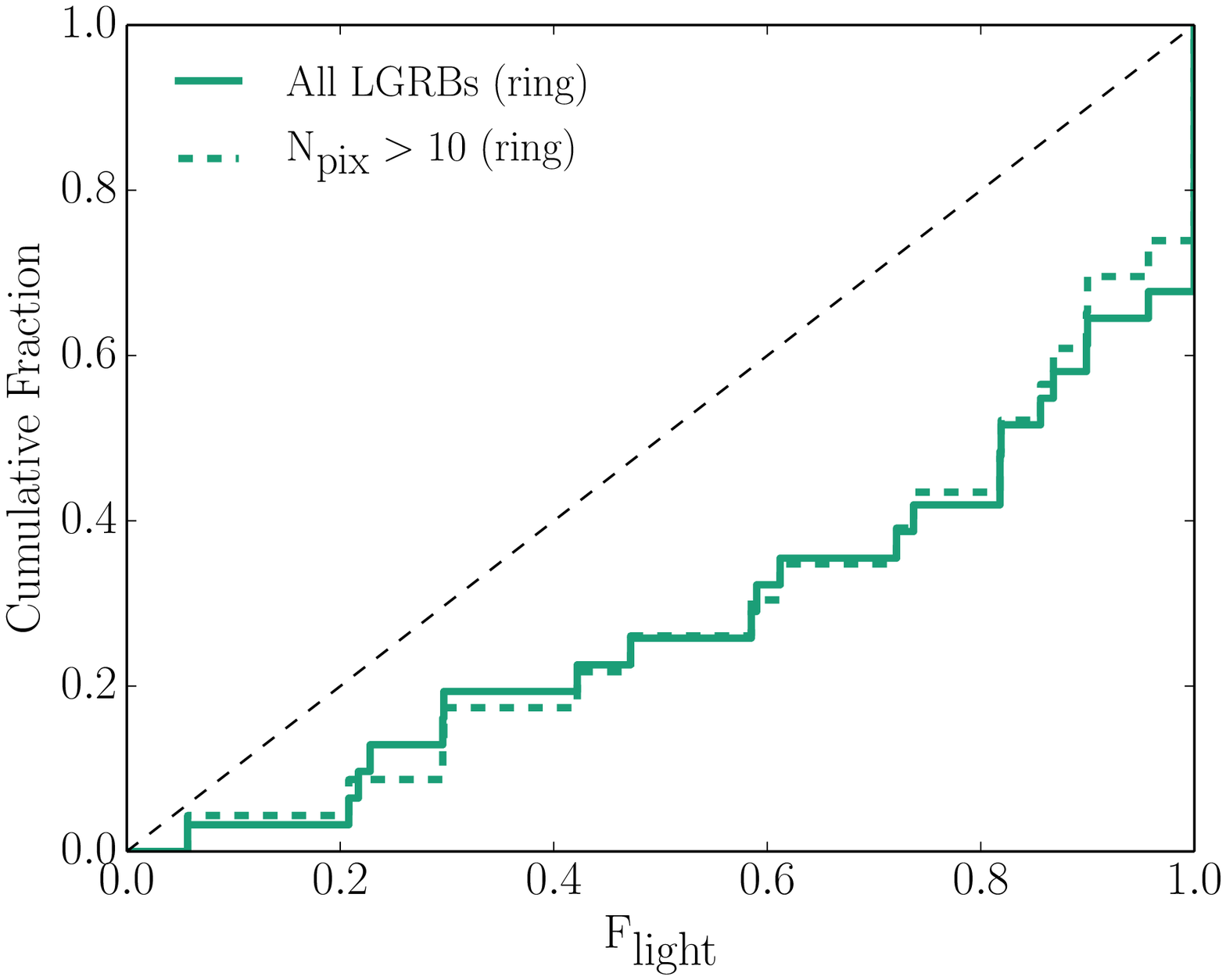}
 \caption{{\em Top:} An example \fl{} heatmap from \cref{fig:segmaps} shown for the whole pixel distribution {\em (left)} and for a `ring' of pixels that are approximately the same apparent offset as the LGRB location {\em (right)}. Markers have the same meaning as \cref{fig:segmaps}. {\em Bottom:} Cumulative distribution of \fl{} calculated for bursts with when including only the `ring' of pixels at similar offset to the LGRB location. For any given offset, LGRBs explode on the brighter regions of their hosts. In order to minimise the effects of discretisation of the \fl{} statistic, we also plot only those values where $> 10$ pixels make up the `ring' distribution, this is indistinguishable from the entire sample.}
 \label{fig:flight_ring}
\end{figure}

To address this, we recalculate the \fl{} values for our offset $> 0.5 \times r_{50}$ bursts after excluding a circular region with radius $0.5 \times r_{50}$ that is centred on the host barycentre.\footnote{The masking was done with whole pixels, where a pixel was excluded if its centre was $< 0.5 \times r_{50}$ from the barycentre of the host.} These new \fl{} values are then a measure of the degree of association that the LGRBs have with the light in the outer regions of their hosts, which is the quantity needed to address the argument in \citetalias{blanchard16}. Our results and a representation of the method is shown in \cref{fig:flight_offset}. We find, as expected, that after accounting for the removal of the inner regions, our \fl{} distribution becomes more weighted to higher values. Indeed we find inconsistency ($p \sim 8\times10^{-4}$) with the uniform distribution -- the case that LGRBs linearly trace the underlying light of their hosts -- confirming that larger offset LGRBs preferentially explode on brighter outer regions of their hosts. Our raw distribution based on the entire pixel distribution of the hosts also appears weighted to higher \fl{} values than the uniform distribution at lower significance, as opposed to the almost uniform distribution found by \citetalias{blanchard16}. We perform a direct comparison of \fl{} results for the overlapping sample between this study and \citetalias{blanchard16} in \cref{sect:flight_compare}, and discuss potential factors that may contribute to this discrepancy. To further investigate the \fl{} distribution of LGRBs we have calculated values based only on the light distribution of pixels at the same offset as the LGRB (\cref{fig:flight_ring}). This gives a direct handle on the association of an LGRB to the morphological features of its host that exist at the same offset. Similar to the whole \fl{} maps (\cref{fig:flight}) and the cut-outs (\cref{fig:flight_offset}), we find LGRBs preferentially explode on the brighter regions of their hosts that are at a similar offset to the LGRB, for any given offset (inconsistent with a uniform distribution at $p \sim 6\times10^{-5}$). This result remains when excluding those LGRBs for which the `ring' pixel distribution is less than 10 pixels, where significant discretisation of the \fl{} statistic will occur.

This strong association shows that LGRBs preferentially occur on bright regions even when significantly offset, as expected if LGRBs arise from very massive stars \citep[e.g.][]{fruchter06}. We note that we use a single radius and that inclination is not accounted for in both prescriptions described above. As such, positions at the same apparent offset from the hosts' centres will represent a spread of intrinsic, deprojected distances. This should not induce a systematic effect in the absence of a preferred orientation of the hosts.

Splitting our \fl{} distribution at $z = 1$, we find marginal evidence ($p = 0.03$) that the low redshift sample is less associated to their host galaxies' light distributions than the high redshift sample (\cref{fig:flight_zsplit}). This would indicate a higher degree of association with relatively younger stellar populations (since the rest-frame wavelength is bluer for the high-redshift sample) rather than with the mass distribution of their hosts (more appropriately traced by redder wavelengths, i.e. for our observations of the $z < 1$ sample). A caveat to consider is that when splitting by distance, individual pixels are probing different physical scales of the hosts. However, aside from the three very low redshift events (GRBs 060218, 060505 and 060614), the extrema of the angular scales are 5.77 ($z = 0.4903$) and 8.10 ($z = 1.608$) kpc~arcsec$^{-1}$, meaning there is only a modest change in the linear pixel size over the majority of the redshift range. After excluding the three very low redshift GRBs, the AD test gives $p = 0.06$ and we cannot thus conclusively state the two populations of our sample differ significantly. The effects of resolution and depth of imaging on such ranked pixel-based statistics are discussed in \citet{kangas16} for more local galaxies. 

For one of our sample, GRB 060729, an extremely long lasting X-ray afterglow was found. It was suggested by \citet{grupe10} that the progenitor must be in a low density environment to explain this long-lived emission. Indeed, we find that the location of GRB 060729 is located on the outskirts of its host, in a faint region -- \fl{} = 0.11, the lowest of our sample.\footnote{Furthermore, the `cut-out' and `ring' \fl{} values (see \cref{sect:discusssites}) are similarly low, 0.15 and 0.21, respectively.} Additionally, the absolute surface luminosity of the explosion site and intrinsic neutral hydrogen column density also mark GRB 060729 as having exploded in a particularly faint and relatively low-density environment compared to the rest of our sample (see \cref{sect:surflum,fig:NH_lum}), in support of the prediction \citep[see also discussion of comparatively remote LGRB environments by][]{decia11}.

\subsubsection{Surface luminosities and N$_\textrm{H}$}
\label{sect:surflum}

\begin{figure}
\centering
 \includegraphics[width=\columnwidth]{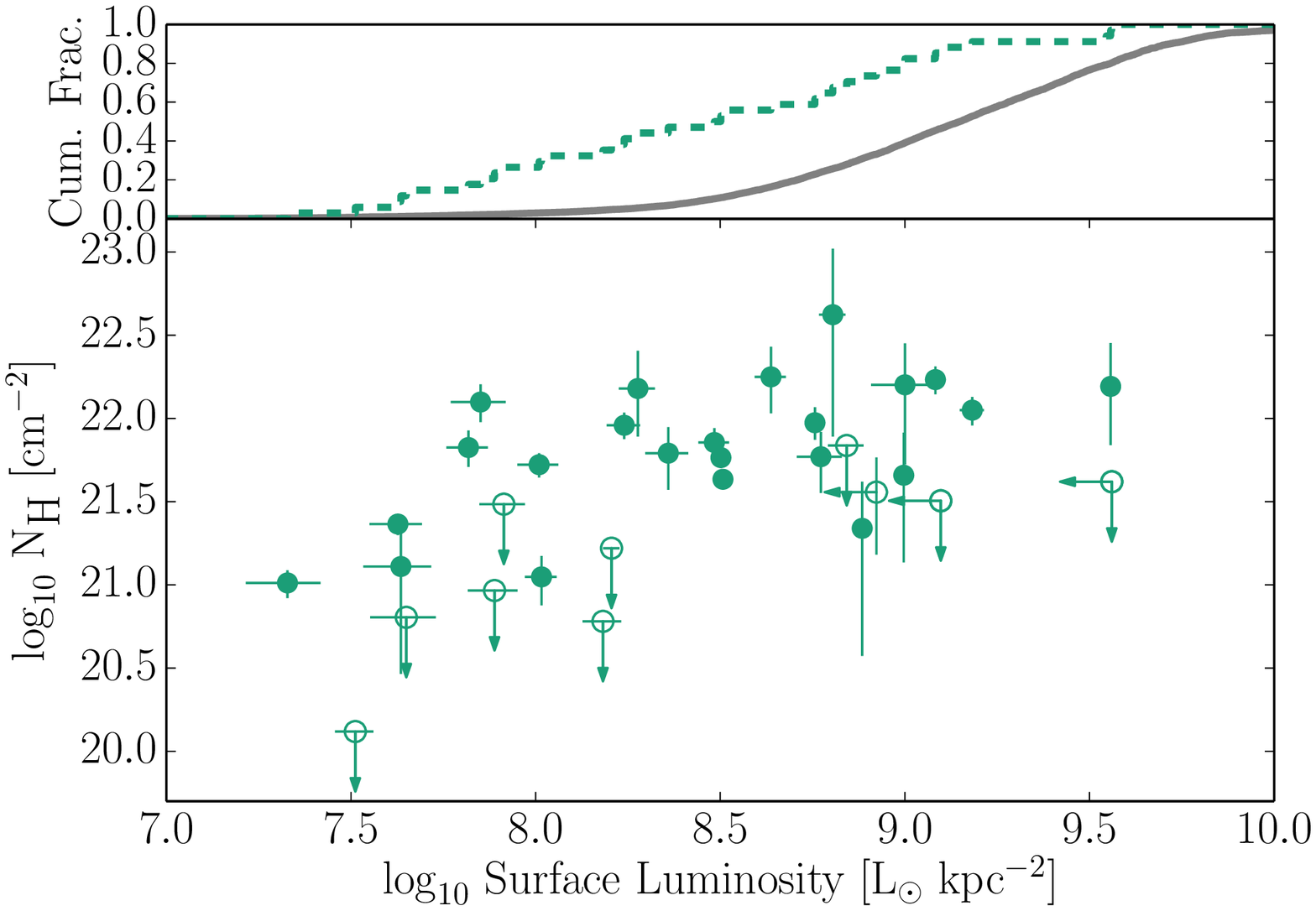}
 \caption{{\em Bottom:} The N$_{\text H}$ values determined from GRBs plotted against the surface luminosity of their explosion site. Upper limits of surface luminosities and/or N$_{\text H}$ are shown as open circles. {\em Top:} The cumulative distribution of the observed LGRB explosion sites (dashed) and that expected from a LGRB population exploding in the most luminous hosts (M$_{\text{F160W}/(1+z)} < -22$) assuming they trace their host light distribution in a similar fashion (see text).}
 \label{fig:NH_lum}
\end{figure}

We have additionally calculated surface luminosities of the hosts at the GRB explosion sites (for which we have accurately located the burst) in order to compare them not only relatively to the individual hosts' light distributions, but also absolutely against each other. The surface luminosity of the GRB explosion sites were determined by converting the pixel flux into units of solar luminosity and then accounting for the physical size of the pixel at the distance of the GRB host. The conversion into solar luminosities was performed using a linearly interpolated broadband filter SED of the Sun, which was sampled at the rest-frame wavelength of the host observation based on the redshift of the host. Uncertainties are quoted for the Poisson uncertainty on the pixel value but do not include uncertainties due to redshift estimates or systematics due to transforming the observations to a monochromatic effective wavelength (which is dependant on the unknown underlying spectral shape). Upper limits for the surface luminosities at the location of our undetected hosts (\cref{sect:undetectedhosts}) were found by taking the 1$\sigma$ background level luminosity. These values are compared to intrinsic neutral hydrogen column density values, N$_\textrm{H}$,  determined by {\em Swift}, where possible. The X-ray column density values were obtained following the general method of \citet{starling13}, with significant refinements in selection criteria and statistical analysis (McGuire et al. in preparation). Values are given in \cref{tab:siteprops}. We find a positive correlation (\cref{fig:NH_lum}; Spearman's rank $= 0.529$,  $p = 9.5\times10^{-3}$, excluding limits.) between the two -- i.e. LGRBs in more luminous regions have higher column densities. This appears to be an incarnation of the Schmidt-Kennicutt law. \citet{wang15} showed that N$_\textrm{H}$ is well correlated to $\Sigma_\textrm{SFR}$, the surface SFR, averaged over the burst hosts. In \cref{fig:NH_lum} this is shown for the explosion sites themselves, taking higher surface luminosity regions as being indicative of a larger presence of younger, massive stars. We note that our initial findings here should be further investigated based on rest-frame UV imaging of the explosion sites, where the relation between luminosity and SFR is more clear-cut. We find that high N$_\textrm{H}$ values, expected for more intense regions of star formation, occur at a relatively wide range of surface luminosities - this may be at least partly due to extinction effects. However, bursts with lower (or only upper limits on) N$\textrm{H}$ ($\log_{10}\textrm{N}_\textrm{H} \simlt 21$~cm$^{-2}$) are almost all located on fainter regions. The sharp drop off above $\log_{10}\textrm{N}_\textrm{H} \sim 22$~cm$^{-2}$ may be the result of a selection bias against more heavily extinguished bursts as our sample is comprised of LGRBs with determined redshifts, almost exclusively from their afterglows \citep[see also discussion in][]{jakobsson06c}. Indeed, a study of {\em Swift}-detected GRBs showed that those with redshift determinations are systematically offset to lower N$\textrm{H}$ values than those without \citep{fiore07}. We note that \citet{arabsalmani15} detected H{\sc i} in 21cm emission for the host galaxy of GRB 980425/SN 1998bw. These authors found, in accordance with our findings here, that the explosion site of the LGRB, nearby a very luminous SF region hosting a Wolf-Rayet population, is coincident with the highest column density region of the galaxy and that the column density inferred is typical of those found for cosmological LGRBs.

Given a positive relation between N$_\textrm{H}$ and surface luminosity, we investigated any potential impact on LGRB detectability in the most luminous galaxies, which seem to be under-represented in LGRB host samples (\cref{fig:absmag}). To address this we used the F160W data of the Cosmic Assembly Near-infrared Deep Extragalactic Legacy Survey \citep[CANDELS,][]{grogin11,koekemoer11} GOOD-S field to find luminous (M$_{\text{F160W}/(1+z)} < -22$) hosts with $z < 3$ in the catalogue of \citet{santini15}. After running our source detection method (\cref{sect:radphot}) on the CANDELS image\footnote{\url{http://candels.ucolick.org/data_access/GOODS-S.html}, v1.0.} we cross matched the luminous catalogue entries with our detections. For each source we determined potential LGRB explosion sites assuming they would follow the same host-normalised offset distribution (\cref{fig:offsetsnorm}); for every offset we found the pixel with \fl{} nearest to 0.8 (the median value for our sample, \cref{fig:flight,fig:flight_ring}) and calculated its surface luminosity. This overall distribution of surface luminosities of potential LGRB-explosion sites in luminous hosts is shown in \cref{fig:NH_lum}. Although weighted towards higher luminosities than our observed distribution (as expected), the vast majority lies within our observed range and, in particular, there is no significant tail at very high luminosities. This indicates that bursts exploding in the most luminous hosts would not be subject to a strong selection effect due to obscuration of the GRB (above that of those bursts in less luminous hosts), which could occur at exceptionally high column density values.

\section{Conclusions}
\label{sect:conclusions}
39 LGRB locations ($z < 3$) have been observed with a SNAPSHOT \hst{} program in F160W filter, of which we detect 35 hosts. We performed astrometric alignment of the bursts to study the explosion sites of 31 bursts within their host galaxy light profiles. From the study of these data we have found:

\begin{itemize}
 \item The detected LGRB host population is significantly fainter than a SFR-weighted field galaxy population over the same redshift range, indicating LGRBs are not unbiasedly tracing the star formation over this redshift range.
 \item When parametrising the hosts' morphologies by concentration ($C$) and asymmetry ($A$), we found the population to be mainly composed of galaxies that share properties with spiral-like and irregular galaxies, with a smaller contribution from elliptical-like and merging systems. We found evidence that LGRB hosts become more concentrated and less luminous at lower redshift, consistent with the cosmic downsizing of star formation.
 \item Our LGRB projected offsets confirm the centralised nature of these explosions, with a median offset of $1.0\pm0.3$~kpc, in agreement with previous work. LGRBs are more centrally concentrated than CCSNe and comparable to SLSNe.
 \item Our host $r_{80}$ distribution is larger than that found by \citetalias{svensson10}, in agreement with \citetalias{blanchard16}, indicating LGRB hosts are  intermediate in size between the hosts of SLSNe and CCSNe.
 \item We found that LGRBs are strongly biased towards exploding in bright regions of their hosts. This bias exists for LGRBs at all offsets (i.e. larger offset bursts preferentially explode on the brighter outer regions of their hosts).
 \item We found a correlation between the surface luminosities of the explosion sites and the column density towards the bursts.
\end{itemize}

\section*{Acknowledgements}

SFR values for the GOODS-MUSIC sample were kindly provided by Paola Santini. Sam Oates is thanked for helpful discussion.

JDL and AJL acknowledge support from the UK Science and Technology Facilities Council (grant ID ST/I001719/1) and the Leverhulme Trust, as part of a Philip Leverhulme Prize award.

Based on observations made with the NASA/ESA Hubble Space Telescope (programmes SNAP 12307 and GO 10909), obtained from the data archive at the Space Telescope Science Institute. STScI is operated by the Association of Universities for Research in Astronomy, Inc. under NASA contract NAS 5-26555. This work is based in part on observations made with the Spitzer Space Telescope (programme 80054), which is operated by the Jet Propulsion Laboratory, California Institute of Technology under a contract with NASA.
Based in part on observations collected at the European Organisation for Astronomical Research in the Southern Hemisphere under ESO programme(s) 075.D-0270(A), 076.D-0015(A), 077.D-0661(C), 077.D-0425(A), 077.D-0691(A), 078.D-0752(A), 078.D-0416(C), 080.A-0398(F) and 081.A-0856(BC).
Based in part on observations obtained at the Gemini Observatory acquired through the Gemini Science Archive, which is operated by the Association of Universities for Research in Astronomy, Inc., under a cooperative agreement with the NSF on behalf of the Gemini partnership: the National Science Foundation (United States), the National Research Council (Canada), CONICYT (Chile), Ministerio de Ciencia, Tecnolog\'{i}a e Innovaci\'{o}n Productiva (Argentina), and Minist\'{e}rio da Ci\^{e}ncia, Tecnologia e Inova\c{c}\~{a}o (Brazil). 
This work made use of data supplied by the UK Swift Science Data Centre at the University of Leicester.
The Liverpool Telescope is operated on the island of La Palma by Liverpool John Moores University in the Spanish Observatorio del Roque de los Muchachos of the Instituto de Astrofisica de Canarias with financial support from the UK Science and Technology Facilities Council. 
Based on observations made with the Nordic Optical Telescope, operated by the Nordic Optical Telescope Scientific Association at the Observatorio del Roque de los Muchachos, La Palma, Spain, of the Instituto de Astrofisica de Canarias.
The WHT is operated on the island of La Palma by the Isaac Newton Group in the Spanish Observatorio del Roque de los Muchachos of the Instituto de Astrofísica de Canarias. 
This work is based in part on observations taken by the CANDELS Multi-Cycle Treasury Program with the NASA/ESA HST, which is operated by the Association of Universities for Research in Astronomy, Inc., under NASA contract NAS5-26555.

\bibliographystyle{mnras}
\bibliography{/home/jdl/references}

\appendix

\section{Host data}
\label{sect:appendix}

In \cref{tab:hostprops} we show the properties of the detected GRB hosts in our sample. \cref{tab:siteprops} contains our results on the explosion sites of the LGRBs, including the \fl{} statistic.

\begin{landscape}
\begin{table}
\centering
\begin{threeparttable}
 \caption{Properties of the LGRB hosts detected with \sex{}}
 \begin{tabular}{lrccccccccc}
GRB          & $r_{20}$                            & $r_{50}$                          & $r_{80}$                            & Plate scale & $E(B-V)$& Host abs mag\tnote{a} & $\sigma_{\text{mag}}$  & $A$\tnote{b}      & $C$\tnote{c}                      & $P_{\text{ch}}$\tnote{d} \\ 
             & (kpc)                               & (kpc)                             & (kpc)                               & (kpc/arcsec)& (mag)   & (mag)                   & (mag)                &                   &                                   &         \\ 
\hline                                                                                                                                                                                                                                                           
050315       & $ 0.85\substack{+0.06 \\ -0.02}$  & $ 1.72\substack{+0.18 \\ -0.06}$   & $ 2.83\substack{+0.63 \\ -0.08}$    & 8.02      & 0.06    & $-20.621 $           & 0.06                 &$  0.22\pm 0.09$   & $  2.60\substack{+0.51 \\ -0.00}$   & 0.0022  \\
050401       & $ 0.62\substack{+0.11 \\ -0.03}$  & $ 1.16\substack{+0.35 \\ -0.07}$   & $ 2.31\substack{+1.15 \\ -0.04}$    & 7.43      & 0.07    & $-19.907 $           & 0.18                 &$  0.23\pm 0.16$   & $  2.85\substack{+0.59 \\ -0.12}$   & 0.0035  \\
050824       & $ 1.39\substack{+0.21 \\ -0.08}$  & $ 3.08\substack{+0.56 \\ -0.19}$   & $ 5.74\substack{+2.10 \\ -0.61}$    & 7.26      & 0.03    & $-18.959 $           & 0.14                 &$  0.11\pm 0.24$   & $  3.08\substack{+0.33 \\ -0.26}$   & 0.0106  \\
051016B      & $ 0.86\substack{+0.01 \\ -0.02}$  & $ 1.66\substack{+0.00 \\ -0.07}$   & $ 2.85\substack{+0.00 \\ -0.14}$    & 7.53      & 0.05    & $-20.741 $           & 0.01                 &$  0.16\pm 0.03$   & $  2.61\substack{+0.03 \\ -0.09}$   & 0.0134  \\
060124       & $ 0.85\substack{+0.08 \\ -0.19}$  & $ 1.55\substack{+0.24 \\ -0.35}$   & $ 2.66\substack{+0.49 \\ -0.79}$    & 7.84      & 0.14    & $-18.844 $           & 0.26                 &$  0.15\pm 0.27$   & $  2.47\substack{+0.56 \\ -0.28}$   & 0.0060  \\
060218       & $ 0.18\substack{+0.00 \\ -0.00}$  & $ 0.36\substack{+0.00 \\ -0.01}$   & $ 0.65\substack{+0.00 \\ -0.02}$    & 0.64      & 0.16    & $-15.958 $           & 0.01                 &$  0.20\pm 0.01$   & $  2.82\substack{+0.00 \\ -0.05}$   & 0.0024  \\
060502A      & $ 0.61\substack{+0.21 \\ -0.03}$  & $ 1.17\substack{+0.66 \\ -0.07}$   & $ 2.25\substack{+1.25 \\ -0.03}$    & 8.09      & 0.04    & $-17.949 $           & 0.20                 &$  0.28\pm 0.13$   & $  2.84\substack{+0.46 \\ -0.25}$   & 0.0031  \\
060505       & $ 0.88\substack{+0.00 \\ -0.02}$  & $ 1.99\substack{+0.00 \\ -0.05}$   & $ 3.70\substack{+0.00 \\ -0.12}$    & 1.59      & 0.02    & $-20.015 $           & 0.01                 &$  0.16\pm 0.01$   & $  3.13\substack{+0.00 \\ -0.04}$   & 0.0077  \\
060602A      & $ 1.36\substack{+0.08 \\ -0.07}$  & $ 3.24\substack{+0.24 \\ -0.25}$   & $ 6.29\substack{+0.74 \\ -0.40}$    & 7.14      & 0.03    & $-19.727 $           & 0.04                 &$  0.12\pm 0.14$   & $  3.32\substack{+0.21 \\ -0.10}$   & 0.0070  \\
060614       & $ 0.40\substack{+0.01 \\ -0.00}$  & $ 0.79\substack{+0.03 \\ -0.01}$   & $ 1.46\substack{+0.15 \\ -0.01}$    & 2.14      & 0.02    & $-16.398 $           & 0.02                 &$  0.13\pm 0.06$   & $  2.80\substack{+0.21 \\ -0.01}$   & 0.0051  \\
060729       & $ 0.80\substack{+0.05 \\ -0.03}$  & $ 1.69\substack{+0.20 \\ -0.06}$   & $ 3.23\substack{+0.57 \\ -0.15}$    & 6.09      & 0.06    & $-18.163 $           & 0.07                 &$  0.13\pm 0.14$   & $  3.04\substack{+0.35 \\ -0.11}$   & 0.0047  \\
060912A      & $ 2.64\substack{+0.00 \\ -0.18}$  & $ 4.82\substack{+0.00 \\ -0.28}$   & $ 7.50\substack{+0.00 \\ -0.69}$    & 7.53      & 0.05    & $-21.412 $           & 0.00                 &$  0.07\pm 0.11$   & $  2.27\substack{+0.05 \\ -0.18}$   & 0.0114  \\
061007       & $ 1.61\substack{+0.12 \\ -0.07}$  & $ 3.00\substack{+0.49 \\ -0.11}$   & $ 6.48\substack{+1.34 \\ -0.56}$    & 7.98      & 0.02    & $-20.013 $           & 0.05                 &$  0.21\pm 0.12$   & $  3.03\substack{+0.29 \\ -0.18}$   & 0.0065  \\
061110A      & $ 0.75\substack{+0.01 \\ -0.14}$  & $ 1.64\substack{+0.00 \\ -0.53}$   & $ 2.94\substack{+0.00 \\ -0.91}$    & 7.04      & 0.10    & $-17.670 $           & 0.15                 &$  0.24\pm 0.22$   & $  2.97\substack{+0.24 \\ -0.48}$   & 0.0053  \\
070318       & $ 0.78\substack{+0.05 \\ -0.02}$  & $ 1.38\substack{+0.12 \\ -0.03}$   & $ 2.28\substack{+0.53 \\ -0.02}$    & 7.29      & 0.02    & $-18.588 $           & 0.05                 &$  0.09\pm 0.08$   & $  2.33\substack{+0.49 \\ -0.01}$   & 0.0024  \\
070521       & $ 0.93\substack{+0.05 \\ -0.04}$  & $ 1.95\substack{+0.10 \\ -0.12}$   & $ 3.91\substack{+0.61 \\ -0.36}$    & 7.96      & 0.03    & $-21.540 $           & 0.03                 &$  0.12\pm 0.07$   & $  3.11\substack{+0.37 \\ -0.18}$   & 0.1000  \\
071010A      & $ 0.58\substack{+0.01 \\ -0.15}$  & $ 1.29\substack{+0.00 \\ -0.49}$   & $ 2.66\substack{+0.02 \\ -1.23}$    & 7.62      & 0.11    & $-17.717 $           & 0.47                 &$  0.14\pm 0.13$   & $  3.29\substack{+0.00 \\ -0.88}$   & 0.0036  \\
071010B      & $ 0.85\substack{+0.03 \\ -0.01}$  & $ 1.63\substack{+0.08 \\ -0.04}$   & $ 2.78\substack{+0.26 \\ -0.07}$    & 7.55      & 0.01    & $-20.255 $           & 0.01                 &$  0.08\pm 0.03$   & $  2.57\substack{+0.24 \\ -0.04}$   & 0.0016  \\
071112C      & $ 1.22\substack{+0.24 \\ -0.02}$  & $ 2.54\substack{+0.80 \\ -0.00}$   & $ 4.91\substack{+1.82 \\ -0.00}$    & 7.24      & 0.12    & $-18.771 $           & 0.14                 &$ -0.05\pm 0.24$   & $  3.03\substack{+0.39 \\ -0.00}$   & 0.0071  \\
071122       & $ 1.33\substack{+0.00 \\ -0.10}$  & $ 2.66\substack{+0.00 \\ -0.31}$   & $ 4.96\substack{+0.04 \\ -0.73}$    & 7.86      & 0.05    & $-20.562 $           & 0.03                 &$  0.06\pm 0.10$   & $  2.86\substack{+0.08 \\ -0.41}$   & 0.0038  \\
080319C      & $ 0.75\substack{+0.12 \\ -0.16}$  & $ 1.23\substack{+0.24 \\ -0.27}$   & $ 1.86\substack{+0.59 \\ -0.55}$    & 8.02      & 0.03    & $-19.367 $           & 0.05                 &$  0.28\pm 0.08$   & $  1.99\substack{+0.86 \\ -0.10}$   & 0.0026  \\
080430       & $ 1.18\substack{+0.01 \\ -0.33}$  & $ 2.42\substack{+0.21 \\ -0.71}$   & $ 5.24\substack{+0.03 \\ -1.93}$    & 7.07      & 0.01    & $-18.010 $           & 0.14                 &$  0.21\pm 0.32$   & $  3.24\substack{+0.15 \\ -0.47}$   & 0.0090  \\
080520       & $ 1.94\substack{+0.06 \\ -0.05}$  & $ 3.66\substack{+0.18 \\ -0.05}$   & $ 5.86\substack{+0.41 \\ -0.00}$    & 8.09      & 0.08    & $-21.727 $           & 0.03                 &$  0.37\pm 0.09$   & $  2.41\substack{+0.21 \\ -0.00}$   & 0.0092  \\
080605       & $ 0.80\substack{+0.00 \\ -0.02}$  & $ 1.50\substack{+0.00 \\ -0.07}$   & $ 2.66\substack{+0.00 \\ -0.17}$    & 8.10      & 0.14    & $-21.731 $           & 0.01                 &$  0.15\pm 0.03$   & $  2.60\substack{+0.08 \\ -0.08}$   & 0.0012  \\
080707       & $ 0.99\substack{+0.03 \\ -0.04}$  & $ 1.96\substack{+0.10 \\ -0.07}$   & $ 3.71\substack{+0.54 \\ -0.18}$    & 7.96      & 0.10    & $-20.615 $           & 0.04                 &$  0.17\pm 0.07$   & $  2.86\substack{+0.29 \\ -0.06}$   & 0.0021  \\
080805       & $ 1.04\substack{+0.04 \\ -0.03}$  & $ 2.16\substack{+0.10 \\ -0.11}$   & $ 3.98\substack{+0.38 \\ -0.27}$    & 8.09      & 0.05    & $-20.775 $           & 0.06                 &$  0.14\pm 0.07$   & $  2.91\substack{+0.21 \\ -0.12}$   & 0.0040  \\
080916A      & $ 0.78\substack{+0.01 \\ -0.03}$  & $ 1.54\substack{+0.02 \\ -0.06}$   & $ 2.68\substack{+0.10 \\ -0.13}$    & 6.78      & 0.02    & $-19.582 $           & 0.01                 &$  0.08\pm 0.03$   & $  2.67\substack{+0.14 \\ -0.06}$   & 0.0017  \\
081007       & $ 0.77\substack{+0.17 \\ -0.02}$  & $ 1.42\substack{+0.54 \\ -0.00}$   & $ 3.11\substack{+0.63 \\ -0.15}$    & 6.01      & 0.02    & $-17.026 $           & 0.11                 &$  0.16\pm 0.19$   & $  3.02\substack{+0.26 \\ -0.22}$   & 0.0051  \\
081121       & $ 0.74\substack{+0.08 \\ -0.04}$  & $ 1.35\substack{+0.16 \\ -0.10}$   & $ 2.10\substack{+0.42 \\ -0.16}$    & 7.71      & 0.05    & $-19.824 $           & 0.10                 &$  0.11\pm 0.12$   & $  2.26\substack{+0.63 \\ -0.07}$   & 0.0078  \\
090418A      & $ 0.83\substack{+0.01 \\ -0.04}$  & $ 1.56\substack{+0.02 \\ -0.10}$   & $ 2.50\substack{+0.07 \\ -0.19}$    & 8.10      & 0.05    & $-20.366 $           & 0.03                 &$  0.15\pm 0.05$   & $  2.39\substack{+0.15 \\ -0.10}$   & 0.0018  \\
090424       & $ 1.32\substack{+0.00 \\ -0.02}$  & $ 2.38\substack{+0.00 \\ -0.04}$   & $ 3.79\substack{+0.02 \\ -0.08}$    & 6.09      & 0.03    & $-20.577 $           & 0.01                 &$  0.19\pm 0.02$   & $  2.29\substack{+0.03 \\ -0.03}$   & 0.0032  \\
090618       & $ 1.38\substack{+0.09 \\ -0.04}$  & $ 2.74\substack{+0.38 \\ -0.06}$   & $ 5.31\substack{+1.31 \\ -0.00}$    & 6.07      & 0.09    & $-19.198 $           & 0.03                 &$  0.13\pm 0.11$   & $  2.93\substack{+0.44 \\ -0.00}$   & 0.0091  \\
091127       & $ 1.09\substack{+0.08 \\ -0.02}$  & $ 2.20\substack{+0.24 \\ -0.00}$   & $ 3.99\substack{+0.88 \\ -0.00}$    & 5.77      & 0.04    & $-18.818 $           & 0.03                 &$  0.13\pm 0.10$   & $  2.82\substack{+0.40 \\ -0.00}$   & 0.0057  \\
091208B      & $ 0.85\substack{+0.09 \\ -0.27}$  & $ 1.58\substack{+0.29 \\ -0.50}$   & $ 3.07\substack{+0.40 \\ -1.26}$    & 7.76      & 0.06    & $-17.116 $           & 0.24                 &$  0.19\pm 0.40$   & $  2.77\substack{+0.24 \\ -0.55}$   & 0.0063  \\
\hline
\end{tabular}
\label{tab:hostprops}
\begin{tablenotes}
 \item [a]{Absolute magnitude of host at $\lambda = 15400/(1+z)$~\AA{} where M$_\lambda = m_\textrm{F160W} - \mu + 2.5\log_{10}(1+z)$. Corrected for Milky Way dust extinction following \citet{schlafly11}.}
 \item [b]{Host asymmetry (\cref{sect:hostmorph}).}
 \item [c]{Host concentration (\cref{sect:hostmorph}).}
 \item [d]{The probability of a chance alignment of the galaxy and burst, see \cref{sect:hostassignment}.}
\end{tablenotes}
\end{threeparttable}
\end{table}
\end{landscape}

\begin{table*}
\centering
\begin{threeparttable}
 \caption{Properties of the LGRBs explosion sites}
 \begin{tabular}{lcccccc}
\hline
GRB         & rms\tnote{a} & Offset$_{\text{cen}}$ & Offset$_{\text{bright}}$ & \fl{} & Surface Lum                       & $\log_{10}$N$_\text{H}$ \\
            & (mas)        & (kpc)                 & (kpc)                    &       & ($\log_{10}$L$_\odot$kpc$^{-2}$)  & cm$^{-2}$ \\
\hline
050315   & 41           & 0.17                & 0.60                 & 0.91     &  $ 9.18\substack{+0.03 \\ -0.03}$  & $ 22.05\substack{+0.08 \\ -0.09}$ \\
050401   & 39           & 1.02                & 1.06                 & 0.13     &  $ 9.00\substack{+0.08 \\ -0.09}$  & $ 22.20\substack{+0.25 \\ -0.48}$ \\
050824   & 60           & 3.33                & 4.07                 & 0.62     &  $ 7.89\substack{+0.06 \\ -0.07}$  & $<20.97                         $ \\
051016B  & 440          & ---                 & ---                  & ---      &  ---                               & ---                               \\
060124   & 88           & 0.93                & 0.69                 & 0.96     &  $ 8.77\substack{+0.06 \\ -0.07}$  & $ 21.77\substack{+0.15 \\ -0.22}$ \\
060218   & 29           & 0.10                & 0.10                 & 0.86     &  $ 8.51\substack{+0.01 \\ -0.02}$  & $ 21.63\substack{+0.06 \\ -0.06}$ \\
060502A  & 39           & 0.44                & 0.46                 & 0.91     &  $ 8.36\substack{+0.05 \\ -0.06}$  & $ 21.79\substack{+0.16 \\ -0.22}$ \\
060505   & 48           & 6.72                & 6.84                 & 0.60     &  $ 8.21\substack{+0.02 \\ -0.02}$  & $<21.22                         $ \\
060602A  & 76           & 1.21                & 1.35                 & 0.78     &  $ 8.12\substack{+0.05 \\ -0.05}$  & ---                               \\
060614   & 40           & 0.77                & 0.71                 & 0.47     &  $ 7.51\substack{+0.05 \\ -0.05}$  & $<20.12                         $ \\
060729   & 14           & 2.27                & 2.48                 & 0.11     &  $ 7.33\substack{+0.09 \\ -0.11}$  & $ 21.01\substack{+0.08 \\ -0.09}$ \\
060912A  & 33           & 4.76                & 4.93                 & 0.60     &  $ 8.18\substack{+0.05 \\ -0.06}$  & $<20.78                         $ \\
061007   & 80           & 2.11                & 0.78                 & 0.97     &  $ 8.48\substack{+0.04 \\ -0.04}$  & $ 21.86\substack{+0.09 \\ -0.10}$ \\
061110A  & 38           & 0.82                & 0.82                 & 0.77     &  $ 7.91\substack{+0.06 \\ -0.07}$  & $<21.48                         $ \\
070318   & 72           & 1.12                & 0.56                 & 0.77     &  $ 8.24\substack{+0.04 \\ -0.05}$  & $ 21.96\substack{+0.08 \\ -0.08}$ \\
070521   & ---          & ---                 & ---                  & ---      &  ---                               & ---                               \\
071010A  & 21           & 0.34                & 0.18                 & 0.89     &  $ 8.28\substack{+0.05 \\ -0.05}$  & $ 22.18\substack{+0.23 \\ -0.29}$ \\
071010B  & 53           & 0.66                & 0.71                 & 0.89     &  $ 8.88\substack{+0.02 \\ -0.02}$  & $ 21.34\substack{+0.28 \\ -0.77}$ \\
071031   & 25           & ---                 & ---                  & ---      &  $<9.56$                           & $<21.62                         $ \\
071112C  & 21           & 1.76                & 1.78                 & 0.43     &  $ 7.65\substack{+0.08 \\ -0.10}$  & $<20.81                         $ \\
071122   & 63           & 0.59                & 0.81                 & 0.96     &  $ 8.71\substack{+0.03 \\ -0.03}$  & ---                               \\
080319C  & 21           & 0.47                & 0.26                 & 0.85     &  $ 8.84\substack{+0.05 \\ -0.05}$  & $<21.84                         $ \\
080430   & 39           & 1.26                & 0.44                 & 0.96     &  $ 8.01\substack{+0.05 \\ -0.06}$  & $ 21.72\substack{+0.07 \\ -0.08}$ \\
080520   & 73           & 4.94                & 5.73                 & 0.78     &  $ 8.80\substack{+0.03 \\ -0.04}$  & $ 22.62\substack{+0.40 \\ -0.73}$ \\
080603B  & 82           & ---                 & ---                  & ---      &  $<9.55$                           & ---                               \\
080605   & 32           & 0.91                & 0.67                 & 0.85     &  $ 9.56\substack{+0.02 \\ -0.02}$  & $ 22.19\substack{+0.26 \\ -0.35}$ \\
080707   & 23           & 0.53                & 0.24                 & 1.00     &  $ 9.00\substack{+0.02 \\ -0.02}$  & $ 21.66\substack{+0.26 \\ -0.52}$ \\
080710   & 25           & ---                 & ---                  & ---      &  $ 7.64\substack{+0.08 \\ -0.10}$  & $ 21.11\substack{+0.26 \\ -0.65}$ \\
080805   & 45           & 3.48                & 3.97                 & 0.63     &  $ 8.64\substack{+0.04 \\ -0.04}$  & $ 22.25\substack{+0.18 \\ -0.22}$ \\
080916A  & 49           & 0.14                & 0.15                 & 1.00     &  $ 8.76\substack{+0.02 \\ -0.02}$  & $ 21.97\substack{+0.09 \\ -0.10}$ \\
080928   & 29           & ---                 & ---                  & ---      &  $<8.92$                           & $ 21.56\substack{+0.21 \\ -0.38}$ \\
081007   & 33           & 1.01                & 0.42                 & 0.94     &  $ 7.82\substack{+0.05 \\ -0.06}$  & $ 21.83\substack{+0.10 \\ -0.12}$ \\
081008   & 23           & ---                 & ---                  & ---      &  $<9.10$                           & $<21.51                         $ \\
081121   & 179          & ---                 & ---                  & ---      &  ---                               & ---                               \\
090418A  & 55           & 0.74                & 0.37                 & 0.93     &  $ 9.08\substack{+0.03 \\ -0.03}$  & $ 22.23\substack{+0.08 \\ -0.09}$ \\
090424   & 30           & 2.09                & 2.33                 & 0.70     &  $ 8.50\substack{+0.03 \\ -0.03}$  & $ 21.76\substack{+0.06 \\ -0.06}$ \\
090618   & 48           & 3.47                & 2.45                 & 0.41     &  $ 7.63\substack{+0.07 \\ -0.08}$  & $ 21.37\substack{+0.06 \\ -0.07}$ \\
091127   & 41           & 2.01                & 1.69                 & 0.76     &  $ 8.02\substack{+0.04 \\ -0.05}$  & $ 21.05\substack{+0.13 \\ -0.17}$ \\
091208B  & 34           & 0.79                & 1.23                 & 1.00     &  $ 7.85\substack{+0.07 \\ -0.08}$  & $ 22.10\substack{+0.11 \\ -0.12}$ \\
\hline
\end{tabular}
\label{tab:siteprops}
\begin{tablenotes}
 \item [a]{Total positional uncertainty of the LGRB: the star-matched geometric alignment in each axis and afterglow centroiding uncertainties added in quadrature.}
\end{tablenotes}
\end{threeparttable}
\end{table*}

\section{Direct comparison of \fl{} values}
\label{sect:flight_compare}

\begin{figure*}
\centering
\includegraphics[width=\linewidth]{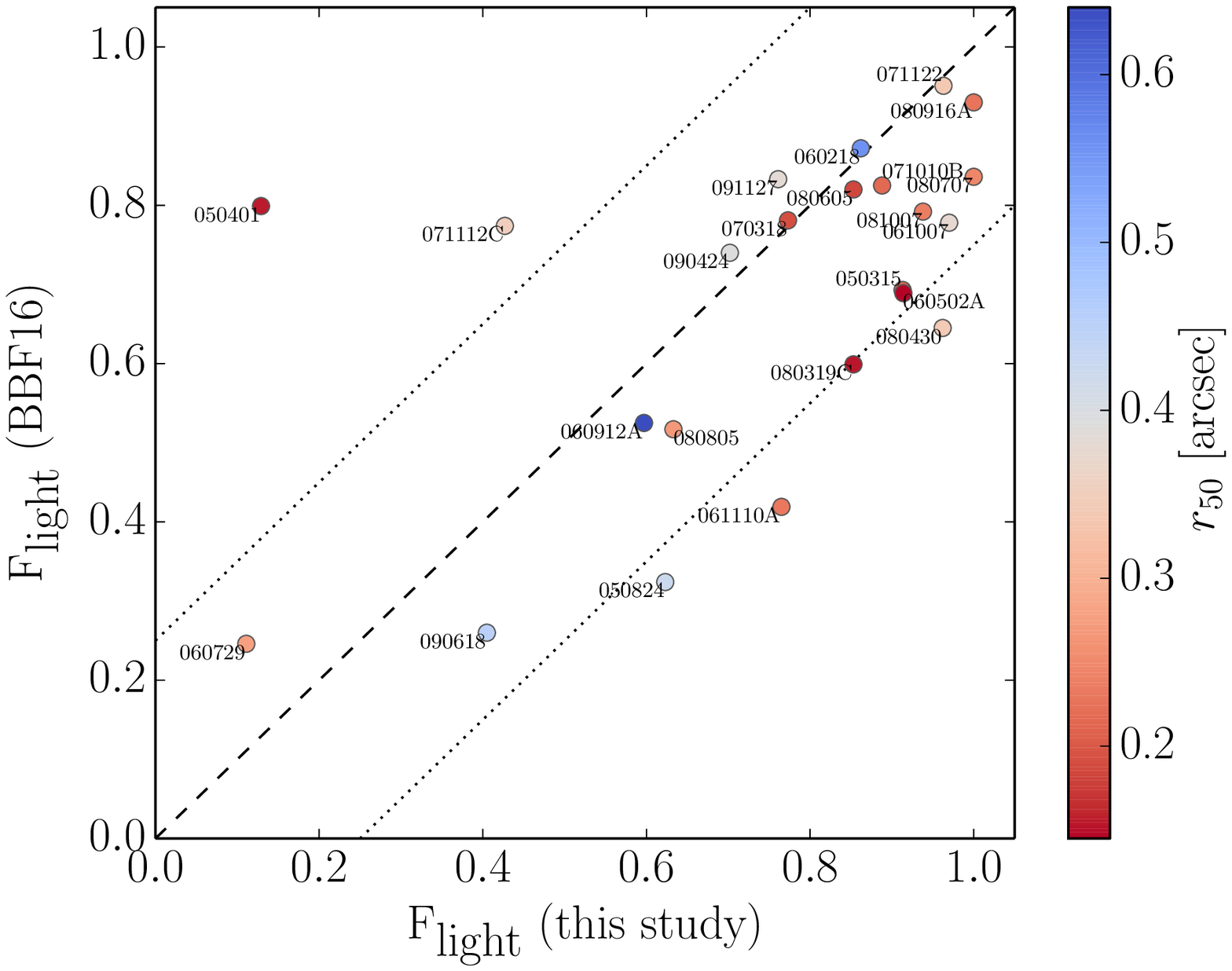}
\caption{Direct comparison of \fl{} determined here and in \citetalias{blanchard16} for the overlapping sample where measurements were made in F160W. The 1:1 relation is plotted with a black dashed line, and black dotted lines denote differences between the studies of 0.3 in \fl{}. A linear regression fit (grey solid line) is given by an intercept and slope of -0.044 and 0.898, respectively. GRBs are labelled, with their colour coding set by their $r_{50}$ value.} 
 \label{fig:blanchard_fl_compare}
\end{figure*}

Recently, \citetalias{blanchard16} have found that the degree of association of LGRBs to the brightest regions of their hosts is less than what has previously been suggest \citepalias{fruchter06, svensson10}, in particular for larger offset LGRBs. In this study we have found a very strong association, in support of previous works and at odds with the distribution of \citetalias{blanchard16}. Here we present a comparison of the \fl{} values determined for LGRBs in this study and that of \citetalias{blanchard16}, where both measurements were made in the F160W filter. We note that the choice of drizzling parameters for the data reduction were the same in each study and we consider only those events where the burst was well localised in each study. In \cref{fig:blanchard_fl_compare} we plot \fl{} values from each study for the overlapping events. Although there is a correlation between the values, there is a tendency to larger \fl{} values in this study compared to those of \citetalias{blanchard16} (for 16/23 events we find a larger \fl{} value). Additionally, there are notable exceptions. Six events have a difference in \fl{} of $\geq 0.25$ between the studies, the causes of which we assess by visually inspecting our \sex{} segmentation maps and the highlighted explosion location with the plots of \citetalias{blanchard16}. 

For GRB 050401 we find a relatively low \fl{} of 0.13, with the LGRB location being just offset from the bright central region of the host.\footnote{We note that the VLT/FORS afterglow image we used for astrometric alignment is slightly under-sampled, and may constitute another source of positional uncertainty.} Although difficult to tell, it appears the location adopted in \citetalias{blanchard16} is more centred on the host giving rise to a larger \fl{}.
GRB 050824 is located close to a bright compact region of the more extended host in each study.
For GRB 061110A, we find a central location, slightly offset west, compared to the somewhat more western offset adopted in \citetalias{blanchard16}, explaining the larger \fl{} value we find. 
The locations for GRB 071112C appear to agree very well, although \citetalias{blanchard16} find a larger \fl{} value. Our location is close to high \fl{} value pixels, but it is unfortunately difficult to assess the precise location adopted in \citetalias{blanchard16}. 
Similar locations for GRB 080319C were determined in each study based on GEMINI/GMOS-N afterglow imaging. The compact nature of the host results in large changes of \fl{} with only a small shift in pixels.
Our location of GRB 080430 is on the brightest knot of the galaxy light. The location in \citetalias{blanchard16} appears to also be centred on this knot however they find a lower \fl{} value. Again, the apparent size of the host means a shift of only a few pixels can drastically alter the \fl{} value. 

A common theme, which has been shown visually in \cref{fig:segmaps,fig:segmaps_noalign}, is the large change in \fl{} over the space of only a few pixels with such distant and thus small apparent-size hosts. The larger apparent-size hosts have \fl{} values that agree very well between the studies. Small differences in adopted explosion locations in the \fl{} distributions of these hosts do not make such drastic changes as the distributions is more smoothly varying with respect to pixel size (see \cref{fig:segmaps}). The bursts with large differences in the \fl{} values are compact hosts -- even in the case of GRB 050824, the large change in \fl{} is the result of the burst being located close to a bright compact knot, mimicking the effect seen for the compact hosts.

\subsection{Weighted \fl{} statistic}
\label{sect:weightedfl}

\begin{figure}
\centering
 \includegraphics[width=0.8\columnwidth]{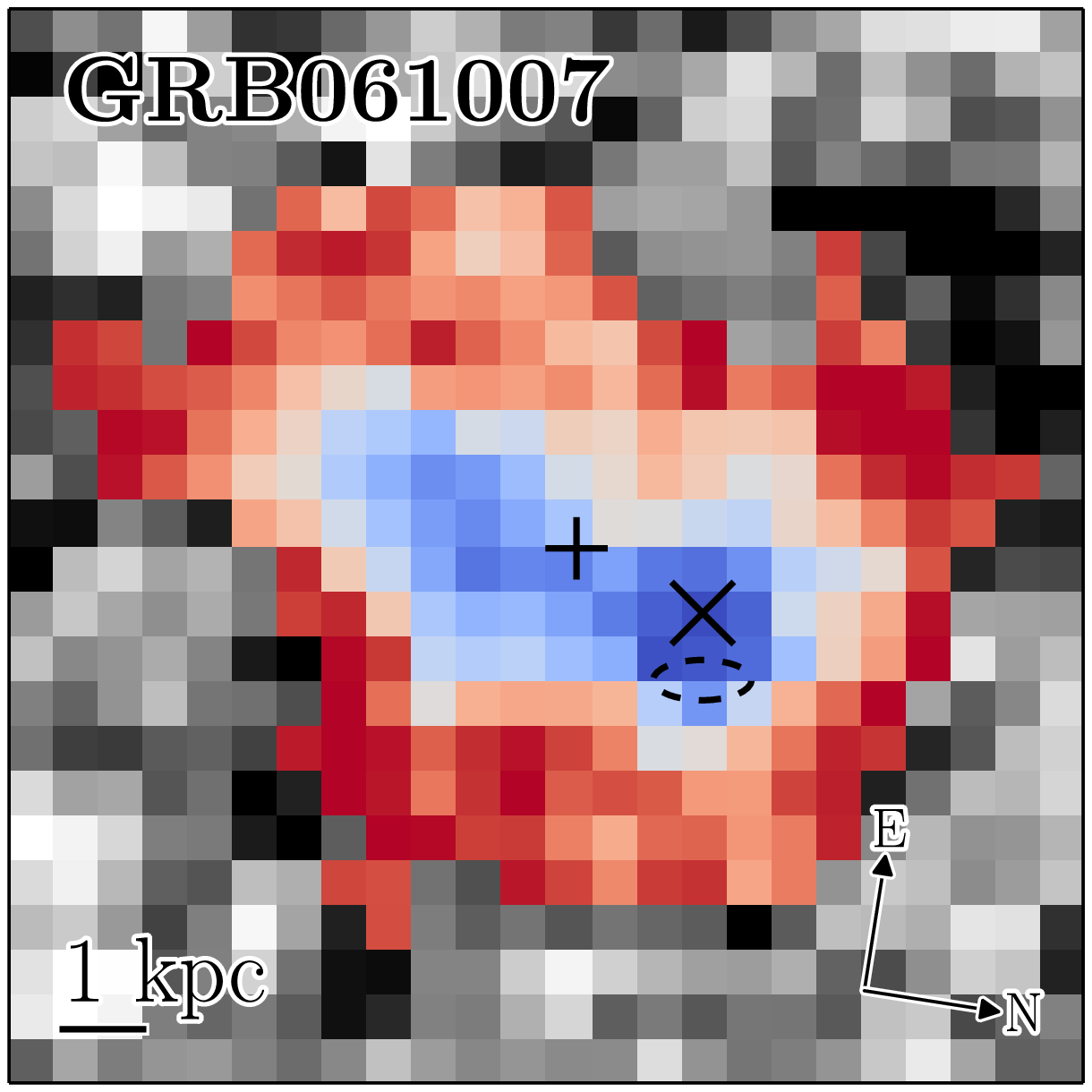}\\
 \includegraphics[width=\columnwidth]{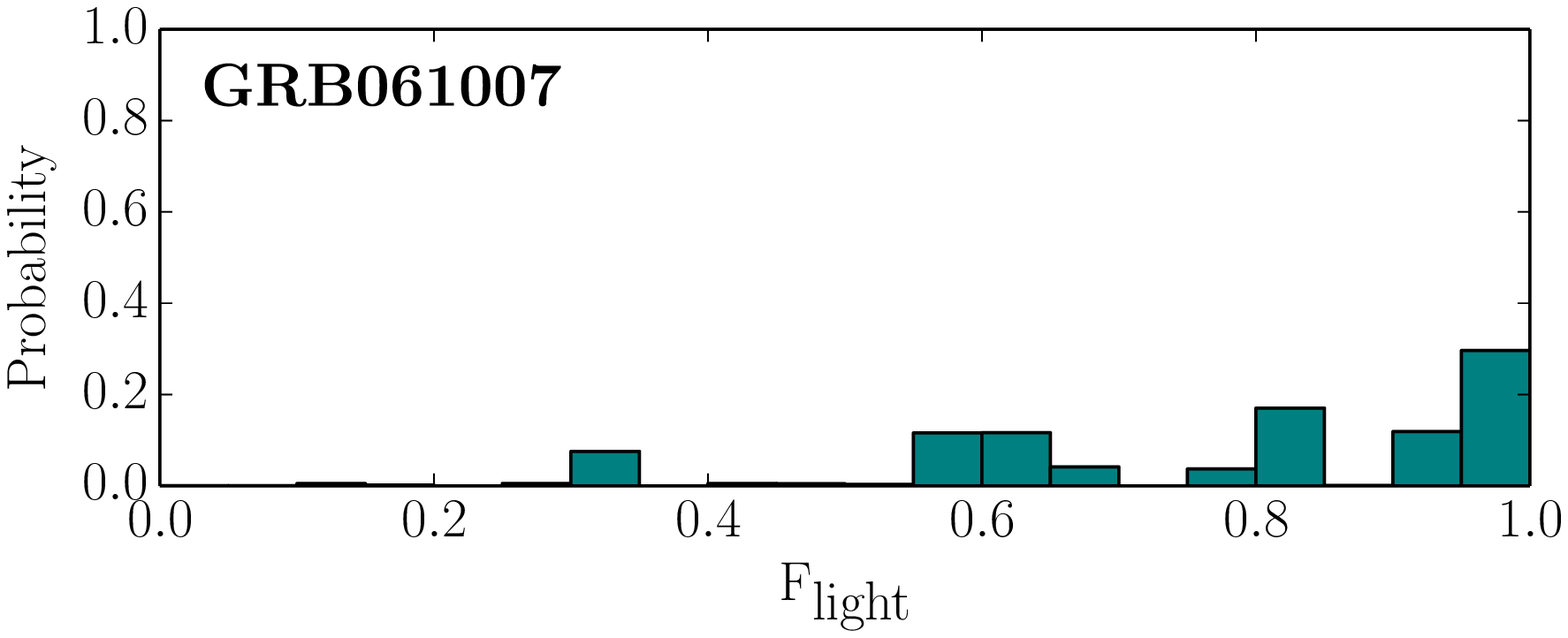}
 \caption{{\em Top:} An example \fl{} heatmap from \cref{fig:segmaps} where the localisation is poorer (notwithstanding GRBs 051016B, 070521 and 081121). The GRB localisation is indicated by the dashed ellipse. {\em Bottom:} The probability distribution of \fl{} values for this GRB (binned in intervals of 0.05). Although the value we determine in the main study (\fl{} = 0.97) is most probable, when considering the location uncertainty, there is significant probability of other \fl{} values.}
 \label{fig:flight_prob}
\end{figure}

\begin{figure}
\centering
 \includegraphics[width=\columnwidth]{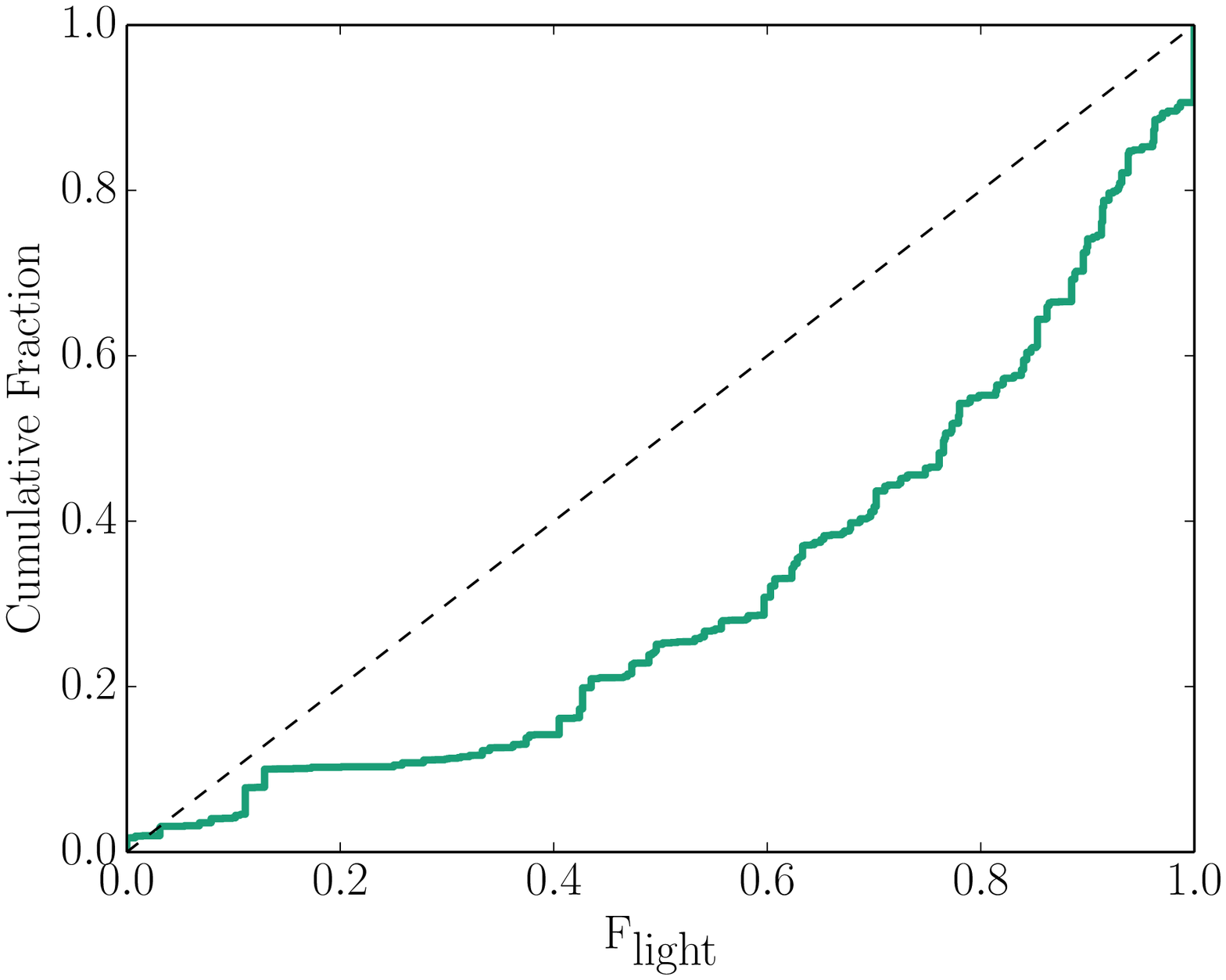}
 \caption{The {\em weighted} \fl{} distribution for LGRBs in this study. Every pixel within 3$\sigma$ of the location for each burst contributes an amount to the distribution determined by the integral of the 2D location uncertainty Gaussian over that pixel. The preference for LGRBs to explode on brighter regions of their hosts remains when using this method.}
 \label{fig:flight_weighted}
\end{figure}

Although in a statistical sense the uncertainties in the overall \fl{} distribution from the choices of the precise pixel value to use should be circumvented with large samples, a detailed investigation into the potential biases and uncertainties on such pixel-based statistics, in particular with respect to alignment uncertainties, is lacking.
When the location rms is less than that of the seeing in the images, one can consider the pixel distribution has already undergone a `smoothing' to account for this uncertainty and the \fl{} value of the pixel underlying the determined location can be taken (as is considered here and in previous works), however this may not be the best approach to determine \fl{} for a given location, which should rather be done in a probabilistic manner. \citetalias{fruchter06} and \citetalias{svensson10}  convolved their images with the positional uncertainty and selected the pixel \fl{} underlying their adopted location whereas \citet{kangas16} used a MC approach to estimate the effect of positional uncertainty on their pixel rank values. \citetalias{blanchard16} used a method for poorly localised bursts whereby the location is modelled as a 2D Gaussian centred on the best-guess location. Each pixel's \fl{} value is then given a probability of being the actual \fl{} value determined by the 2D Gaussians probability density function (pdf) integrated over that pixel. The \fl{} value is taken as the mean of this distribution, with an uncertainty given by the standard deviation. We have implemented a similar procedure (\cref{fig:flight_prob}) but do not reduce the probability distribution to a single \fl{} value and uncertainty (cf. \citetalias{blanchard16}), since the resulting probability distributions are not well described by a Gaussian distribution. Instead we allow every pixel in a 3$\sigma$ box centred on the determined locations to contribute to the LGRB cumulative \fl{} distribution by an amount given by its probability (\cref{fig:flight_weighted}). This was done for all bursts for which we calculated an \fl{} previously, since the procedure is also applicable for well-localised bursts (in the limit of zero uncertainty on the location, the pdf becomes a delta function and the procedure reduces to that described in \cref{sect:methods}).\footnote{We still do not include the three bursts (GRBs 051016B, 070521, 081121) for which we only have very poor localisation. As was noted by \citetalias{blanchard16}, even in the case of correctly accounting for the relative probability of each pixel's contribution, the \fl{} value is of little use with such poor localisation since most or all of the host is included, as well as significant portions outside the \sex{} segmentation map, which weights the \fl{} distribution to zero.}
Using this weighted \fl{} statistic we confirm the preference for LGRBs to explode on the brighter regions of their hosts; $\sim 50$~per~cent of the total summed location pdf lies on the brightest $\sim 20$~per~cent of the hosts. Thus, since our alignments are generally quite accurate, we are not particularly sensitive to the choice of method for determining \fl{}. The use of such methods becomes necessary for more poorly localised events, so long as the uncertainty is small relative to the apparent size of the host system..

\label{lastpage}
\end{document}